\documentclass[preprints,article,accept,pdftex,moreauthors]{Definitions/mdpi}

\firstpage{1}
\pubvolume{1}
\issuenum{1}
\articlenumber{0}
\pubyear{2025}
\copyrightyear{2025}
\datereceived{ }
\daterevised{ } 
\dateaccepted{ }
\datepublished{ }
\hreflink{https://doi.org/} 

\usepackage{dcolumn}

\DeclareUnicodeCharacter{2212}{-}

\Title{FIMP Dark Matter in Bulk Viscous Non-Standard Cosmologies}

\TitleCitation{Title}

\Author{Esteban González $^{1,\ddagger}$\orcidA{}, Carlos Maldonado $^{2,*,\ddagger}$\orcidB{}, N. Stefanía Mite $^{1,\ddagger}$ \orcidC{}, and Rodrigo Salinas $^{1,\ddagger}$\orcidD{}}

\AuthorNames{Firstname Lastname, Firstname Lastname and Firstname Lastname}

\isAPAStyle{%
       \AuthorCitation{Lastname, F., Lastname, F., \& Lastname, F.}
         }{%
        \isChicagoStyle{%
        \AuthorCitation{Lastname, Firstname, Firstname Lastname, and Firstname Lastname.}
        }{
        \AuthorCitation{Lastname, F.; Lastname, F.; Lastname, F.}
        }
}

\address{%
$^{1}$ \quad Departamento de Física, Universidad Católica del Norte, Avenida Angamos 0610, Casilla 1280, Antofagasta 1270709, Chile; esteban.gonzalez@ucn.cl (E.G.); nelly.mite@ucn.cl (N.S.M.); rodrigo.salinas@alumnos.ucn.cl (R.S.)\\
$^{2}$ \quad Facultad de Medicina y Ciencia, Universidad San Sebasti\'an, Lago Panguipulli 1390, Puerto Montt 5501842, Chile; carlos.maldonado@uss.cl}

\corres{Correspondence: carlos.maldonado@uss.cl}

\secondnote{These authors contributed equally to this work.}

\abstract{In this paper, we revisit the extension of the classical non-standard cosmological model in which dissipative processes are considered through a bulk viscous term in the new field $\phi$, which interacts with the radiation component during the early universe. Specifically, we consider an interaction term of the form $\Gamma_{\phi} \rho_{\phi}$, where $\Gamma_{\phi}$ represents the decay rate of the field and $\rho_{\phi}$ denotes its energy density and a bulk viscosity described by $\xi=\xi_{0}\rho_{\phi}^{1/2}$, within the framework of Eckart's theory. This extended non-standard cosmology is employed to explore the parameter space for the production of Feebly Interacting Massive Particles (FIMPs) as Dark Matter candidates, assuming a constant thermal averaged Dark Matter production cross-section ($\langle\sigma v\rangle$), as well as a preliminary analysis of the non-constant case. In particular, for certain combinations of the model and Dark Matter parameters, namely ($T_\text{end}$,$\kappa$) and $(m_\chi,\langle\sigma v\rangle)$, where $T_\text{end}$ corresponds to the temperature at which $\phi$ decays, $\kappa$ is the ratio between the initial energy density of $\phi$ and radiation, and $m_\chi$ is the Dark Matter mass, we identify extensive new parameter regions where Dark Matter can be successfully established while reproducing the currently observed relic density, in contrast to the predictions of $\Lambda$CDM and classical non-standard cosmological scenarios.}


\begin{document}

\section{Introduction}\label{sec:Introduction}
Since the publication of the results obtained by the Supernova Search Team \cite{SupernovaSearchTeam:1998fmf} and the Supernova Cosmology Project \cite{SupernovaCosmologyProject:1998vns}, which consistently provided evidence that the universe is currently undergoing an accelerating expansion phase, the $\Lambda$CDM model has become the most successful cosmological framework for describing the evolution of the universe and fitting observational cosmological data. This model aligns well with various cosmological observations, including Type Ia supernovae \cite{Scolnic:2021amr}, measurements of the Hubble parameter \cite{Moresco:2012jh,Zhang:2012mp,Moresco:2015cya}, baryon acoustic oscillations \cite{SDSS:2005xqv}, the cosmic microwave background \cite{WMAP:2012fli,Planck:2018vyg}, among others. Regardless of its success, the model faces significant observational challenges, such as the $H_{0}$ tension, where local measurements of the Hubble constant $H_{0}$ using Cepheid host distance anchors (model-independent) show a discrepancy of $5\sigma$ with the value inferred from the Planck CMB assuming the $\Lambda$CDM model \cite{Riess:2021jrx}. This tension is further supported by the H0LiCOW collaboration, which finds a discrepancy of $5.3\sigma$ with the value inferred from the Planck CMB \cite{Wong:2019kwg}. From a more fundamental point of view, the nature and evolution of Dark Matter (DM) and Dark Energy (DE) remain open questions without definitive answers. The former, an as-yet-undetermined non-baryonic component of the universe, is estimated to be approximately five times more abundant than ordinary matter \cite{Planck:2018vyg}.

All we know about DM is that, if it is a particle, it must be dark, neutral, non-baryonic matter, non-relativistic (Cold DM or CDM), and stable. Within this context, various theoretical frameworks propose potential DM candidates with these characteristics, such as Supersymmetry \cite{Jungman:1995df} and string theory \cite{King:2006cu}. Some of these DM particles are generally classified into two categories: Weakly Interacting Massive Particles (WIMPs) \cite{Steigman:1984ac, Bertone:2004pz, Arcadi:2017kky, Roszkowski:2017nbc, Arcadi:2024ukq, Singh:2024wdn}, and Feebly Interacting Massive Particles (FIMPs) \cite{Bernal:2017kxu, Chu:2011be, Hall:2009bx}.

WIMPs are massive particles, possibly in the GeV-TeV range. They are characterized by their interaction only through the electroweak force and gravity. The genesis mechanism for WIMPs in the early universe is the freeze-out process. This occurs when a non-negligible initial DM particle population is in thermal equilibrium with the Standard Model (SM) bath. As the universe expands and cools, the interactions between DM and SM particles gradually decrease, eventually ceasing and leaving behind the current observed DM relic density, with a value of $\Omega_{c}h^{2}=0.12$ according to Planck \cite{Planck:2018vyg}. Understanding DM interactions is fundamental to identifying the most feasible methods for their detection. Among the most well-known experimental approaches to search for DM particles are direct detection \cite{DUNE:2020lwj}, indirect detection \cite{Feng:2010gw}, and collider experiments \cite{LHCb:2008vvz}. To be consistent with observations within the $\Lambda$CDM framework, the total thermally averaged DM production cross-section for this category must be approximately $\langle\sigma v\rangle_0 = \text{few} \times 10^{-9}$ GeV \cite{Steigman:2012nb}. Unfortunately, no favorable results have been obtained to date in the search for WIMP DM, which has led to the exploration of alternative production mechanisms for these elusive particles. In this context, one of the most recently studied mechanisms is freeze-in, which gives rise to the second category of DM candidates, the FIMPs.

FIMPs are massive particles in the keV-TeV range. These particles interact so weakly with SM particles that they never reach thermal equilibrium with the surrounding medium in the early universe. Consequently, these particles undergo the Freeze-In process to establish their population. In this mechanism, the DM abundance is gradually built up as the universe cools, until the production ceases, leaving behind the current observed DM relic density. These DM candidates can emerge from various particle physics models that require the introduction of new symmetries beyond the SM. One example is the Higgs portal mechanism, which incorporates a scalar field as the DM component, stabilized by a $Z_2$ odd symmetry to prevent its decay. Alternatively, in scenarios where the DM candidate is fermionic, the associated vector boson corresponds to a gauge boson of a hidden gauge symmetry, such as $SU(2)$ \cite{Bernal:2017kxu}. Another well-motivated DM candidate is the pseudo-Nambu-Goldstone boson, exemplified by Majoron DM, which arises when the $U(1)_{B-L}$ symmetry is spontaneously broken \cite{Frigerio:2011in}. Additional scenarios include the spontaneous breaking of $Z_4$ symmetry within an inverse seesaw model, leading to neutrino mass generation alongside a FIMP DM candidate \cite{Wang:2024qhe}, or the implementation of the Froggatt-Nielsen mechanism via the introduction of a $U(1)_{FN}$ symmetry \cite{Babu:2023zni}. Even though the feeble interactions of FIMPs make it challenging to detect them with current experimental instruments, they are of special interest because these candidates, which are not being generated through thermal equilibrium, allow us to explore a less restricted parameter space compared to WIMPs. Following this purpose, many studies suggest that these candidates could be found in colliders and direct detection experiments as they can couple to long-lived particles due to the feeble couplings or may also be produced from the decay of heavy particles \cite{Junius:2022vzl,Barman:2024nhr,Belanger:2023azf, Blinnikov:1982eh, Blinnikov:1983gh, Khlopov:1989fj}. Furthermore, due to the mass range of FIMPs, they can be the origin of Warm DM (WDM) \cite{DEramo:2020gpr,McDonald:2015ljz,Klasen:2013ypa,Hall:2009bx,Frigerio:2011in,deGiorgi:2022yha,Kuo:2018fgw}. The possibility of this type of DM has an impact on the structure formation in the universe. While the CDM can explain the observed structure on scales above $\sim 1 Mpc$, it faces challenges in explaining small-scale structure observations \cite{Bode:2000gq,deVega:2011si}, such as the missing satellite problem \cite{Klypin:1999uc,Moore:1999nt}. This issue can be alleviated by WDM instead of a CDM \cite{Weinberg:2013aya}, a possibility that is supported by many studies related to the number of satellites in the Milky Way and small halos with dwarf galaxies \cite{deVega:2011si,Viel:2005qj,Newton:2020cog}. Therefore, FIMPs are more versatile than WIMPs, as they are less constrained and can alleviate other cosmological tensions.

The DM relic density depends on the era in which its abundance is set. For instance, in the $\Lambda$CDM model, it is assumed that this relic density is established during the radiation-dominated era. Some interesting cosmological scenarios include an additional field ($\phi$) in the early universe (scalar or fermion), which generates different domination eras due to the decay of the field into SM particles, leading to entropy injection into the SM bath. This behavior expands the parameter space available for detecting DM particles and can modify DM production if it occurs during these new domination eras. These alternative scenarios, known as Non-Standard Cosmologies (NSCs), offer new avenues for DM detection and may re-open parameter spaces previously excluded in the $\Lambda$CDM model, potentially revealing viable DM candidates and mechanisms within these non-standard frameworks. Therefore, if DM is experimentally detected, its particle physics properties and characteristics—such as mass, interactions with SM particles, and coupling constants—must be determined in a way that is consistent with the $\Lambda$CDM framework. Conversely, it is necessary to explore new NSC scenarios where these quantities align with the cosmological model.

In the effort to address tensions within the $\Lambda$CDM model, one approach is to consider that this model is a particular approximation of a more general cosmological framework. In this context, a natural extension is to incorporate non-perfect fluids, which offer a more realistic description of the cosmic medium \cite{Maartens:1996vi}. This is especially relevant given that, in the $\Lambda$CDM model, all matter components of the universe are typically modeled as perfect fluids. When non-perfect fluids are taken into account, phenomena like viscosity arise, whose effects play a crucial role in many cosmological processes, including the reheating of the universe, the decoupling of neutrinos from the cosmic plasma, nucleosynthesis, and others. Even more, the effects of viscosity are significant in various astrophysical mechanisms, such as the collapse of radiating stars into neutron stars or black holes, and the accretion of matter around compact objects \cite{Maartens:1996vi}. More recently and following the main idea, viscosity has been proposed as an alternative to mitigate some of the recent tensions in the $\Lambda$CDM model. For instance, a decaying DM scenario increases the expansion rate relative to $\Lambda$CDM, potentially alleviating the $H_{0}$ tension \cite{Pandey:2019plg}, with other viable solutions for this tension considering viscosity studied in Refs. \cite{Yang:2019qza,DiValentino:2021izs,Normann:2021bjy} (for a comprehensive review of viscous cosmology in the early and late universe, see Ref. \cite{Brevik:2017msy}). It is important to highlight that these tensions (particularly the late-time tensions) are mentioned solely as a motivation to explore alternatives to the $\Lambda$CDM model by considering bulk viscous fluids. In this sense, we also include this extension in the early universe to study its impact on FIMP dark matter production as a potential dark matter candidate. Nonetheless, it is important to note that our model is not incompatible with these late-time approaches. Therefore, one could develop a comprehensive model that could alleviate the issues of the $\Lambda$CDM model at both early and late times.

Working with non-perfect fluids requires a relativistic thermodynamic theory out of equilibrium, such as the Eckart \cite{Eckart:1940zz, PhysRev.58.919} or Israel-Stewart \cite{1977RSPSA.357...59S, 1979RSPSA.365...43I} formalisms. Although Eckart's theory is non-causal \cite{Israel:1976tn}, it remains widely studied due to its mathematical simplicity compared to the full Israel-Stewart formalism. Nevertheless, Eckart's approach serves as a convenient starting point for understanding dissipative effects in the universe, as the Israel-Stewart theory reduces to Eckart’s theory when the relaxation time for transient viscous effects is negligible \cite{Maartens:1995wt}. On the other hand, viscosity can manifest in two forms: shear viscosity and bulk viscosity. While the former can play a significant role in certain contexts \cite{Floerchinger:2014jsa}, our focus will be on bulk viscosity, which can emerge due to the presence of mixtures in the universe \cite{Zimdahl:1996fj}, the decaying of DM into relativistic particles \cite{Wilson:2006gf,Mathews:2008hk}, the energy transfer between CDM and the radiation fluid \cite{Hofmann:2001bi}, and might exist in a hidden sector, reproducing various observational properties of disk galaxies \cite{Foot:2014uba,Foot:2016wvj}. The recent advancements in gravitational wave detection have further opened up the possibility of observing dissipative effects in DM and DE through the dispersion and dissipation experienced by these waves as they propagate through a non-perfect fluid \cite{Goswami:2016tsu}. Additionally, bulk viscosity could contribute significantly to the emission of gravitational waves during neutron star mergers \cite{Alford:2017rxf}. It is interesting to remark that until now there has been no fundamental origin for bulk viscosity from particle physics; it is instead treated as an effective description.

Bulk viscous effects in the matter components of the universe have been deeply and widely studied in the literature, particularly about viscous DE components \cite{Nojiri:2004pf,Capozziello:2005pa,Nojiri:2005sr,Cataldo:2005qh,Brevik:2006wa,Cruz:2016rqi}. However, a more common approach is to study a DM component that undergoes dissipative processes during its cosmic evolution \cite{Velten:2012uv,Acquaviva:2014vga,Cruz:2017bcv,Cruz:2018yrr,Cruz:2022wme,Cruz:2022zxe,Gomez:2022qcu,Cruz:2023dzn}. This approach is of particular interest because it can account for the recent accelerated expansion of the universe without requiring an explicit DE component, leading to unified DM models \cite{Fabris:2005ts,Avelino:2008ph,Li:2009mf,Avelino:2010pb,Hipolito-Ricaldi:2010wrq,Velten_2011,Gagnon:2011id,Bruni:2012sn,Cruz:2017lbu,Cruz:2018psw,Cruz:2019uya}. Previous studies have explored dissipative stiff matter fluids within the framework of the full Israel-Stewart theory \cite{Mak:2003gw}. Furthermore, bulk viscous DM has been analyze in various contexts, including inflation \cite{Padmanabhan:1987dg,Barrow:1988yc,Maartens:1995wt,Maartens:1996dk}, interacting fluids \cite{Zimdahl:1996fj,Avelino:2013wea,Hernandez-Almada:2020ulm}, and modified gravity \cite{Brevik:2005ue,Brevik:2006wa}. These effects have also been studied concerning singularities such as the Big Rip and Little Rip, across both classical and quantum regimes \cite{Nojiri:2004pf,Nojiri:2005sr,Brevik:2005bj,Brevik:2006wa,Brevik:2008xv,Brevik:2010okp,Brevik:2010jv,Brevik:2011mm,Contreras:2015ooa,Contreras:2018two,Cruz:2021knz}. Further research has investigated the role of bulk viscosity in the radial oscillations of relativistic stars as well as its cosmological implications for Quark-Gluon plasma-filled universes \cite{Barta:2019tpv,BravoMedina:2019han}.

In Ref. \cite{Gonzalez:2024dtb} the authors propose a novel NSC scenario in which the early universe is dominated by two interacting fluids, namely the new field $\phi$ and radiation, by considering that $\phi$ experiences dissipative processes during their cosmic evolution in the form of a bulk viscosity. Working in the framework of Eckart's theory for non-perfect fluids, the authors consider an interaction term of the form $\Gamma_{\phi}\rho_{\phi}$, where $\Gamma_{\phi}$ represents the decay rate of the field and $\rho_{\phi}$ denotes its energy density, and a bulk viscosity described by $\xi=\xi_{0}\rho_{\phi}^{1/2}$. The latter has the advantage that, when the field fully decays, the dissipation is negligible and the standard $\Lambda$CDM cosmology is fully recovered. This novel NSC scenario was applied to the study of the parameter space for WIMP DM candidate production, showing that it is possible to explore even wider regions in which the model can account for the actual DM relic abundance in the universe. This result is obtained by comparing the classical NSC scenario with their bulk viscous counterpart, considering the same values for the NSC parameters, namely, $\kappa$ (ratio between the initial energy density of $\phi$ and radiation), $T_\text{end}$ (the temperature at which $\phi$ decays), $\xi_0$, and $\omega$ (the barotropic index of $\phi$); and for the WIMP DM candidate $m_\chi$ (the DM mass) and $\langle\sigma v\rangle$. From the study, it can be noted that in the bulk viscous NSC, there is a boost in the production of the field $\phi$ in comparison with the NSC case and, therefore, an increased injection of entropy to the SM bath. On the other hand, for certain values of the parameter space $(T_\text{end},\kappa)$, it can be observed that the model with viscosity allows a considerable range of lower values of $\kappa$ to reproduce the DM relic density compared with the NSC. The latter means that for larger values of $\kappa$ both cases, NSC with and without bulk viscosity, are comparable due to the contribution of the viscosity being almost neglected, and the addition of viscosity enables a bit larger values of $T_\text{end}$. Also, in the $(m_\chi, \langle\sigma v\rangle)$ space for the WIMP DM candidate in the same range of masses, NSC and NSC with bulk viscosity allow higher values of $m_\chi$ unlike the $\Lambda$CDM model, and the effects of dissipative processes let us to explore smaller total thermally averaged DM production cross-section for the candidate. These results raise the following question: How do dissipative processes influence the parameter space for FIMP DM production?

The aim of this paper is to revisit the bulk viscous NSC studied in \cite{Gonzalez:2024dtb} by investigating the impact of these dissipative effects on the production and relic density of FIMP DM candidates. Following the same scheme as in \cite{Gonzalez:2024dtb} for WIMP DM candidates, we investigate the parameter space that can accurately reproduce the current DM relic density by adjusting both the model and DM parameters within this novel NSC framework for FIMP DM candidates.

This paper is organized as follows: Section \ref{sec:originalmodel} provides a brief review of the original NSC scenario. Subsection \ref{sec:FIMPS} examines its applicability to FIMP DM candidates. In section \ref{sec:Viscousmodel}, we introduce the bulk viscous extension to the original NSC model, being compared the two models for FIMP DM candidates with a constant total thermally averaged DM production cross-section in subsection \ref{sec:Comparison}. The analysis of the parameter space for DM production that results in the current relic density is presented in Subsection \ref{sec:Parameters}. In Subsection \ref{sec:Non-constant}, we briefly study this bulk viscous NSC for FIMP DM candidates with a non-constant total thermally averaged DM production cross-section. Finally, in Section \ref{sec:Conclusions}, we present some conclusions and future directions. Throughout this paper, we use $c=1$ units.

\section{Non-standard cosmologies}\label{sec:originalmodel}
As discussed in Refs. \cite{Giudice:2000ex, Hamdan:2017psw, DEramo:2017ecx, DEramo:2017gpl, Visinelli:2017qga, Drees:2018dsj, Bernal:2018ins, Bernal:2018kcw, Maldonado:2019qmp, Arias:2019uol, Bernal:2019mhf, Arias:2021rer,Bernal:2022wck, Silva-Malpartida:2024emu}, a straightforward approach to generate NSC scenarios is adding in the early universe a field ($\phi$) which eventually decays into SM plasma. The evolution of the energy density, $\rho_{\phi}$, for this field is given by
\begin{equation}
    \dot{\rho}_\phi+3(\omega+1)H\rho_\phi = -\Gamma_\phi \rho_\phi,\label{phi}
\end{equation}
where ``dot'' denotes the derivative with respect to the cosmic time $t$, $\Gamma_\phi$ is the decay rate of $\phi$, $\omega\equiv p_\phi/\rho_\phi$ is the barotropic index with $p_\phi$ the pressure of the field, and $H=(\rho_\phi+\rho_\gamma)/(3M_p^2)$ is the Hubble parameter with $M_p=2.48\times10^{18}$ GeV the reduced Planck mass. In the same way as for $\phi$, the equation for the radiation energy density, $\rho_\gamma$, can also be written. However, to be more precise, it is necessary to incorporate the relativistic degrees of freedom for radiation. Therefore, the evolution of the Standard Model (SM) bath is more accurately described by the entropy density, due to the entropy injection resulting from the decay of $\phi$ \cite{Drees:2017iod}. Consequently, the evolution of the SM bath is given by
\begin{equation}
    \dot{s}+3Hs= \frac{\Gamma_\phi \rho_\phi}{T}\label{dots},
\end{equation}
where the entropy density can be defined as $s=2\pi^2 g_{\star s}(T) T^3/45$, with $g_{\star s}$ the degrees of freedom that contribute to the entropy density and $T$ the temperature of the SM bath. Note that the interaction term between the field and the SM bath corresponds to the usual expression $\Gamma_{\phi}\rho_{\phi}$. Using the definition for the entropy, the Eq. \eqref{dots} can be rewritten in terms of the temperature as follow
\begin{equation} \dot{T}=\left(-HT+\frac{\Gamma_\phi \rho_\phi}{3s}\right)\left(\frac{dg_{\star s}(T)}{dT}\frac{T}{3g_{\star s}(T)}+1\right)^{-1}. \label{temp}
\end{equation}
The SM bath temperature is related with the radiation energy density by $\rho_\gamma=\pi^2 g_\star (T)T^4/30$, with $g_\star$ the degrees of freedom that contribute to this energy density. It is important to mention that this NSC scenario must not alter the Big Bang Nucleosynthesis (BBN) epoch due to the highly accurate observations based on the $\Lambda$CDM model \cite{Sarkar:1995dd, Hannestad:2004px, DeBernardis:2008zz}. Therefore, the $\phi$ field must decay before the beginning of BBN, i.e, at $T_\text{end}\gtrsim T_{BBN}\sim 4$ MeV. Also, the temperature $T_\text{end}$ is defined as the moment when $H=\Gamma_\phi$ and can be related to the decay rate as
\begin{equation}
    T_\text{end}^4=\frac{90}{\pi^2 g_\star (T_\text{end})}M_p^2\, \Gamma_\phi^2. \label{tend}
\end{equation}
Finally, the evolution of the number density $n_{\chi}$ for the DM component is governed by the Boltzmann equation
\begin{equation}
    \dot{n_\chi}+3H n_\chi=-\langle\sigma v\rangle\left(n_\chi^2-n_{eq}^2\right) \label{boltzdm},
\end{equation}
where $\langle\sigma v\rangle$ is the total thermally averaged DM production cross-section and $n_\text{eq}$ is the equilibrium number density defined as $n_\text{eq}=m_\chi^2\, T\, K_2(m_\chi/T)/\pi^2$, with $K_2$ the Bessel function of second kind and $m_\chi$ the DM mass. The DM energy density is related to its mass by $\rho_\chi=m_\chi\, n_\chi$. The contribution of the DM candidate to the total energy budget can be neglected due to its negligible contribution compared to the field and radiation components. To compute the DM relic density it is useful to define the Yield as $Y\equiv n_\chi/s$.

\begin{figure}
\centering
\includegraphics[scale=0.80]{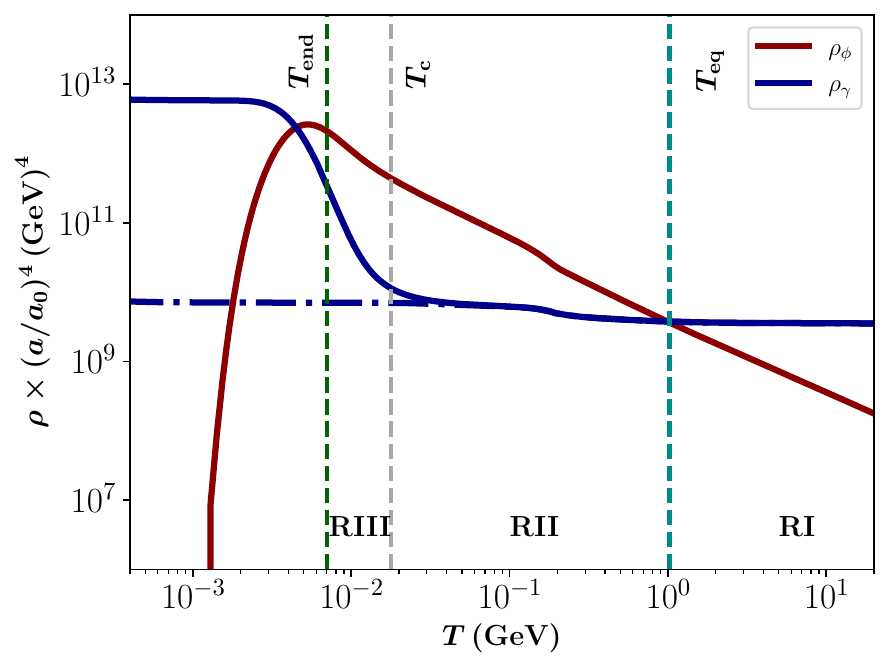}
\caption{Evolution of $\rho\times (a/a_{0})^{4}$ as a function of the temperature $T$ for $\kappa=10^{-2}$, $T_\text{end}=7\times 10^{-3}$ GeV, $m_\chi=100$ GeV, and $\omega=0$. The solid lines correspond to the NSC scenario and the dashed-dotted line to the standard $\Lambda$CDM scenario, while the red line represents the new field $\phi$ and the blue lines the radiation component. The dashed lines correspond to the values of $T_\text{eq}$ (cyan), $T_\text{c}$ (grey), and $T_\text{end}$ (green line).}
\label{figenergydensitystd}
\end{figure}

There are four regions of interest in these kinds of scenarios which are related to different moments in the universe expansion. Region I (RI) occurs when $T_\text{eq}<T$ where $T_\text{eq}$ is the temperature at which $\rho_\phi=\rho_\gamma$. Note that this region does not exist if $\rho_\phi$ is initially higher than radiation. Region II (RII) and III (RIII) take place when $T_\text{eq}>T>T_\text{c}$ and $T_\text{c}>T>T_\text{end}$, respectively, with $T_\text{c}$ the temperature at which the decay of the field becomes relevant. Finally, Region IV (RIV) corresponds to $T_\text{end}>T$, that is, when the $\Lambda$CDM cosmology is recovered. Consequently, the NSC scenario is characterized by the parameters $\kappa \equiv \rho_{\phi,\text{ini}}/\rho_{\gamma,\text{ini}}$, the temperature $T_\text{end}$, and the barotropic index $\omega$. For illustration, Fig. \ref{figenergydensitystd} shows an NSC scenario with $\kappa = 10^{-2}$, $T_\text{end} = 7 \times 10^{-3}$ GeV, $m_\chi = 100$ GeV, and $\omega = 0$. From now on, the subscripts ``0'' and ``ini'' refer to the current and initial time, respectively.

\begin{figure}
\centering
\includegraphics[scale=0.80]{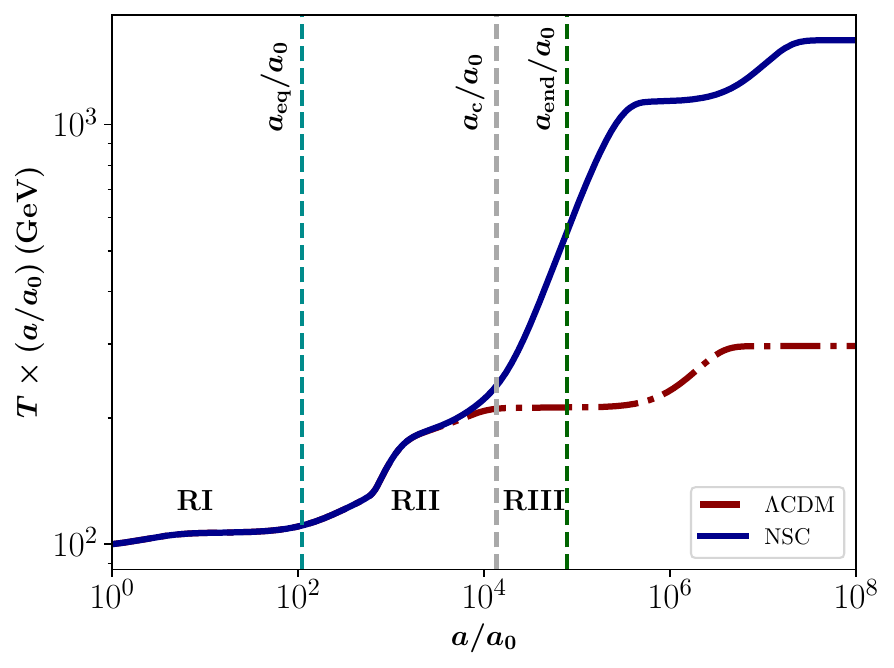}
\caption{Evolution of $T\times (a/a_{0})$ as a function of $a/a_0$ for $\kappa=10^{-2}$, $T_\text{end}=7\times 10^{-3}$ GeV, $m_\chi=100$ GeV, and $\omega=0$. The solid blue line represents the NSC scenario and the dashed-dotted red line the standard $\Lambda$CDM scenario. The dashed lines correspond to the values of $T_\text{eq}$ (cyan), $T_\text{c}$ (grey), and $T_\text{end}$ (green line).}
\label{figtempstd}
\end{figure}

Assuming constant degrees of freedom, the energy density of the field $\phi$ behaves as $\rho_\phi \propto a^{-3(\omega+1)}$. On the other hand, in RI and RII, where $T > T_\text{c}$, the temperature evolves as $T \propto a^{-1}$, while in RIII, the temperature behaves as $T \propto a^{-3(\omega+1)/8}$. Nevertheless, after the complete decay of $\phi$, the temperature takes the usual form $T \propto a^{-1}$, recovering the $\Lambda$CDM cosmology in RIV. This temperature behavior is illustrated in Fig.~\ref{figtempstd} for an NSC scenario with $\kappa = 10^{-2}$, $T_\text{end} = 7 \times 10^{-3}$~GeV, $m_\chi = 100$~GeV, and $\omega = 0$. The variations in temperature and radiation energy density in the figure are derived from the full numerical integration, which accounts for entropy and radiation degrees of freedom.

It is important to mention that the addition of the field can influence the inflation and reheating epochs, as studied in various works \cite{Kofman:1997yn, Ackerman:2010he, Garcia:2020eof, Garcia:2020wiy, Bernal:2022wck, Silva-Malpartida:2023yks, Haque:2024zdq}. While some propose that $\phi$ acts as the inflationary field within the NSC framework, others suggest that $\phi$ solely generates the reheating epoch \cite{Gelmini:2006pw, Gelmini:2006pq, Visinelli:2017qga, Maldonado:2019qmp}.

\subsection{FIMPs in non-standard cosmologies}\label{sec:FIMPS}
The Boltzmann equation described in eq. \eqref{boltzdm} is valid for both WIMP and FIMP candidates. The main differences between them are the magnitude of the thermally averaged cross-section and the initial conditions: WIMPs have an initial number density, as they were in equilibrium with the Standard Model (SM) bath in the early universe, whereas FIMPs are produced via decays of other particles in the early universe and therefore have an initial number density of zero. This production of FIMPs particles is generated by a non-thermal production mechanism called Freeze-In. This DM genesis is based on the assumption that if the coupling of DM particles to the SM bath is small, meaning that their interactions with these particles are very feebly, then the DM particle never achieves the thermal equilibrium in the early universe with the bath particles. With the ongoing expansion of the universe, the production of FIMPs come to an end as their interaction rates become increasingly inefficient relative to the expansion rate of the universe, leading to the observed DM abundance.

These candidates can establish their abundance through two distinct mechanisms. The first one is UltraViolet (UV) freeze-in, and the second one is InfraRed (IR) freeze-in. In the UV freeze-in scenario, the process occurs when the mass of the mediator particle of the interaction with FIMPs exceeds the temperature of the thermal bath. This interaction involves higher-order, non-renormalizable operators, making it highly sensitive to the reheating temperature in the early universe. On the other hand, the IR freeze-in mechanism takes place at temperatures comparable to the mass of DM, through renormalizable operators. For example, in Ref. \cite{Freese:2024ogj}  it is explore DM UV freeze-in from a thermal bath during inflation (a warm inflation phase), showing that the DM yield has a non-trivial dependence on the temperature and the Hubble rate during inflation.

In general, the non-renormalizable operators that connect the DM with the SM bath are operators with mass dimension $5+n/2$, for $n=0$ or even values of $n$. This implies that a total thermally averaged DM production cross-section in the UV Freeze-in can be parametrized as
\begin{equation}
    \langle\sigma v\rangle \propto \frac{T^n}{\Lambda^{n+2}},
    \label{svnocte}
\end{equation}
with $\Lambda$ an energy dimension quantity related to the mass scale of the DM and SM bath mediator. Following this line, operators of mass dimension 5 ($n=0$) give constant values of $\langle\sigma v\rangle$, with some examples presented in Refs. \cite{Chowdhury:2018tzw, Elahi:2014fsa, Hall:2009bx, Biswas:2019iqm, Shakya:2015xnx, McDonald:2008ua}. Meanwhile, examples of operators with mass dimensions 6 ($n=2$) or 7 ($n=4$) are shown in Refs. \cite{Lebedev:2011iq, Bernal:2018qlk, Elahi:2014fsa, Krauss:2013wfa}. Along this paper, we mainly analyze the case with a constant $\langle\sigma v\rangle$, and briefly discuss the cases with $n=2$ and $n=4$ in Subsec. \ref{sec:Non-constant}.

\begin{figure}
\centering
\includegraphics[scale=0.80]{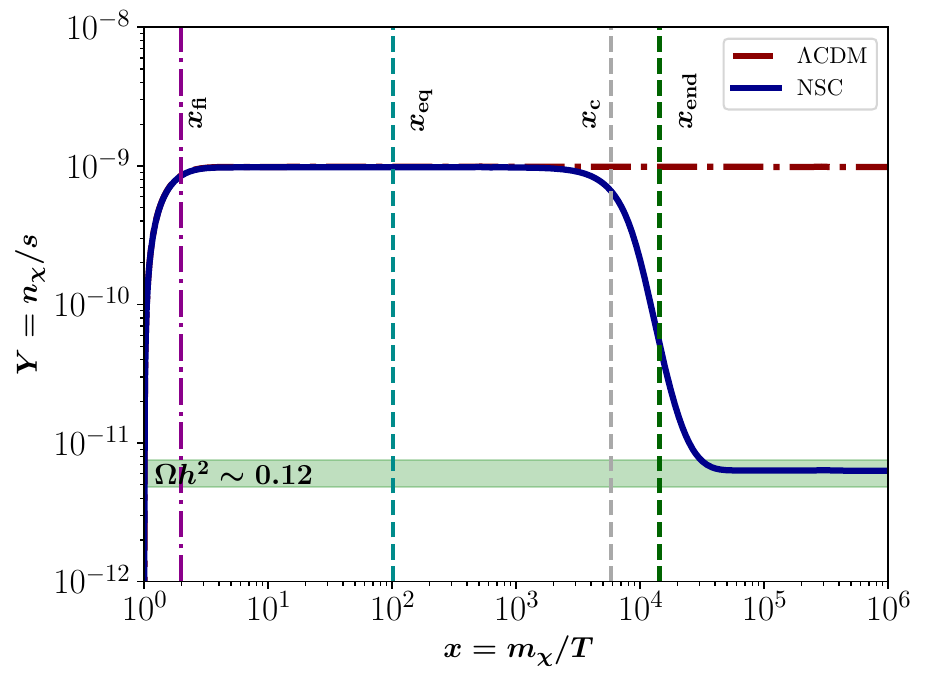}
\caption{Comparison in the yield production between the NSC (solid blue line) and the $\Lambda$CDM (dashed-dotted red line) scenarios for $\kappa=10^{-2}$, $T_\text{end}=7\times 10^{-3}$ GeV, $m_\chi=100$ GeV, $\langle\sigma v\rangle=5\times10^{-26}$ GeV$^{-2}$, and $\omega=0$. The dashed lines correspond to $x_\text{eq}$ (cyan), $x_\text{c}$ (grey), and $x_\text{end}$ (green); while the dashed-dotted magenta line is the time when the DM candidate Freeze-In their number at $x_\text{fi}$. The green zone represents the current DM relic density according to Ref. \cite{Planck:2018vyg}, observing that the parameter space considered in this case gives us the right amount of DM in the NSC scenario.}
\label{figyieldR1}
\end{figure}

\begin{figure}
\centering
\includegraphics[scale=0.80]{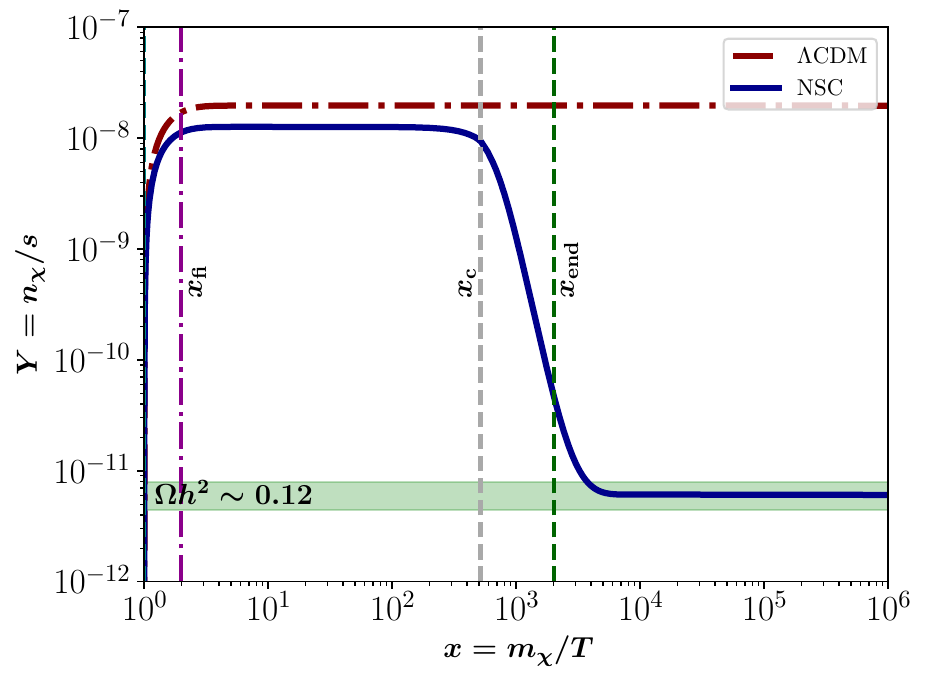}
\caption{Comparison in the yield production between the NSC (solid blue line) and the $\Lambda$CDM  (dashed-dotted red line) scenarios for  $\kappa=1$, $T_\text{end}=5\times 10^{-2}$ GeV, $m_\chi=100$ GeV, $\langle\sigma v\rangle=10^{-24}$ GeV$^{-2}$, and $\omega=0$. The dashed lines correspond to $x_\text{c}$ (grey) and $x_\text{end}$ (green), while the dashed-dotted magenta line is the time when the DM candidate Freeze-In their number at $x_\text{fi}$. The green zone represents the current DM relic density according to Ref. \cite{Planck:2018vyg}, observing that the parameter space considered in this case gives us the right amount of DM in the NSC scenario.}
\label{figyieldR2}
\end{figure}

\begin{figure}
\centering
\includegraphics[scale=0.80]{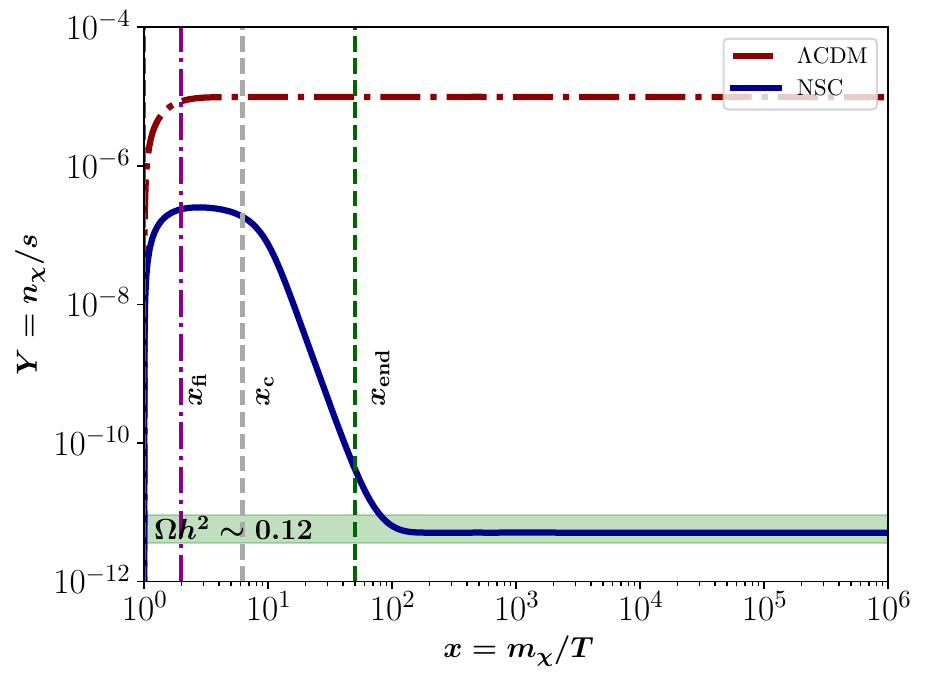}
\caption{Comparison in the yield production between the NSC (solid blue line) and the $\Lambda$CDM (dashed-dotted red line) scenarios for $\kappa=10^{3}$, $T_\text{end}=2$ GeV, $m_\chi=100$ GeV, $\langle\sigma v\rangle=5\times10^{-22}$ GeV$^{-2}$, and $\omega=0$. The dashed lines correspond to $x_\text{c}$ (grey) and $x_\text{end}$ (green), while the dashed-dotted magenta line is the time when the DM candidate Freeze-In their number at $x_\text{fi}$. The green zone represents the current DM relic density according to Ref. \cite{Planck:2018vyg}, observing that the parameter space considered in this case gives us the right amount of DM in the NSC scenario.}
\label{figyieldR3}
\end{figure}

In the case of the NSC, the decay of the new field, introduced at early times, generates an increment in the temperature of the universe and, consequently, in the entropy and radiation energy density. This change in the temperature can also be seen as a faster or slower expansion rate of the universe. The entropy injection can be parametrized by $D\equiv s(T_\text{end})/s(m_\chi)=\left(T_\text{end}/m_\chi\right)^3$, i.e, the entropy density before and after the decays of the new field $\phi$. This left significant imprints on the DM production, considering that the Yield depends explicitly on the entropy density. Therefore, an increment in $s(T)$ will dilute the DM relic density. The latter means that DM parameters ($m_\chi$, $\langle \sigma v\rangle$) that overproduce the relic density in the $\Lambda$CDM model can, in the NSC scenario, reproduce the current DM relic density. The establishment of DM can occur in the four regions described in Section \ref{sec:originalmodel}:
\begin{itemize}
    \item RI: In this region, the radiation component of the universe still dominates its expansion, i.e., $H\sim\sqrt{\rho_\gamma/3M_p^2}\propto T^2$ is almost the same as in the $\Lambda$CDM model. The condition $\kappa<1$ ($\rho_{\gamma,\text{ini}}>\rho_{\phi,\text{ini}}$) is necessary for this region to exist. This behavior is illustrated in Fig. \ref{figyieldR1}, for a NSC with $\kappa=10^{-2}$, $T_\text{end}=7\times 10^{-3}$ GeV, $m_\chi=100$ GeV, $\langle\sigma v\rangle=5\times 10^{-26}$ GeV$^{-2}$, and $\omega=0$. It can be observed that the DM freezes its abundance before the decay of $\phi$, and as the decays become significant, the DM relic density is diluted to reach its current value, allowing the parameters  $\left(m_\chi,\,\langle\sigma v\rangle\right)=\left(100\, \text{GeV},\,5\times 10^{-26}\, \text{GeV}^{-2}\right)$, which were previously ruled out in the $\Lambda$CDM scenario.
    \item RII: In this region, the expansion of the universe is dominated by the energy density of the field, i.e.,  $H\sim\sqrt{\rho_\phi/3M_p^2}\propto T^{3(\omega+1)/2}$. Fig. \ref{figyieldR2} shown the evolution of the DM Yield for $\kappa=1$, $T_\text{end}=5\times 10^{-2}$ GeV, $m_\chi=100$ GeV, $\langle\sigma v\rangle=10^{-24}$ GeV$^{-2}$, and $\omega=0$. In this case, the freeze-in happens at different but closer times, and it can be seen that the DM Yield in the $\Lambda$CDM model is slightly higher than the NSC scenario. This is produced by the expansion rate of the universe different to $H\propto T^2$. Nevertheless, after the decay of $\phi$, the entropy injection dilutes the DM relic density, bringing it to its current value.
    \item RIII: For this case, the universe expansion is still dominated by the $\phi$ field, but the decays begin to inject entropy to the SM bath. The expansion rate can be approximated as $H\sim\sqrt{\rho_\phi/3M_p^2}\propto T^4$, which means a decaying epoch. Fig. \ref{figyieldR3} shows the evolution of the DM Yield for $\kappa=10^3$, $T_\text{end}=2$ GeV, $m_\chi=100$ GeV, $\langle\sigma v\rangle=5\times 10^{-22}$ GeV$^{-2}$, and $\omega=0$. Once again, the particles freeze their abundance earlier in the NSC scenario compared to $\Lambda$CDM model, resulting in a lower DM yield. As the decay of $\phi$ becomes significant, the resulting entropy injection dilutes the DM relic density, bringing it to its present value.
    \item RIV: Finally, in this region the $\phi$ filed has already fully decay and the $\Lambda$CDM model is recovered. This region is not of our interest.
\end{itemize}

It is important to note that in the case of UV Freeze-in, the DM production can also be generated by other mechanisms, such as the decay of $\phi$ or gravitational production. According to Ref. \cite{Bernal:2019mhf}, considering couplings of $\mathcal{O}(1)$, $\omega\sim 1$, and $m_\phi<M_{p}$, the contribution of the Branching ratio of $\phi$ decaying in DM is negligible compared to the UV Freeze-in. In the case of gravitational production, the interaction rate density is proportional to $T^8/M_p^4$, meaning that higher initial temperatures ($T_\text{ini}>10^{11}$ GeV) and higher reheating temperatures $(T_\text{rh}>10^7$ GeV) would provide a perfect mechanism for generating FIMP DM \cite{Barman:2022tzk, Bernal:2025fdr}. In the case we are analyzing, the form of $\langle\sigma v\rangle$ is more efficient.

\section{Bulk viscous non-standard cosmologies}\label{sec:Viscousmodel}
In the bulk viscous extension of the classical NSC, it is assumed that the field $\phi$ experiences a dissipative process in the form of a bulk viscosity during its cosmic evolution. In this extension, and working in the relativistic thermodynamic framework out of equilibrium of Eckart's theory \cite{PhysRev.58.919}, the equations for the Hubble parameter and the evolution of the SM bath \eqref{dots} remain unchanged, while the acceleration equation and conservation equation for the field \eqref{phi} become
\begin{eqnarray}
    &&2\dot{H}+3H^{2}=-p_{\gamma}-p_{\phi}-\Pi, \label{Dissacceleration} \\ &&\dot{\rho_\phi}+3\left(\omega+1\right)H\rho_\phi=-\Gamma_\phi\rho_\phi-3H\Pi, \label{Dissphi}
\end{eqnarray}
where $P_{\text{eff},\phi}=p_{\phi}+\Pi$ is the effective pressure of the field, with $p_{\phi}$ being the equilibrium pressure, $\Pi=-3H\xi$ the bulk viscous pressure, and $\xi$ is the bulk viscosity. In particular, we assume that the dissipative fluid does not experience heat flow or shear viscosity. For more details on the procedure used to obtain this novel NSC, we refer the reader to Ref. \cite{Gonzalez:2024dtb}, where the Bulk viscous NSCs were studied for the first time. It is important to note that the bulk viscosity influences the evolution of the universe through the bulk viscous pressure. In particular, in an expanding universe, the expression $\Pi=-3H\xi$ is always negative ($\xi>0$ to maintain consistency with the second law of thermodynamics \cite{SWeinberg}). Consequently, this viscosity leads to an acceleration in the expansion of the universe, as seen from Eq. \eqref{Dissacceleration}.

In general, bulk viscosity can depend on the temperature and pressure of the dissipative fluid \cite{SWeinberg}. Hence, a natural and most extensively studied expression for the bulk viscosity is to assume a proportional dependence to the power of its energy density, $\xi=\xi_{0}\rho_{\phi}^{1/2}$, where $\xi_{0}=\hat{\xi_{0}}M_{p}$, allowing $\hat{\xi_0}$ to be a dimensionless parameter. With this consideration, Eq. \eqref{Dissphi} takes the form
\begin{equation}\label{eqphibulk}
    \dot{\rho}_{\phi}+3(\omega+1)H\rho_{\phi}=-\Gamma_\phi\rho_\phi+9M_{p}\hat{\xi_0}H^{2}\rho_\phi^{1/2}.
\end{equation}
The chosen parameterization for the bulk viscosity has the advantage that, once the field $\phi$ fully decays into SM plasma, the dissipation becomes negligible, and the standard $\Lambda$CDM scenario without viscosity is recovered. Therefore, to compare the classical NSC scenario with its bulk viscous counterpart, we numerically integrate Eqs. \eqref{phi}, \eqref{temp}, \eqref{boltzdm}, and \eqref{eqphibulk}, taking into account that $H=(\rho_\phi+\rho_\gamma)/(3M_p^2)$. Note that as we incorporate the field in a model-independent manner, we do not assume that $\phi$ is the inflationary field. Therefore, the initial temperature will be $T_\text{ini}=m_\chi$ to explore a wide range of maximum temperatures in the DM parameter space and study its impact on the DM production, under the assumption that $T_\text{end}$ (or the reheating temperature) must be higher than $4$ MeV. The results of this analysis are presented in the following subsection.

\subsection{Comparison between scenarios}\label{sec:Comparison}
In this section, we will compare all the features discussed in Sections \ref{sec:originalmodel} and \ref{sec:Viscousmodel}, between the NSC and the bulk viscous NSC scenarios.

\begin{figure}
\centering
\includegraphics[scale=0.80]{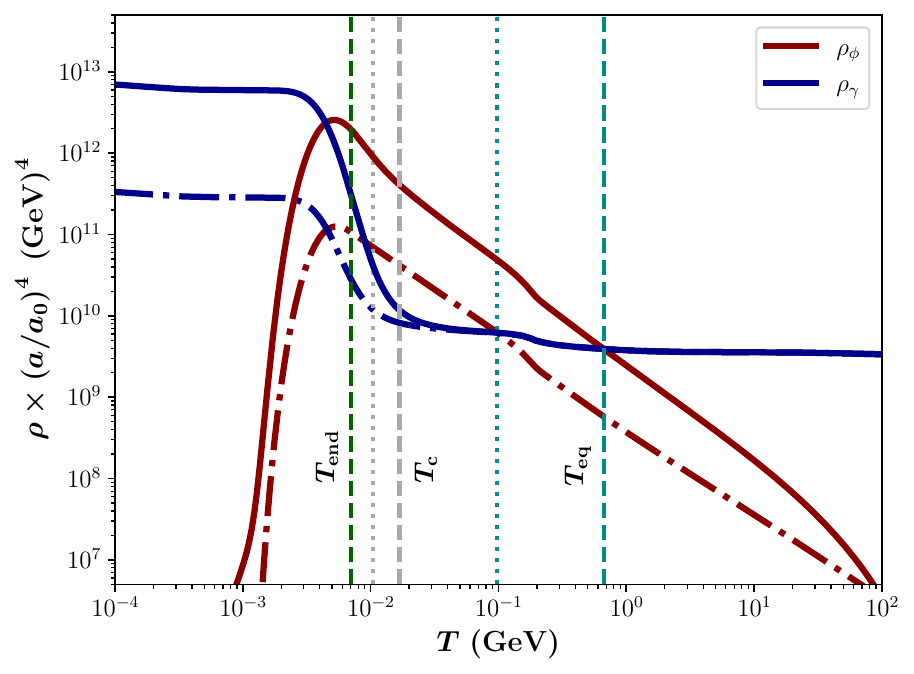}
\caption{Evolution of $\rho\times (a/a_{0})^{4}$ as a function of the temperature $T$ for $\kappa=10^{-3}$, $T_\text{end}=7\times 10^{-3}$ GeV, $\hat{\xi_0}=10^{-2}$, $m_\chi=100$~GeV, and $\omega=0$. The solid and dashed-dotted lines correspond to the NSC with and without bulk viscosity, respectively; while the red lines represent the new field $\phi$ and the blue lines the radiation component. We also depict the values of $T_\text{eq}$ (cyan), $T_\text{c}$ (grey), and $T_\text{end}$ (green) for the NSC with bulk viscosity (dashed lines) and the classical NSC (dotted lines).}
\label{figenergydensity}
\end{figure}

\begin{figure}
\centering
\includegraphics[scale=0.80]{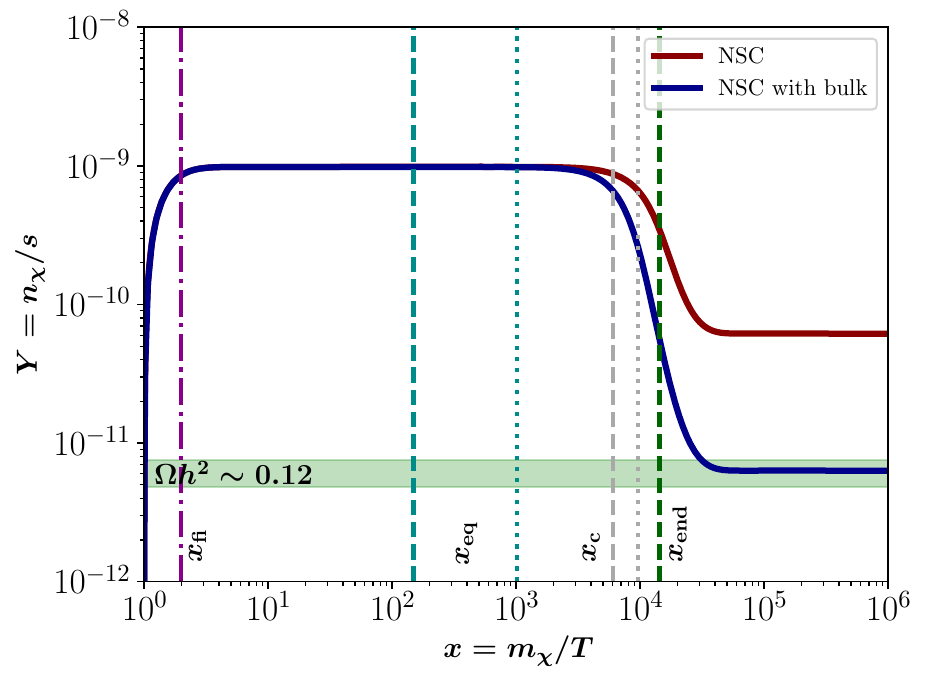}
\caption{Comparison in the yield production between the classical NSC (solid red line) and the NSC with bulk viscosity (solid blue line) for $\kappa=10^{-3}$, $T_\text{end}=7\times 10^{-3}$ GeV, $\hat{\xi_0}=10^{-2}$, $m_\chi=100$ GeV, $\langle\sigma v\rangle=5\times10^{-26}$ GeV$^{-2}$, and $\omega=0$. The dashed and dotted lines correspond to $x_\text{eq}$ (cyan), $x_\text{c}$ (grey), and $x_\text{end}$ (green) for the NSC with and without bulk viscosity, respectively; while the dashed-dotted magenta line is the time when the DM candidate Freeze-In their number at $x_\text{fi}$. The green zone represents the current DM relic density according to Ref. \cite{Planck:2018vyg}, observing that the parameter space considered in this case gives us the right amount of DM in the bulk viscous NSC scenario.}
\label{figyield}
\end{figure}

In Fig. \ref{figenergydensity}, we depict the evolution of $\rho\times (a/a_{0})^{4}$ as a function of the temperature $T$ for $\kappa=10^{-3}$, $T_\text{end}=7\times 10^{-3}$ GeV, $\hat{\xi_0}=10^{-2}$, $m_\chi=100$~GeV, and $\omega=0$. The solid and dashed-dotted lines correspond to the NSC with and without bulk viscosity, respectively; while the red lines represent the new field $\phi$ and the blue lines the radiation component. We also depict the values of $T_\text{eq}$ (cyan), $T_\text{c}$ (grey), and $T_\text{end}$ (green) for the NSC with bulk viscosity (dashed lines) and the classical NSC (dotted lines). From the figure, it can be observed that the bulk viscous NSC leads to an enhanced production of the field $\phi$ compared to the classical NSC scenario, resulting in a greater increase in the radiation energy density due to the decay of the field, and consequently, a higher entropy injection into the SM bath. An interesting feature is that the overall temperature-dependent behavior of the fluids remains largely unchanged between the two scenarios, allowing us to conclude that the field $\phi$ becomes negligible in the bulk viscous NSC scenario as in the classical NSC. Nevertheless, this conclusion is only valid for bulk viscous NSC models with $\omega>-1$, since viscosity can lead to obtaining an effective barotropic index where $\omega_{\text{eff}}\leq-1$ for an $\omega>-1$ but close to $-1$, i.e., we can obtain a behavior where the field $\phi$ never decays. Meanwhile, Figure \ref{figyield} shows the comparison in the yield production between the classical NSC (solid red line) and the NSC with bulk viscosity (solid blue line) for the same values of $\kappa$, $T_\text{end}$, $\hat{\xi_0}$, $m_\chi$, and $\omega$ as in Figure \ref{figenergydensity}, considering $\langle\sigma v\rangle=5\times10^{-26}$ GeV$^{-2}$. The dashed and dotted lines correspond to $x_\text{eq}$ (cyan), $x_\text{c}$ (grey), and $x_\text{end}$ (green) for the NSC with and without bulk viscosity, respectively; while the dashed-dotted magenta line is the time when the DM candidate Freeze-In their number at $x_\text{fi}$. The green zone represents the current DM relic density according to Ref. \cite{Planck:2018vyg}. Firstly, note that the lines for $x_\text{end}$ and $x_\text{fi}$ are the same for both models. Secondly, note that the behavior of the Yield production is the same in both models until the decay of $\phi$. After that, it is observed that the higher entropy injection generated by the viscous term leads to lower values of the DM Yield. This means that the DM relic density agrees with the current observation for a NSC with bulk viscosity in contrast with the standard NSC case which overproduces the DM relic density.

\begin{figure}
\centering
\includegraphics[scale=0.77]{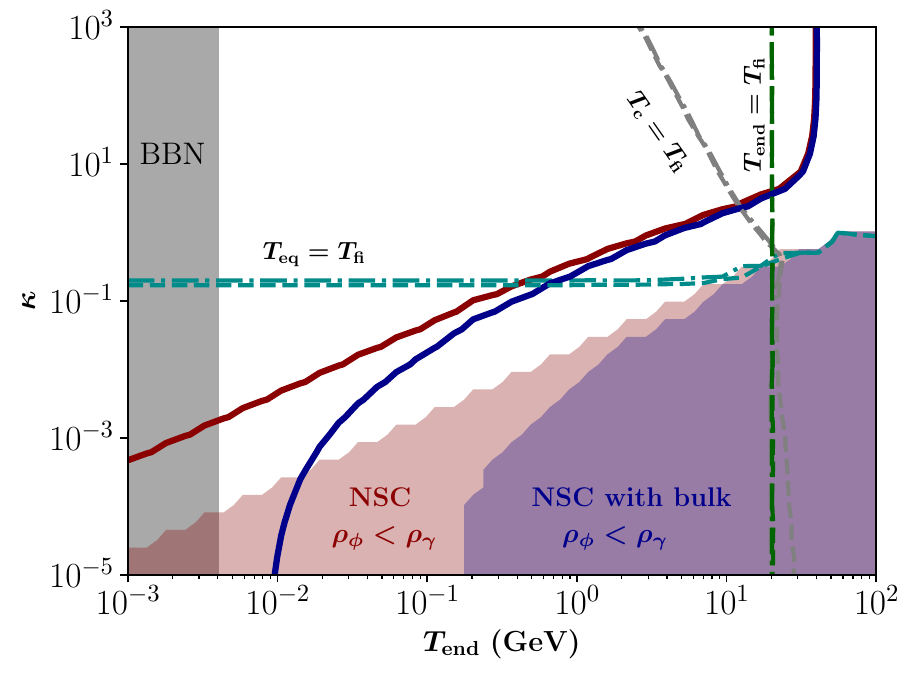}
\caption{Parameter space ($T_\text{end}$,$\kappa$) for the NSC models with $\hat{\xi_0}=10^{-2}$, $m_\chi=100$ GeV, $\langle\sigma v\rangle=10^{-26}$ GeV$^{-2}$, and $\omega=0$. The solid red and blue lines correspond to the parameter space that reproduces the current DM relic density for the classical NSC and the NSC with bulk viscosity, respectively. We also delimit the three regions of interest through the equalities  $T_\text{eq}=T_\text{fi}$ (cyan), $T_\text{c}=T_\text{fi}$ (grey), and $T_\text{end}=T_\text{fi}$ (green) for the NSC with bulk viscosity (dashed lines) and the classical NSC (dashed-dotted lines). The red and blue zones represent their respective parameter space in which $\rho_\phi<\rho_\gamma$, $\forall t$. The grey zone corresponds to the forbidden BBN epoch that starts at $T_\text{BBN}\sim 4\times 10^{-3}$ GeV. Note that the bulk viscous NSC scenario allows the model to take lower values of $\kappa$.}
\label{figkappaTendw0}
\end{figure}

\begin{figure}
\centering
\includegraphics[scale=0.80]{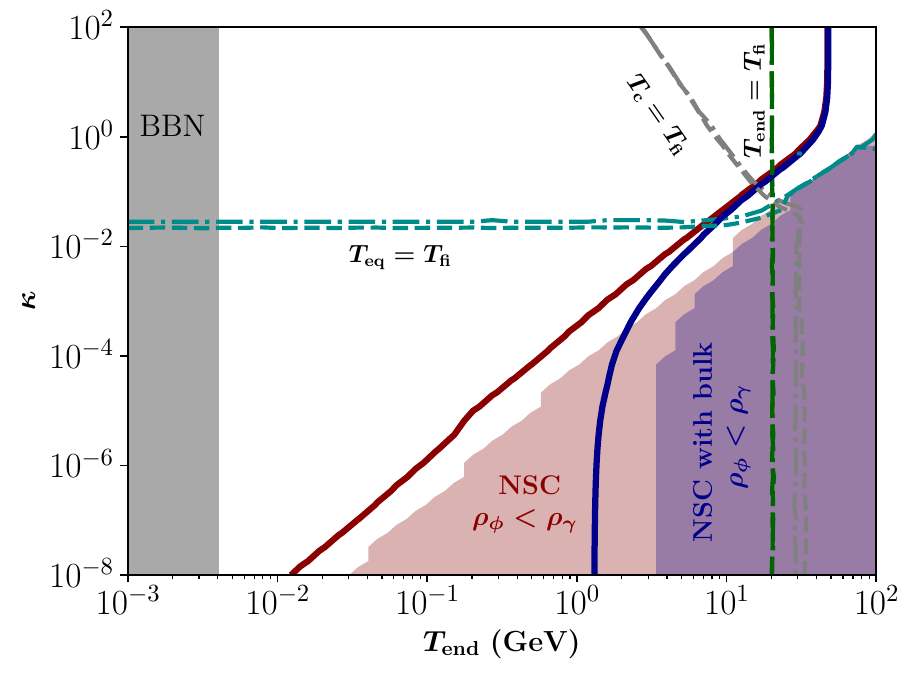}
\caption{Parameter space ($T_\text{end}$,$\kappa$) for the NSC models with $\hat{\xi_0}=10^{-2}$, $m_\chi=100$ GeV, $\langle\sigma v\rangle=10^{-26}$ GeV$^{-2}$, and $\omega=-2/5$. The solid red and blue lines correspond to the parameter space that reproduces the current DM relic density for the classical NSC and the NSC with bulk viscosity, respectively. We also delimit the three regions of interest through the equalities  $T_\text{eq}=T_\text{fi}$ (cyan), $T_\text{c}=T_\text{fi}$ (grey), and $T_\text{end}=T_\text{fi}$ (green) for the NSC with bulk viscosity (dashed lines) and the classical NSC (dashed-dotted lines). The red and blue zones represent their respective parameter space in which $\rho_\phi<\rho_\gamma$, $\forall t$. The grey zone corresponds to the forbidden BBN epoch that starts at $T_\text{BBN}\sim 4\times 10^{-3}$ GeV.}
\label{figkappaTendw-25}
\end{figure}

\begin{figure}
\centering
\includegraphics[scale=0.80]{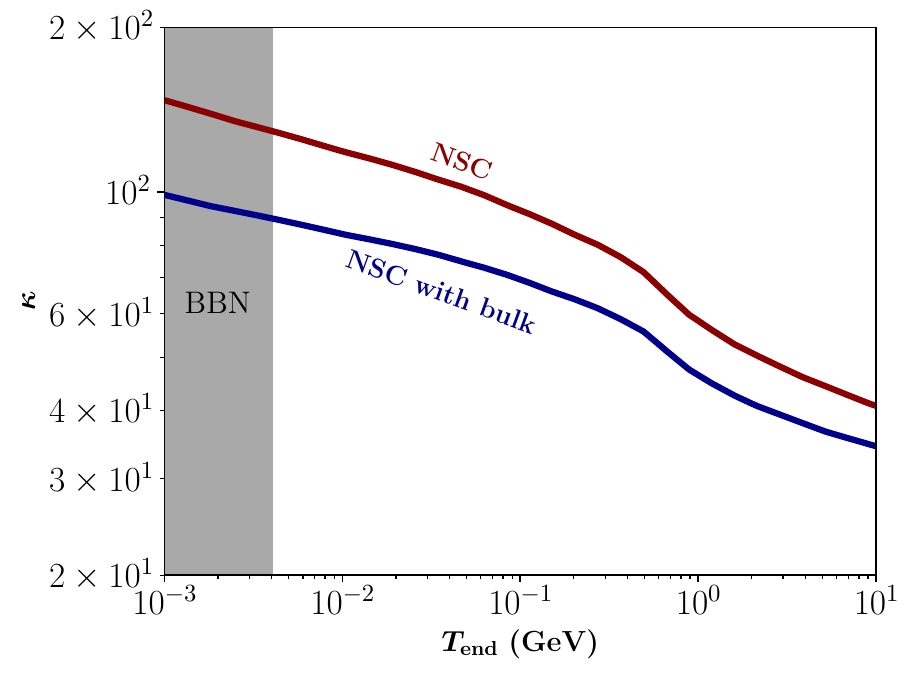}
\caption{Parameter space ($T_\text{end}$, $\kappa$) for the NSC models with $\hat{\xi_0}=10^{-2}$, $m_\chi=100$ GeV, $\langle\sigma v\rangle=10^{-26}$ GeV$^{-2}$, and $\omega=2/5$. The solid red and blue lines correspond to the parameter space that reproduces the current DM relic density for the classical NSC and the NSC with bulk viscosity, respectively. The grey zone represents the forbidden BBN epoch that starts at $T_\text{BBN}\sim 4\times 10^{-3}$ GeV. Note that the bulk viscous NSC scenario allows the model to take lower values of $\kappa$.}
\label{figkappaTendw25}
\end{figure}

The comparison in the parameter space of the NSC models, namely $\kappa$ and $T_\text{end}$, with a fixed $\omega$, is depicted in Figures \ref{figkappaTendw0}, \ref{figkappaTendw-25}, and \ref{figkappaTendw25}. The figures were obtained for the values $\hat{\xi_0}=10^{-2}$, $m_\chi=100$ GeV, and $\langle\sigma v\rangle=10^{-26}$ GeV$^{-2}$. The solid red and blue lines correspond to the parameter space that reproduces the current DM relic density for the classical NSC and the NSC with bulk viscosity, respectively. We also delimit the three regions of interest (see Section \ref{sec:FIMPS}) through the equalities $T_\text{eq}=T_\text{fi}$ (cyan), $T_\text{c}=T_\text{fi}$ (grey), and $T_\text{end}=T_\text{fi}$ (green) for the NSC with bulk viscosity (dashed lines) and the classical NSC (dashed-dotted lines). The red and blue zones represent their respective parameter space in which $\rho_\phi<\rho_\gamma$, $\forall t$, i.e., the regions in which the Hubble parameter is always approximately similar to the $\Lambda$CDM model ($H\sim T^2$). The grey zone corresponds to the forbidden BBN epoch that starts at $T_\text{BBN}\sim 4\times 10^{-3}$ GeV. In Figure \ref{figkappaTendw0}, obtained for $\omega=0$, it can be seen that for higher values of $\kappa$ and $T_\text{end}$ (RIII), the parameter space is slightly the same, meanwhile in RII the curves separate from each other. However, the greatest differences between the NSC with and without bulk viscosity cases are presented in RI, exhibiting an approximate independent behavior of the $\kappa$ parameter, which can be explained by the dominance of the viscous term over the decaying term. On the other hand, Figure \ref{figkappaTendw-25}, obtained for $\omega=-2/5$, exhibits a similar behavior to the $\omega=0$ case, with the difference that the approximate independent $\kappa$ values in RI are reached to higher temperatures and, therefore, there is a smaller window of $T_\text{end}$ to reproduce the current DM relic density. Finally, in Figure \ref{figkappaTendw25}, obtained for $\omega=2/5$, there is no existence of RI since the values of $\kappa$ are greater than $1$. In other words, $\rho_{\phi,\text{ini}}$ is always greater than $\rho_{\gamma,\text{ini}}$. The parameters for NSC with and without bulk viscosity are very close (less than one order of magnitude) and, therefore, the FIMP DM candidates are not so sensitive to the inclusion of the bulk viscosity when the value of $\omega$ is higher compared these two cosmological scenarios. This also explains the absence of the red and blue zones, and the lines for $T_\text{eq}$, $T_\text{c}$, and $T_\text{end}$.

\begin{figure}
\centering
\includegraphics[scale=0.80]{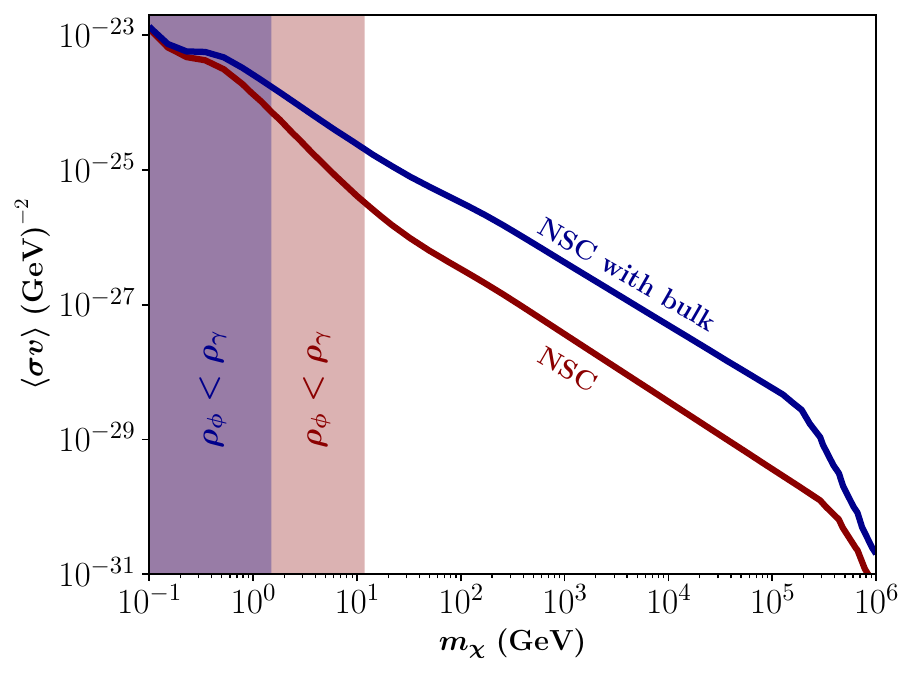}
\caption{Parameter space $(m_\chi,\langle\sigma v\rangle)$ for the FIMP DM candidate with $\kappa=10^{-3}$, $T_\text{end}=7\times 10^{-3}$ GeV, $\hat{\xi_0}=10^{-2}$, and $\omega=0$. The solid red and blue lines correspond to the parameter space that reproduces the current DM relic density for the classical NSC and the NSC with bulk viscosity, respectively. The red and blue zones represent their respective parameter space in which $\rho_\phi<\rho_\gamma$, $\forall t$. Note that the bulk viscous NSC scenario allows to achieve larger values of $\langle\sigma v\rangle$.}
\label{figmsigma}
\end{figure}

Last but not least, Figure \ref{figmsigma} shows the FIMP DM candidate parameter space, namely $m_\chi$ and $\langle\sigma v\rangle$, for $\kappa=10^{-3}$, $T_\text{end}=7\times 10^{-3}$ GeV, $\hat{\xi_0}=10^{-2}$, and $\omega=0$. The solid red and blue lines correspond to the parameter space that reproduces the current DM relic density for the classical NSC and the NSC with bulk viscosity, respectively. The red and blue zones represent their respective parameter space in which $\rho_\phi<\rho_\gamma$, $\forall t$. From the figure, a similar behavior can be observed between the two scenarios for small values of FIMP DM candidate mass and high values of the thermally averaged DM production cross-section (and vice-versa). Also, there is a gap between the curves, shifting the case with bulk viscosity upward with respect to the standard NSC scenario, reaching larger values of $\langle\sigma v\rangle$ (approximately two orders of magnitude) in the same range of $m_\chi$.

\subsection{Parameter spaces in the bulk viscous non-standard cosmologies}\label{sec:Parameters}
From now on, we focus the analysis on the study of the parameter spaces ($T_\text{end}$,$\kappa$) of the model and $(m_\chi,\langle\sigma v\rangle)$ of the FIMP DM candidate for the bulk viscous NSC scenario alone. It is important to mention that, from this point onward, the choice of values for the free parameters of the model and for the DM is solely intended to illustrate the behavior of DM production within this new cosmological scenario and it is not a preference for an arbitrary value.

\begin{figure}
\centering
\includegraphics[scale=0.80]{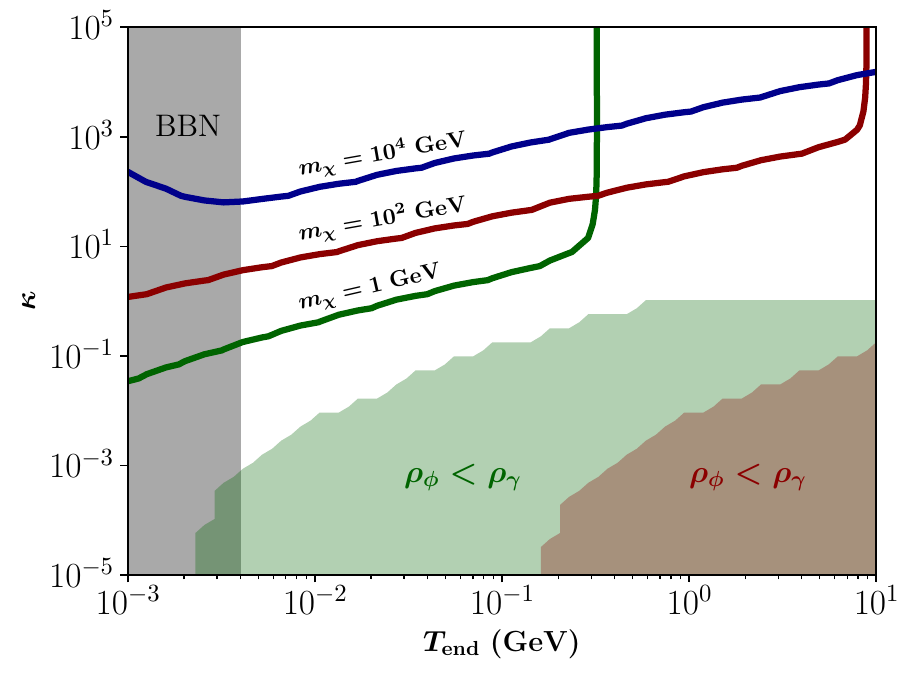}
\caption{Parameter space $(T_\text{end},\kappa)$ for the bulk viscous NSC with $\hat{\xi_0}=10^{-2}$, $\langle\sigma v\rangle=10^{-22}$ GeV$^{-2}$, and $\omega=0$. The solid blue, red, and green lines correspond to the parameter space that reproduces the current DM relic density for $m_{\chi}=10^{4}$, $10^{2}$, and $1$ GeV, respectively. The green and red zones represent the parameter space in which $\rho_\phi<\rho_\gamma$, $\forall t$, for $m_{\chi}=1$ and $10^{2}$ GeV, respectively. The grey zone corresponds to the forbidden BBN epoch that starts at $T_\text{BBN}\sim 4\times 10^{-3}$ GeV. Note that lower values of $m_{\chi}$ allow the model to take lower values of $\kappa$.}
\label{figvarmdm}
\end{figure}

\begin{figure}
\centering
\includegraphics[scale=0.80]{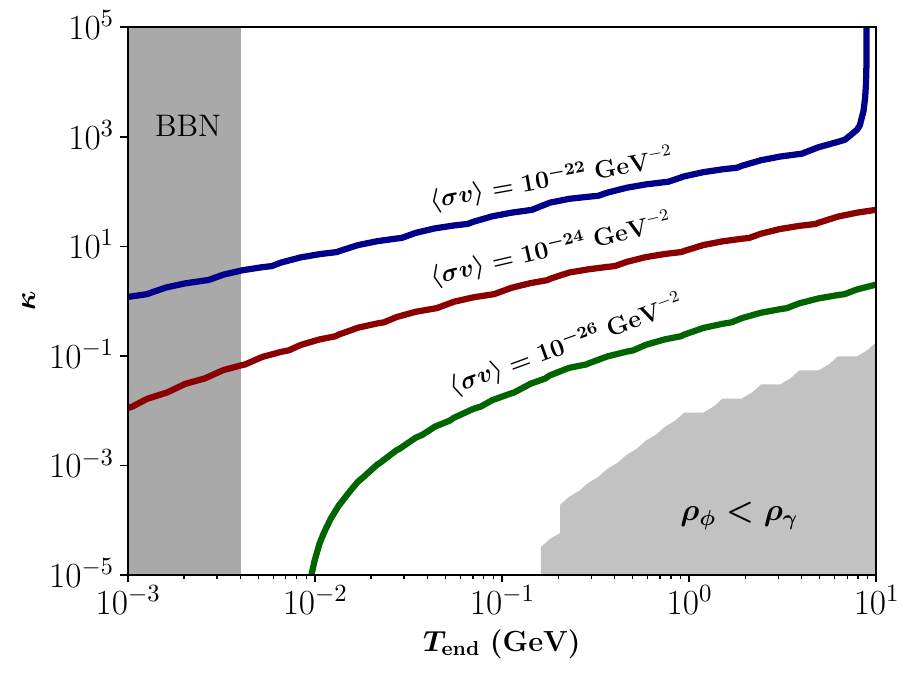}
\caption{Parameter space $(T_\text{end},\kappa)$ for the bulk viscous NSC with $\hat{\xi_0}=10^{-2}$, $m_\chi=100$ GeV, and $\omega=0$. The solid blue, red, and green lines correspond to the parameter space that reproduces the current DM relic density for $\langle\sigma v\rangle=10^{-22}$, $10^{-24}$, and $10^{-26}$ GeV$^{-2}$, respectively. The grey zones represent the parameter space in which $\rho_\phi<\rho_\gamma$, $\forall t$ (for the three cases), and the forbidden BBN epoch that starts at $T_\text{BBN}\sim 4\times 10^{-3}$ GeV. Note that a decrease in the total thermally averaged DM production cross-section allows the model to take lower values of $\kappa$.}
\label{figvarsv}
\end{figure}

\begin{figure}
\centering
\includegraphics[scale=0.80]{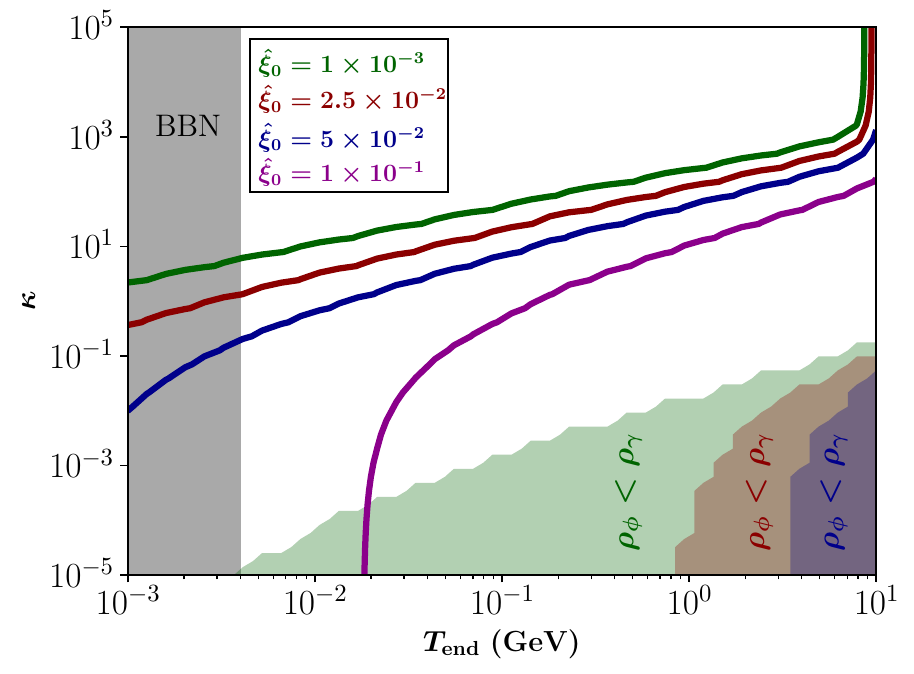}
\caption{Parameter space $(T_\text{end},\kappa)$ for the bulk viscous NSC with $m_\chi=100$ GeV, $\langle\sigma v\rangle=10^{-22}$ GeV$^{-2}$, and $\omega=0$. The solid green, red, blue and magenta lines correspond to the parameter space that reproduces the current DM relic density for $\hat{\xi}_{0}=10^{-3}$, $2.5\times10^{-2}$, $5\times10^{-2}$ and $10^{-1}$, respectively. The green, red, and blue zones represent their respective parameter space in which $\rho_\phi<\rho_\gamma$, $\forall t$. The grey zone corresponds the forbidden BBN epoch that starts at $T_\text{BBN}\sim 4\times 10^{-3}$ GeV. Note that higher values of $\hat{\xi}_0$ allow the model to take lower values of $\kappa$.}
\label{figvarxi0}
\end{figure}

In Figure \ref{figvarmdm}, we depict the parameter space $(T_\text{end},\kappa)$ for $\hat{\xi_0}=10^{-2}$, $\langle\sigma v\rangle=10^{-22}$ GeV$^{-2}$, and $\omega=0$. The solid blue, red, and green lines correspond to the parameter space that reproduces the current DM relic density for $m_{\chi}=10^{4}$, $10^{2}$, and $1$ GeV, respectively. The green and red zones represent the parameter space in which $\rho_\phi<\rho_\gamma$, $\forall t$, for $m_{\chi}=1$ and $10^{2}$ GeV, respectively. The grey zone corresponds to the forbidden BBN epoch that starts at $T_\text{BBN}\sim 4\times 10^{-3}$ GeV. From the figure, it can be observed that lower values of $m_\chi$ shift the curves downward to the left, while the regions in which $\rho_\phi<\rho_\gamma$ are shifted upward to the left. Hence, the blue zone is achieved at higher values of $T_\text{end}$ and lower values of $\kappa$. Interestingly, for $m_\chi=10^4$~GeV, exists a region in which the curve that reproduces the current DM relic density goes to higher values of $\kappa$. While this behavior occurs in the forbidden BBN zone, this parameter space may be accessible for
higher values of the DM mass. Otherwise, Figure \ref{figvarsv} shows the same parameter space as Figure \ref{figvarmdm} for $\hat{\xi_0}=10^{-2}$, $m_\chi=100$ GeV, and $\omega=0$. The solid blue, red, and green lines correspond to the parameter space that reproduces the current DM relic density for $\langle\sigma v\rangle=10^{-22}$, $10^{-24}$, and $10^{-26}$ GeV$^{-2}$, respectively. The grey zones represent the parameter space in which $\rho_\phi<\rho_\gamma$, $\forall t$ (for the three cases), and the forbidden BBN epoch. As it can be noted, a decrease in the values for the thermally averaged DM production cross-section shifts the parameters that reproduce the current DM relic density downward to the right, allowing access to the ``approximated independent $\kappa$-zone'' described in Section \ref{sec:Comparison}. Also, it is interesting to analyze the behavior of varying the DM mass (Figure \ref{figvarmdm}) and its interactions (Figure \ref{figvarsv}). For instance, for higher values of $m_\chi$ appear a zone allowed to reproduce the current DM relic density in which the values of $\kappa$ increase when $T_\text{end}$ is decreasing. On the other hand, smaller values of $\langle\sigma v\rangle$ allow lower values of $\kappa$ when $T_\text{end}$ is diminishing up to the independent zone mentioned above. Finally, Figure \ref{figvarxi0} shows the parameter space mentioned before for $m_\chi=100$ GeV, $\langle\sigma v\rangle=10^{-22}$ GeV$^{-2}$, and $\omega=0$. The solid green, red, blue and magenta lines correspond to the parameter space that reproduces the current DM relic density for $\hat{\xi}_{0}=10^{-3}$, $2.5\times10^{-2}$, $5\times10^{-2}$ and $10^{-1}$, respectively. The green, red, and blue zones represent their respective parameter space in which $\rho_\phi<\rho_\gamma$, $\forall t$. As before, the grey zone corresponds to the forbidden BBN epoch. As it is possible to see, higher values of $\hat{\xi_0}$ shift the curves and the regions, in which $\rho_\phi<\rho_\gamma$, downward to the right. Hence, the magenta zone is achieved at higher values of $T_\text{end}$ and lower values of $\kappa$. Also, this increment in the $\hat{\xi_0}$-values generates more prominent curvatures in the ``approximately independent $\kappa$-zone", reducing the $T_\text{end}$ parameter space that could reproduce the current DM relic density.

The above discussion shows how certain values of $m_\chi$ and $\langle\sigma v\rangle$ open the possibility to new combinations of ($T_\text{end}$, $\kappa$) parameters to reproduce the current DM relic density and how combining different DM parameters could reach higher or lower values in the model parameters. For instance, for higher values of $\hat{\xi_0}$ and small values of $\langle\sigma v\rangle$ it is possible to access small values for $\kappa$ and shorter range for $T_\text{end}$. Meanwhile, a large range of $T_\text{end}$ and lower values of $\kappa$ could be reached for small values of ($m_\chi$, $\langle\sigma v\rangle$) and higher values of $\hat{\xi_0}$.

\begin{figure}
\centering
\includegraphics[scale=0.80]{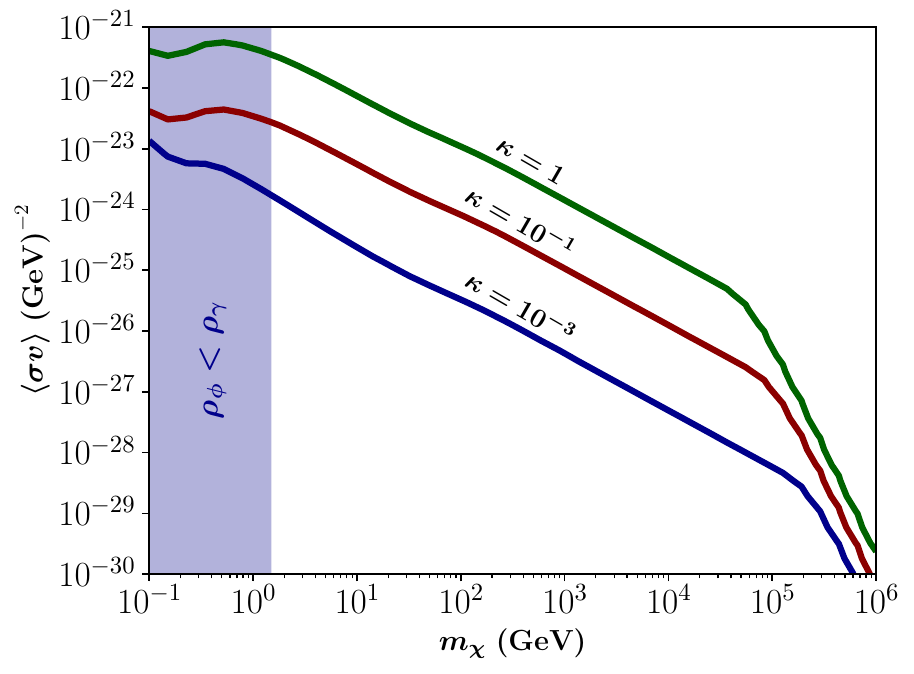}
\caption{Parameter space $(m_\chi,\langle\sigma v\rangle)$ for the FIMP DM candidate with $T_\text{end}=7\times 10^{-3}$ GeV, $\hat{\xi_0}=10^{-2}$, and $\omega=0$. The solid green, red, and blue lines correspond to the parameter space that reproduces the current DM relic density for $\kappa=1$, $10^{-1}$, and $10^{-3}$, respectively. The blue zone represents the parameter space in which $\rho_\phi<\rho_\gamma$, $\forall t$, for $\kappa=10^{-3}$. Note that a decrease in the $\kappa$-values allow to achieve smaller values of $\langle\sigma v\rangle$.}
\label{figvarkappa}
\end{figure}

\begin{figure}
\centering
\includegraphics[scale=0.80]{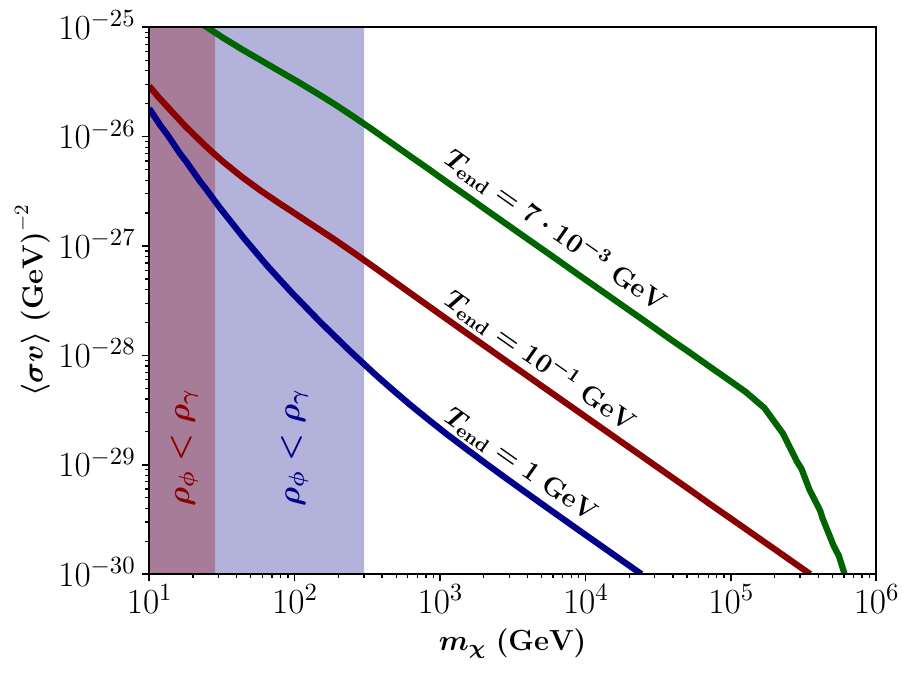}
\caption{Parameter space $(m_\chi,\langle\sigma v\rangle)$ for the FIMP DM candidate with $\kappa=10^{-3}$, $\hat{\xi_0}=10^{-2}$, and $\omega=0$. The solid green, red, and blue lines correspond to the parameter space that reproduces the current DM relic density for $T_\text{end}=7\times10^{-3}$, $10^{-1}$, and $1$ GeV, respectively. The blue and red zones represent the parameter space in which $\rho_\phi<\rho_\gamma$, $\forall t$, for $T_\text{end}=1$ and $10^{-1}$ GeV, respectively. Note that higher values of $T_{\text{end}}$ allow to achieve smaller values of $\langle\sigma v\rangle$.}
\label{figvartend}
\end{figure}

\begin{figure}
\centering
\includegraphics[scale=0.80]{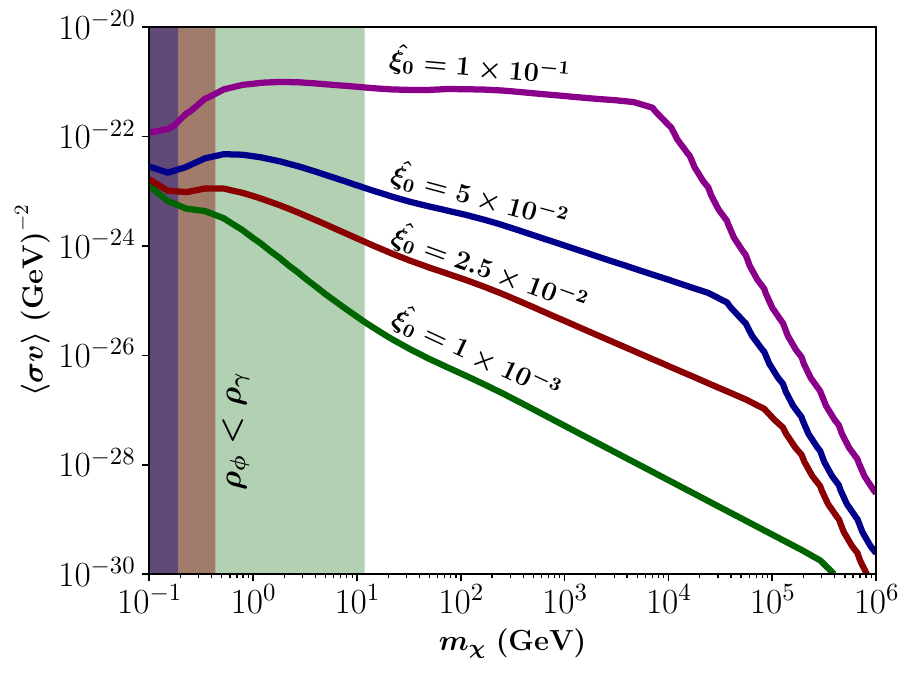}
\caption{Parameter space $(m_\chi,\langle\sigma v\rangle)$ for the FIMP DM candidate with $\kappa=10^{-3}$, $T_\text{end}=7\times10^{-3}$ GeV, and $\omega=0$. The solid magenta, blue, red, and green lines correspond to the parameter space that reproduces the current DM relic density for $\hat{\xi}_0=10^{-1}$, $5\times10^{-2}$, $2.5\times10^{-2}$, and $10^{-3}$, respectively. The blue, red, and green zones represent the parameter space in which $\rho_\phi<\rho_\gamma$, $\forall t$, for $\hat{\xi}_0=5\times10^{-2}$, $2.5\times10^{-2}$, and $10^{-3}$, respectively. Note that higher values of $\hat{\xi}_{0}$ allow to achieve larger values of $\langle\sigma v\rangle$.}
\label{figvarxi0mdm}
\end{figure}

On the other hand, in Figure \ref{figvarkappa}, we depict the parameter space $(m_\chi,\langle\sigma v\rangle)$ for $T_\text{end}=7\times 10^{-3}$ GeV, $\hat{\xi_0}=10^{-2}$, and $\omega=0$. The solid green, red, and blue lines correspond to the parameter space that reproduces the current DM relic density for $\kappa=1$, $10^{-1}$, and $10^{-3}$, respectively. The blue zone represents the parameter space in which $\rho_\phi<\rho_\gamma$, $\forall t$, for $\kappa=10^{-3}$. From the figure, it can be observed that higher values of $\kappa$ shift the curves upward, reaching larger values of $\langle\sigma v\rangle$ in the same DM mass range. Otherwise, Figure \ref{figvartend} shows the same parameter space as Figure \ref{figvarkappa} for $\kappa=10^{-3}$, $\hat{\xi_0}=10^{-2}$, and $\omega=0$. The solid green, red, and blue lines correspond to the parameter space that reproduces the current DM relic density for $T_\text{end}=7\times10^{-3}$, $10^{-1}$, and $1$ GeV, respectively. The blue and red zones represent the parameter space in which $\rho_\phi<\rho_\gamma$, $\forall t$, for $T_\text{end}=1$ and $10^{-1}$ GeV, respectively. As it can be noted, an increment in the $T_\text{end}$-values shifts downward to the right of the parameter space that reproduces the DM relic density, reaching smaller values of $\langle\sigma v\rangle$ in the same $m_\chi$ range. Note that the regions in which $\rho_\phi<\rho_\gamma$ are shifted to the right and, therefore, the green zone is achieved at lower values of $m_{\chi}$. Finally, Figure \ref{figvarxi0mdm} shows the parameter space mentioned before for $\kappa=10^{-3}$, $T_\text{end}=7\times10^{-3}$ GeV, and $\omega=0$. The solid magenta, blue, red, and green lines correspond to the parameter space that reproduces the current DM relic density for $\hat{\xi}_0=10^{-1}$, $5\times10^{-2}$, $2.5\times10^{-2}$, and $10^{-3}$, respectively. The blue, red, and green zones represent the parameter space in which $\rho_\phi<\rho_\gamma$, $\forall t$, for $\hat{\xi}_0=5\times10^{-2}$, $2.5\times10^{-2}$, and $10^{-3}$, respectively. As it is possible to see, higher values of $\hat{\xi_0}$ shift the curves upward, reaching larger values of $\langle\sigma v\rangle$. Also, this increment generates a more prominent slope due to the bulk viscosity term. In this case, the regions in which $\rho_\phi<\rho_\gamma$ are shifted to the left and, therefore, the magenta zone is achieved at lower values of $m_{\chi}$.

The aforementioned shows how certain parameters of the bulk viscosity NSC vary the FIMP DM candidates parameters to reproduce the current DM relic density. For instance, a combination of small values for $\hat{\xi_0}$, $\kappa$, and $T_\text{end}$ could reach smaller values for $\langle\sigma v\rangle$ and a shorter range for $m_\chi$. On the other hand, for stronger DM interaction, it is useful to increase the value of the quantities $\hat{\xi_0}$, $\kappa$, and $T_\text{end}$, which also expands the DM mass range. Notice that the bulk viscous term dominates in these shifts for the DM parameter space.

\begin{figure}
\centering
\includegraphics[scale=0.80]{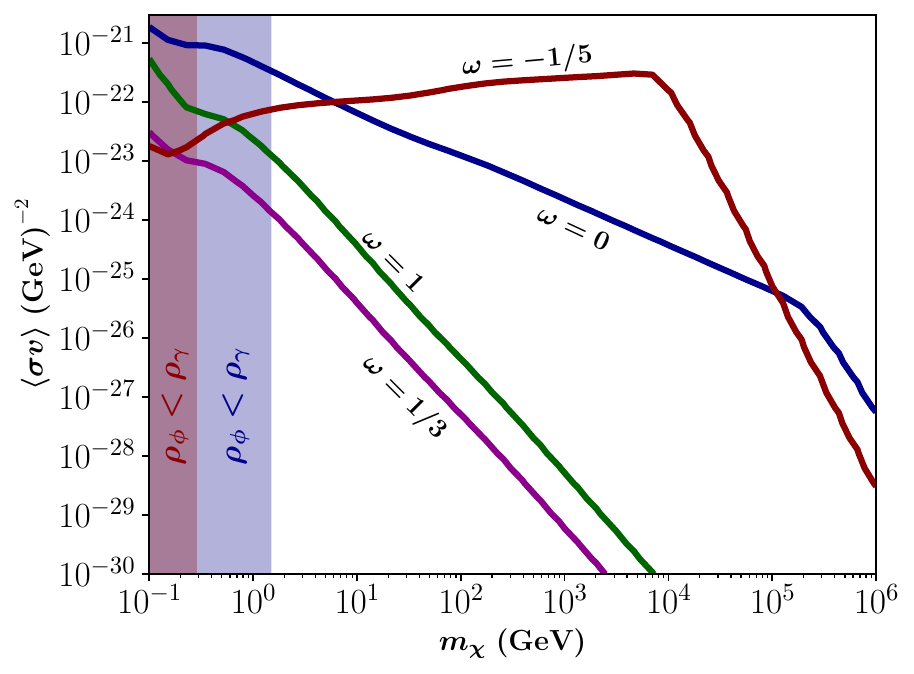}
\caption{Parameter space $(m_\chi,\langle\sigma v\rangle)$ for the FIMP DM candidate with $T_\text{end}=7\times10^{-3}$ GeV and $\hat{\xi}_0=10^{-2}$. The solid red, blue, green, and magenta lines correspond to the parameter space that reproduces the current DM relic density for $\omega=-1/5$, $0$, $1$, and $1/3$, respectively. The blue and red zones represent the parameter space in which $\rho_\phi<\rho_\gamma$, $\forall t$, for $\omega=0$ and $-1/5$, respectively. For illustrative purposes, the curves corresponding to $\omega=-1/5$ and $\omega=0$ were obtained with $\kappa=10^{-3}$, while the curve for $\omega=1/3$ was obtained using $\kappa=1$, and the curve for $\omega=1$ with $\kappa=10^3$.}
\label{figvaromega}
\end{figure}

Last but not least, we present in Figure \ref{figvaromega} the parameter space $(m_\chi,\langle\sigma v\rangle)$ for the FIMP DM candidate with $T_\text{end}=7\times10^{-3}$ GeV and $\hat{\xi}_0=10^{-2}$. The solid red, blue, green, and magenta lines correspond to the parameter space that reproduces the current DM relic density for $\omega=-1/5$, $0$, $1$, and $1/3$, respectively. The blue and red zones represent the parameter space in which $\rho_\phi<\rho_\gamma$, $\forall t$, for $\omega=0$ and $-1/5$, respectively. For illustrative purposes, the curves corresponding to $\omega=-1/5$ and $\omega=0$ were obtained with $\kappa=10^{-3}$, while the curve for $\omega=1/3$ was obtained using $\kappa=1$, and the curve for $\omega=1$ with $\kappa=10^3$. Notice that small values of $\omega$ can be interpreted as a counterclockwise rotation (and vice versa), leading to a maximum $\langle\sigma v\rangle$ in the cases with small values of $\omega$. On the other hand, the larger range for $m_\chi$ is reaching for the case $\omega=0$ (CDM or dust case). Also, it was shown that larger values in $\kappa$ shift the curves upward. This last possibly explains why the parameters that reproduce the current DM relic density for $\omega=1/3$ (radiation) are below $\omega=1$ (kination or stiff matter). It is important to mention that this analysis must be taken carefully because some curves were obtained using different values of $\kappa$.

\begin{figure}
\centering
\includegraphics[scale=0.80]{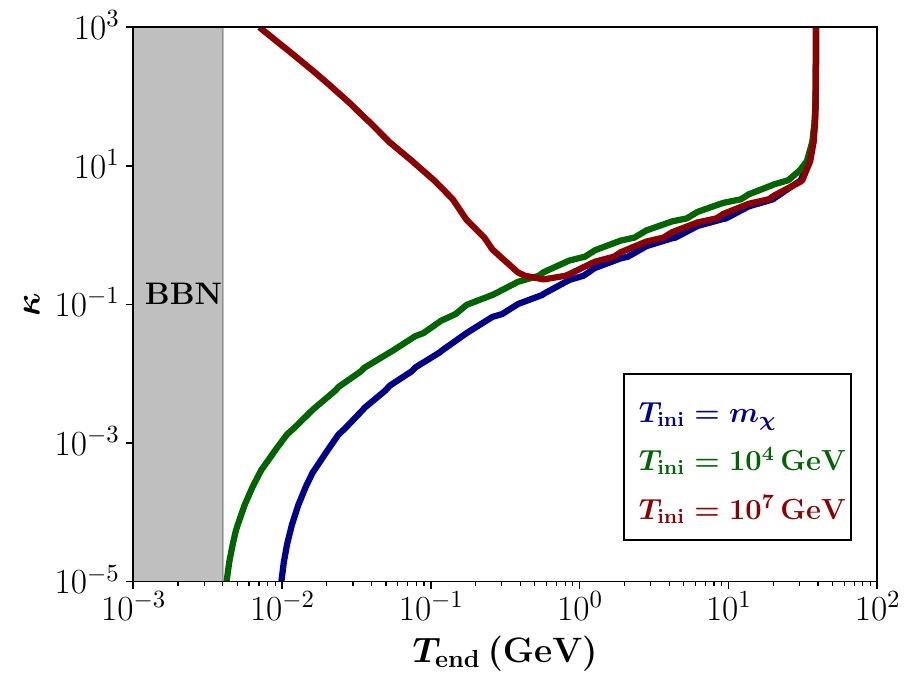}
\caption{Parameter space $(T_\text{end},\kappa)$ for the bulk viscous NSC with $m_\chi=100$ GeV, $\langle\sigma v\rangle=10^{-26}$ GeV$^{-2}$, and $\omega=0$. The solid blue, green and red lines correspond to the parameter space that reproduces the current DM relic density for different intial temperatures $T_\text{ini}=m_\chi$, $T_\text{ini}=10^4$ GeV, and $T_\text{ini}=10^7$ GeV respectively. The grey zone corresponds the forbidden BBN epoch that starts at $T_\text{BBN}\sim 4\times 10^{-3}$ GeV.}
\label{figvarTiktend}
\end{figure}

Another interesting feature to analyze is the maximum temperature of the SM bath, i.e., the initial condition for the temperature that we set in the numerical integration in the previous analysis. Figure \ref{figvarTiktend} shows the parameter space $(T_\text{end}\, \kappa)$ for the bulk viscous NSC with $m_\chi=100$ GeV, $\langle\sigma v\rangle=10^{-26}$ GeV$^{-2}$, and $\omega=0$. The solid blue, green, and red lines correspond to the parameter space that reproduces the current DM relic density for the initial temperatures $T_\text{ini}=m_\chi$, $T_\text{ini}=10^4$ GeV, and $T_\text{ini}=10^5$ GeV, respectively. The grey zone corresponds to the forbidden BBN epoch which begins at $T_\text{BBN}\sim 4\times 10^{-3}$ GeV. From the figure, it is possible to see that higher initial temperatures change the parameter space, which was previously observed in Figure \ref{figvarmdm} with other parameters for the DM candidate. In the latter, a region was observed in which the parameters that reproduce the current DM relic density go up when the $\kappa$-values are lower. This is exactly what we observed in Figure \ref{figvarTiktend}, where for higher temperatures the range of values changes for lower values of $\kappa$. Note that the case for $T_\text{ini}=10^4$ is slightly separated from the case of $T_\text{ini}=m_\chi=100$ GeV and, for higher initial temperatures, the curves do not slope downward but rather upward as $\kappa$ decreases.

\begin{figure}
\centering
\includegraphics[scale=0.80]{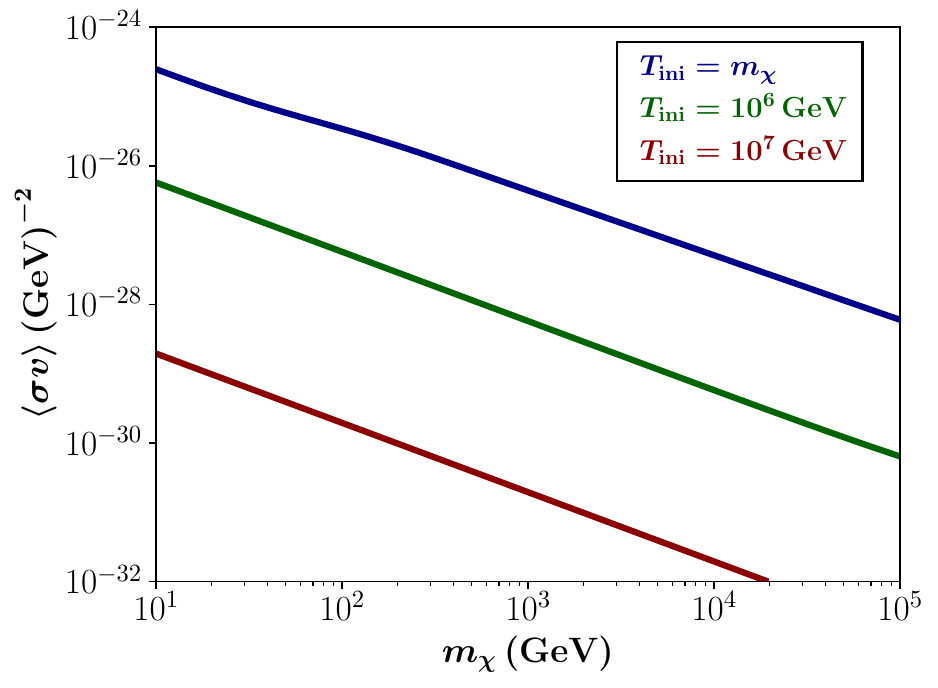}
\caption{Parameter space $(m_\chi,\langle\sigma v\rangle)$ for the bulk viscous NSC with $\kappa=10^{-3}$, $T_\text{end}=7\times10^{-3}$ GeV, and $\omega=0$. The solid blue, green, and red lines correspond to the parameter space that reproduces the current DM relic density for initial temperature $T_\text{ini}=m\chi$, $T_\text{ini}=10^6$ GeV, and $T_\text{ini}=10^7$ GeV, respectively.}
\label{figvarTisvmdm}
\end{figure}

Following the same line, Figure \ref{figvarTisvmdm} shows the parameter space $(m_\chi\, \langle\sigma v\rangle)$ for the bulk viscous NSC with $\kappa=10^{-3}$, $T_\text{end}=7\times 10^{-3}$ GeV, and $\omega=0$. The solid blue, green, and red lines correspond to the parameter space that reproduces the current DM relic density for the initial temperatures $T_\text{ini}=m_\chi$, $T_\text{ini}=10^6$ GeV, and $T_\text{ini}=10^7$ GeV, respectively. In this case, the parameters that reproduce the current DM relic density vary with the initial temperature, with higher values of $T_\text{ini}$ shifting the curves downward. It is important to note that the curve for $T_\text{ini}=m_\chi$ changes for every point in the plot.

\subsection{FIMPs for a non-constant totally thermal averaged DM production cross-section}\label{sec:Non-constant}
\begin{figure}
\centering
\includegraphics[scale=0.80]{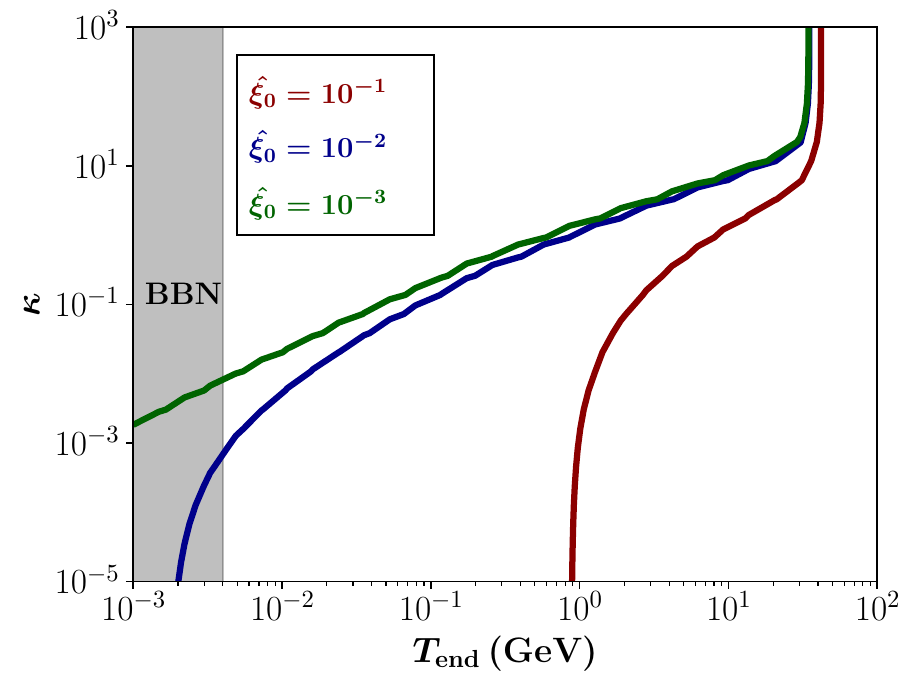}
\caption{Parameter space $(T_\text{end}$, $\kappa$) for the FIMP DM candidate with non-constant $\langle\sigma v \rangle=T^2/\Lambda^4$, considering $\Lambda = 1.8\times 10^7$ GeV, $n=2$, $m_\chi=100$ GeV, and $\omega=0$. The solid red, blue, and green lines correspond to the parameter space that reproduces the current DM relic density for $\hat{\xi}_0=10^{-1}$, $5\times10^{-2}$ and $10^{-3}$, respectively. The grey zone corresponds the forbidden BBN epoch that starts at $T_\text{BBN}\sim 4\times 10^{-3}$ GeV.}
\label{figvartendkappan2}
\end{figure}

In the case of a total thermally averaged DM production cross-section such as the one described by Eq. \eqref{svnocte}, exists a temperature dependence related to the order of the mass dimension operator $(n\neq0)$ and the energy scale of the interaction $(\Lambda)$. Figure \ref{figvartendkappan2} shows the model parameter space $(T_\text{end},\, \kappa)$ for the FIMP DM candidate with non-constant $\langle\sigma v\rangle=T^2/\Lambda^4$, considering $\Lambda=1.8\times10^7$ GeV, $n=2$, $m_\chi=100$ GeV and $\omega=0$. The solid red, blue, and green lines correspond to the parameter space that reproduces the current DM relic density for $\hat{\xi}_0=10^{-1}$, $5\times10^{-2}$ and $10^{-3}$, respectively. The behavior is similar to the one depicted in Figure \ref{figvarxi0}, showing that for higher values of $\hat{\xi_0}$ the curve leans downward, achieving the $\kappa$-independent zone. In contrast, lower values of $\hat{\xi_0}$ tend the curve to the classical NSC.

\begin{figure}
\centering
\includegraphics[scale=0.80]{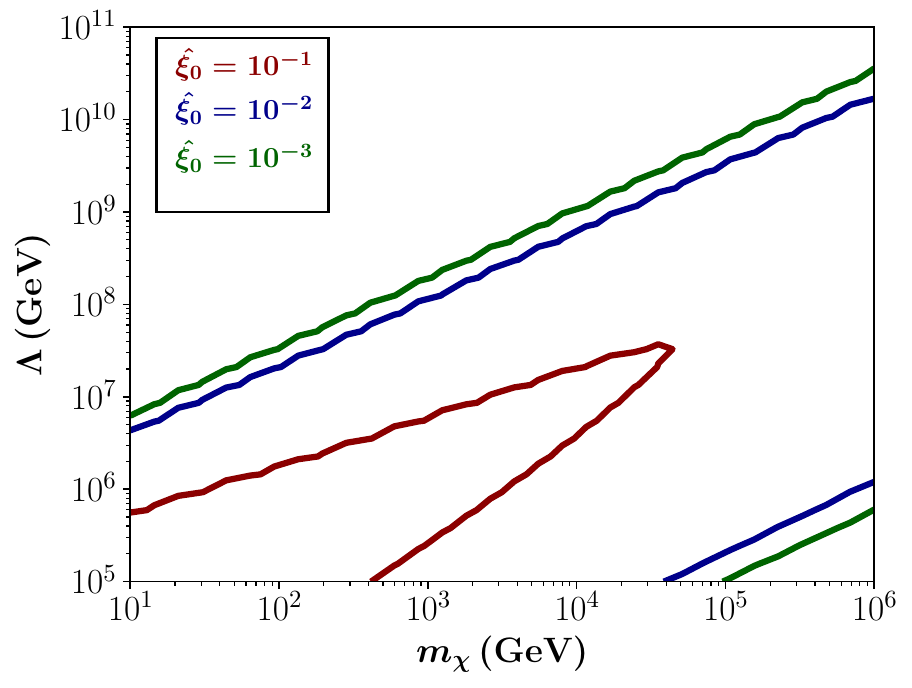}
\caption{Parameter space $(m_\chi,\Lambda)$ for the FIMP DM candidate with non-constant $\langle\sigma v \rangle=T^2/\Lambda^4$, considering $\kappa=10^{-3}$, $T_\text{end}=7\times10^{-3}$ GeV, $n=2$ and $\omega=0$. The solid red, blue, and green lines correspond to the parameter space that reproduces the current DM relic density for $\hat{\xi_0}=10^{-1}$, $10^{-2}$ and $10^{-3}$, respectively. Note that in the case when $\langle\sigma v\rangle\propto T^2$ the scale energy $\Lambda$ has two possible values for the same DM mass in a large scale of $m_\chi$.}
\label{figvarxi0n2}
\end{figure}

Figure \ref{figvarxi0n2} depicts the parameter space $(m_\chi,\, \Lambda)$ for the FIMP DM candidate with a non-constant $\langle\sigma v \rangle=T^2/\Lambda^4$, with $\kappa=10^{-3}$, $T_\text{end}=7\times10^{-3}$ GeV and $\omega=0$. The solid red, blue, and green lines correspond to the parameter space that reproduces the current DM relic density for $\hat{\xi_0}=10^{-1}$, $10^{-2}$ and $10^{-3}$, respectively. Notably, for $\langle\sigma v\rangle \propto T^2$, the energy scale $\Lambda$ can take two distinct values for the same DM mass over a broad range of $m_\chi$, a behavior that does not occur when $\langle\sigma v\rangle$ is constant. For higher values of $\hat{\xi_0}$, the parameter space that reproduces the current DM relic density shifts downward to the left. On the contrary, for lower values of $\hat{\xi_0}$ the curves shift upward to the right. It is important to note that lower values of $\hat{\xi_0}$ achieve wider windows in the same parameter space of searching.

\begin{figure}
\centering
\includegraphics[scale=0.80]{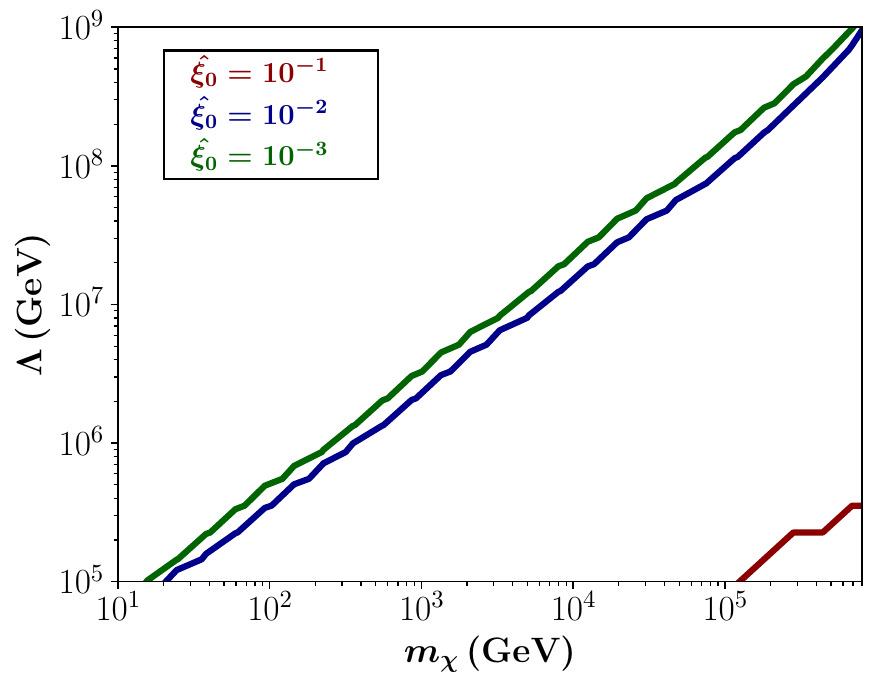}
\caption{Parameter space $(m_\chi,\Lambda)$ for the FIMP DM candidate with non-constant $\langle\sigma v \rangle=T^4/\Lambda^6$, considering $\kappa=10^{-3}$, $T_\text{end}=7\times10^{-3}$ GeV, $n=2$ and $\omega=0$. The solid red, blue, and green lines correspond to the parameter space that reproduces the current DM relic density for $\hat{\xi_0}=10^{-1}$, $10^{-2}$ and $10^{-3}$, respectively.}
\label{figvarxi0n4}
\end{figure}

The case of the mass-dimension 7 operator ($n=4$) is depicted in Figure \ref{figvarxi0n4}, showing the parameter space $(m_\chi,\Lambda)$ for the FIMP DM candidate with non-constant $\langle\sigma v \rangle=T^4/\Lambda^6$ considering $\kappa=10^{-3}$, $T_\text{end}=7\times10^{-3}$ GeV, $n=2$ and $\omega=0$. The solid red, blue, and green lines correspond to the parameter space that reproduces the current DM relic density for $\hat{\xi_0}=10^{-1}$, $10^{-2}$ and $10^{-3}$, respectively. The effect of higher values of $\hat{\xi_0}$ shift the curves downward to the right, showing slight differences between the $\hat{\xi_0}=10^{-2}$ and $10^{-3}$ cases, as it occurs in Figure \ref{figvarxi0n2}. However, it does not show the double value of $\Lambda$ for the same DM mass, at least at the same range of parameters.

Larger couplings would typically be excluded in the standard $\Lambda$CDM scenario, but remain viable within our viscous NSC framework can also enhance the rate of decay processes. If the DM candidates are not absolutely stable but have lifetimes exceeding the age of the Universe, they may decay into SM particles such as photons or neutrinos. This opens the possibility for indirect detection, provided the signal is sufficiently strong, through gamma-ray lines or, in models involving neutrino final states, via detectors such as IceCube or Super-Kamiokande \cite{Barranco:2017lug,Lambiase:2018yql}.

\section{Conclusions}\label{sec:Conclusions}
In this paper, we revisit the bulk viscous NSC scenario studied in \cite{Gonzalez:2024dtb} in which the early universe is dominated by two interacting fluids, namely the new field $\phi$ and radiation, by considering that $\phi$ experiences dissipative processes during their cosmic evolution in the form of bulk viscosity. Working in the framework of Eckart's theory for non-perfect fluids, we consider an interaction term of the form $\Gamma_{\phi}\rho_{\phi}$ and a bulk viscosity described by the expression $\xi=\xi_{0}\rho_{\phi}^{1/2}$. The latter has the advantage that when the field fully decays due to the interaction, the dissipation is negligible, and the standard $\Lambda$CDM cosmology is fully recovered. Also, when $\xi_{0}\to 0$, we recover the classical NSC scenario. Following the same scheme as in \cite{Gonzalez:2024dtb}, we investigate the parameter space that can accurately reproduce the current DM relic density by adjusting both, the model and DM parameters within this novel NSC framework for FIMP DM candidates.

The main finding of our work is that the inclusion of bulk viscosity in NSCs brings to light DM parameters that are disregarded in both, $\Lambda$CDM and the classical NSC scenarios, as it was shown in Figure \ref{figyield}, giving us new windows to search for FIMP DM candidates. In particular, modifying the values of $m_\chi$ and $\langle\sigma v\rangle$ can shift the parameters ($T_\text{end}$, $\kappa$) that reproduce the current DM relic density. For example, an increment in the values of $m_\chi$ and $\langle\sigma v\rangle$ can shift the parameter space upward to the right (Figure \ref{figvarmdm}) and upward to the left (Figure \ref{figvarsv}), respectively. These displacements can be combined to reach new windows of allowed parameters if a signal of DM is experimentally detected. In the same line, larger values of $\hat{\xi_0}$ (Figure \ref{figvarxi0}) make more prominent the curvature to the approximately independent $\kappa$-zone (as it was described in Section \ref{sec:Comparison}) and shift the parameters downward to the right. On the other hand, modifying the model parameters shifts the ($m_\chi$, $\langle\sigma v \rangle$) values that reproduce the current DM relic density. In particular, increasing the values of $\kappa$ and $T_\text{end}$ shift the parameter space upward (Figure \ref{figvarkappa}) and downward to the right (Figure \ref{figvartend}), respectively, reaching larger values of $\langle\sigma v\rangle$ and $m_\chi$. In this sense, larger values of $\hat{\xi_0}$ (Figure \ref{figvarxi0mdm}) make significant imprints in the DM parameters reaching larger values of $\langle\sigma v\rangle$.

On the other hand, the maximum temperature of the thermal bath leaves imprints in the FIMP DM production, as was shown in Figures \ref{figvarTiktend} and \ref{figvarTisvmdm}. In the first one, higher initial temperatures lead to a change in the trend of the parameters that reproduce the current DM relic density, shifting from downward to upward. In the second one, higher values of initial temperatures shift the parameter space that reproduces the current DM relic density downward to the left.

Finally, the case with non-constant $\langle\sigma v\rangle$ is studied for two scenarios: 1) for an operator of mass dimension 6 $(n=2)$ and 2) for an operator of mass dimension 7 $(n=4)$. In the first case, Figure \ref{figvartendkappan2} shows the same behavior previously observed for a constant value of $\langle\sigma v\rangle$ when the value of $\hat{\xi_0}$ changes. In contrast, Figure~\ref{figvarxi0n2} shows that $\Lambda$ can take two distinct values for the same DM mass, a feature that does not arise in the case of a constant total thermally averaged DM production cross-section. Additionally, higher values of $\hat{\xi_0}$ shift the curves downward to the left. In the $n=4$ case (Figure~\ref{figvarxi0n4}), increasing $\hat{\xi}_0$ shifts the curves downward to the right.
The larger couplings that are needed to set the current relic density opens the possibility for indirect detection in next-generation experiments.

Therefore, this paper is a further step in the study of FIMPs as DM candidates and a first approximation to highlight the imprints that the bulk viscosity can leave in these particles and their relic density in the early universe through a NSC scenario. In this sense, this work, in collaboration with Ref. \cite{Gonzalez:2024dtb}, provides a comprehensive set of possibilities for opening new avenues of research in the study of FIMPs and WIMPs as DM candidates. On the other hand, including various cosmological observations, such as the CMB, LSS, or BBN, would allow for tighter constraints on the dark matter parameters. This will be explored in future work.

\vspace{6pt} 

\authorcontributions{Conceptualization, E.G., C.M., N.S.M. and R.S.; methodology, E.G., C.M., N.S.M. and R.S.; software, C.M., N.S.M. and R.S.; validation, E.G. and C.M.; formal analysis, E.G., C.M., N.S.M. and R.S.; investigation, E.G., C.M., N.S.M. and R.S.; data curation, C.M., N.S.M. and R.S.; writing---original draft preparation, E.G. and C.M.; writing---review and editing, E.G., C.M., N.S.M. and R.S.; visualization, E.G., C.M., N.S.M. and R.S.; supervision, E.G. and C.M.; project administration, E.G. and C.M. All authors have read and agreed to the published version of the manuscript.}

\funding{This research received no external funding.}

\dataavailability{No new data were generated during the development of this work.} 

\acknowledgments{E.G. acknowledges the scientific support of Núcleo de Investigación No. 7 UCN-VRIDT 076/2020, Núcleo de Modelación y Simulación Científica (NMSC).}

\conflictsofinterest{The authors declare no conflicts of interest.} 

\abbreviations{Abbreviations}{
The following abbreviations are used in this manuscript:
\\

\noindent 
\begin{tabular}{@{}ll}
DM & Dark Matter\\
DE & Dark Energy\\
CDM & Cold Dark Matter\\
WIMPs & Weakly Interacting Massive Particles\\
FIMPs & Feebly Interacting Massive Particles\\
SM & tandard Model\\
WDM & Warm Dark Matter\\
NSCs & Non-Standard Cosmologies\\
BBN & Big Bang Nucleosynthesis\\
RI & Region I\\
RII & Region II\\
RIII & Region III\\
RVI & Region VI\\
UV & UltraViolet\\
IR &  InfraRed
\end{tabular}
}

\isPreprints{}{
\begin{adjustwidth}{-\extralength}{0cm}
} 

\reftitle{References}


\bibliography{bibliography.bib}

\begin{thebibliography}{999}

\bibitem[Riess et~al.(1998)]{SupernovaSearchTeam:1998fmf}
Riess, A.G.;  et~al.
\newblock {Observational evidence from supernovae for an accelerating universe and a cosmological constant}.
\newblock {\em Astron. J.} {\bf 1998}, {\em 116},~1009--1038,  \href{http://arxiv.org/abs/astro-ph/9805201}{{\normalfont [astro-ph/9805201]}}.
\newblock {\url{https://doi.org/10.1086/300499}}.

\bibitem[Perlmutter et~al.(1999)]{SupernovaCosmologyProject:1998vns}
Perlmutter, S.;  et~al.
\newblock {Measurements of $\Omega$ and $\Lambda$ from 42 High Redshift Supernovae}.
\newblock {\em Astrophys. J.} {\bf 1999}, {\em 517},~565--586,  \href{http://arxiv.org/abs/astro-ph/9812133}{{\normalfont [astro-ph/9812133]}}.
\newblock {\url{https://doi.org/10.1086/307221}}.

\bibitem[Scolnic et~al.(2022)]{Scolnic:2021amr}
Scolnic, D.;  et~al.
\newblock {The Pantheon+ Analysis: The Full Data Set and Light-curve Release}.
\newblock {\em Astrophys. J.} {\bf 2022}, {\em 938},~113,  \href{http://arxiv.org/abs/2112.03863}{{\normalfont [arXiv:astro-ph.CO/2112.03863]}}.
\newblock {\url{https://doi.org/10.3847/1538-4357/ac8b7a}}.

\bibitem[Moresco et~al.(2012)]{Moresco:2012jh}
Moresco, M.;  et~al.
\newblock {Improved constraints on the expansion rate of the Universe up to z\textasciitilde{}1.1 from the spectroscopic evolution of cosmic chronometers}.
\newblock {\em JCAP} {\bf 2012}, {\em 08},~006,  \href{http://arxiv.org/abs/1201.3609}{{\normalfont [arXiv:astro-ph.CO/1201.3609]}}.
\newblock {\url{https://doi.org/10.1088/1475-7516/2012/08/006}}.

\bibitem[Zhang et~al.(2014)Zhang, Zhang, Yuan, Zhang, and Sun]{Zhang:2012mp}
Zhang, C.; Zhang, H.; Yuan, S.; Zhang, T.J.; Sun, Y.C.
\newblock {Four new observational $H(z)$ data from luminous red galaxies in the Sloan Digital Sky Survey data release seven}.
\newblock {\em Res. Astron. Astrophys.} {\bf 2014}, {\em 14},~1221--1233,  \href{http://arxiv.org/abs/1207.4541}{{\normalfont [arXiv:astro-ph.CO/1207.4541]}}.
\newblock {\url{https://doi.org/10.1088/1674-4527/14/10/002}}.

\bibitem[Moresco(2015)]{Moresco:2015cya}
Moresco, M.
\newblock {Raising the bar: new constraints on the Hubble parameter with cosmic chronometers at z \ensuremath{\sim} 2}.
\newblock {\em Mon. Not. Roy. Astron. Soc.} {\bf 2015}, {\em 450},~L16--L20,  \href{http://arxiv.org/abs/1503.01116}{{\normalfont [arXiv:astro-ph.CO/1503.01116]}}.
\newblock {\url{https://doi.org/10.1093/mnrasl/slv037}}.

\bibitem[Eisenstein et~al.(2005)]{SDSS:2005xqv}
Eisenstein, D.J.;  et~al.
\newblock {Detection of the Baryon Acoustic Peak in the Large-Scale Correlation Function of SDSS Luminous Red Galaxies}.
\newblock {\em Astrophys. J.} {\bf 2005}, {\em 633},~560--574,  \href{http://arxiv.org/abs/astro-ph/0501171}{{\normalfont [astro-ph/0501171]}}.
\newblock {\url{https://doi.org/10.1086/466512}}.

\bibitem[Bennett et~al.(2013)]{WMAP:2012fli}
Bennett, C.L.;  et~al.
\newblock {Nine-Year Wilkinson Microwave Anisotropy Probe (WMAP) Observations: Final Maps and Results}.
\newblock {\em Astrophys. J. Suppl.} {\bf 2013}, {\em 208},~20,  \href{http://arxiv.org/abs/1212.5225}{{\normalfont [arXiv:astro-ph.CO/1212.5225]}}.
\newblock {\url{https://doi.org/10.1088/0067-0049/208/2/20}}.

\bibitem[Aghanim et~al.(2020)]{Planck:2018vyg}
Aghanim, N.;  et~al.
\newblock {Planck 2018 results. VI. Cosmological parameters}.
\newblock {\em Astron. Astrophys.} {\bf 2020}, {\em 641},~A6,  \href{http://arxiv.org/abs/1807.06209}{{\normalfont [arXiv:astro-ph.CO/1807.06209]}}.
\newblock [Erratum: Astron.Astrophys. 652, C4 (2021)], {\url{https://doi.org/10.1051/0004-6361/201833910}}.

\bibitem[Riess et~al.(2022)]{Riess:2021jrx}
Riess, A.G.;  et~al.
\newblock {A Comprehensive Measurement of the Local Value of the Hubble Constant with 1 km s$^{−1}$ Mpc$^{−1}$ Uncertainty from the Hubble Space Telescope and the SH0ES Team}.
\newblock {\em Astrophys. J. Lett.} {\bf 2022}, {\em 934},~L7,  \href{http://arxiv.org/abs/2112.04510}{{\normalfont [arXiv:astro-ph.CO/2112.04510]}}.
\newblock {\url{https://doi.org/10.3847/2041-8213/ac5c5b}}.

\bibitem[Wong et~al.(2020)]{Wong:2019kwg}
Wong, K.C.;  et~al.
\newblock {H0LiCOW \textendash{} XIII. A 2.4 per cent measurement of H0 from lensed quasars: 5.3\ensuremath{\sigma} tension between early- and late-Universe probes}.
\newblock {\em Mon. Not. Roy. Astron. Soc.} {\bf 2020}, {\em 498},~1420--1439,  \href{http://arxiv.org/abs/1907.04869}{{\normalfont [arXiv:astro-ph.CO/1907.04869]}}.
\newblock {\url{https://doi.org/10.1093/mnras/stz3094}}.

\bibitem[Jungman et~al.(1996)Jungman, Kamionkowski, and Griest]{Jungman:1995df}
Jungman, G.; Kamionkowski, M.; Griest, K.
\newblock {Supersymmetric dark matter}.
\newblock {\em Phys. Rept.} {\bf 1996}, {\em 267},~195--373,  \href{http://arxiv.org/abs/hep-ph/9506380}{{\normalfont [hep-ph/9506380]}}.
\newblock {\url{https://doi.org/10.1016/0370-1573(95)00058-5}}.

\bibitem[King and Roberts(2007)]{King:2006cu}
King, S.F.; Roberts, J.P.
\newblock {Natural Dark Matter from Type I String Theory}.
\newblock {\em JHEP} {\bf 2007}, {\em 01},~024,  \href{http://arxiv.org/abs/hep-ph/0608135}{{\normalfont [hep-ph/0608135]}}.
\newblock {\url{https://doi.org/10.1088/1126-6708/2007/01/024}}.

\bibitem[Steigman and Turner(1985)]{Steigman:1984ac}
Steigman, G.; Turner, M.S.
\newblock {Cosmological Constraints on the Properties of Weakly Interacting Massive Particles}.
\newblock {\em Nucl. Phys. B} {\bf 1985}, {\em 253},~375--386.
\newblock {\url{https://doi.org/10.1016/0550-3213(85)90537-1}}.

\bibitem[Bertone et~al.(2005)Bertone, Hooper, and Silk]{Bertone:2004pz}
Bertone, G.; Hooper, D.; Silk, J.
\newblock {Particle dark matter: Evidence, candidates and constraints}.
\newblock {\em Phys. Rept.} {\bf 2005}, {\em 405},~279--390,  \href{http://arxiv.org/abs/hep-ph/0404175}{{\normalfont [hep-ph/0404175]}}.
\newblock {\url{https://doi.org/10.1016/j.physrep.2004.08.031}}.

\bibitem[Arcadi et~al.(2018)Arcadi, Dutra, Ghosh, Lindner, Mambrini, Pierre, Profumo, and Queiroz]{Arcadi:2017kky}
Arcadi, G.; Dutra, M.; Ghosh, P.; Lindner, M.; Mambrini, Y.; Pierre, M.; Profumo, S.; Queiroz, F.S.
\newblock {The waning of the WIMP? A review of models, searches, and constraints}.
\newblock {\em Eur. Phys. J. C} {\bf 2018}, {\em 78},~203,  \href{http://arxiv.org/abs/1703.07364}{{\normalfont [arXiv:hep-ph/1703.07364]}}.
\newblock {\url{https://doi.org/10.1140/epjc/s10052-018-5662-y}}.

\bibitem[Roszkowski et~al.(2018)Roszkowski, Sessolo, and Trojanowski]{Roszkowski:2017nbc}
Roszkowski, L.; Sessolo, E.M.; Trojanowski, S.
\newblock {WIMP dark matter candidates and searches\textemdash{}current status and future prospects}.
\newblock {\em Rept. Prog. Phys.} {\bf 2018}, {\em 81},~066201,  \href{http://arxiv.org/abs/1707.06277}{{\normalfont [arXiv:hep-ph/1707.06277]}}.
\newblock {\url{https://doi.org/10.1088/1361-6633/aab913}}.

\bibitem[Arcadi et~al.(2024)Arcadi, Cabo-Almeida, Dutra, Ghosh, Lindner, Mambrini, Neto, Pierre, Profumo, and Queiroz]{Arcadi:2024ukq}
Arcadi, G.; Cabo-Almeida, D.; Dutra, M.; Ghosh, P.; Lindner, M.; Mambrini, Y.; Neto, J.P.; Pierre, M.; Profumo, S.; Queiroz, F.S.
\newblock {The Waning of the WIMP: Endgame?},  2024,  \href{http://arxiv.org/abs/2403.15860}{{\normalfont [arXiv:hep-ph/2403.15860]}}.

\bibitem[Singh et~al.(2024)Singh, Bharadwaj, and Devisharan]{Singh:2024wdn}
Singh, E.; Bharadwaj, H.; Devisharan.
\newblock {Exploring new physics via effective interactions}.
\newblock {\em Int. J. Mod. Phys. A} {\bf 2024}, {\em 39},~2450038,  \href{http://arxiv.org/abs/2406.14389}{{\normalfont [arXiv:hep-ph/2406.14389]}}.
\newblock {\url{https://doi.org/10.1142/S0217751X24500386}}.

\bibitem[Bernal et~al.(2017)Bernal, Heikinheimo, Tenkanen, Tuominen, and Vaskonen]{Bernal:2017kxu}
Bernal, N.; Heikinheimo, M.; Tenkanen, T.; Tuominen, K.; Vaskonen, V.
\newblock {The Dawn of FIMP Dark Matter: A Review of Models and Constraints}.
\newblock {\em Int. J. Mod. Phys. A} {\bf 2017}, {\em 32},~1730023,  \href{http://arxiv.org/abs/1706.07442}{{\normalfont [arXiv:hep-ph/1706.07442]}}.
\newblock {\url{https://doi.org/10.1142/S0217751X1730023X}}.

\bibitem[Chu et~al.(2012)Chu, Hambye, and Tytgat]{Chu:2011be}
Chu, X.; Hambye, T.; Tytgat, M.H.G.
\newblock {The Four Basic Ways of Creating Dark Matter Through a Portal}.
\newblock {\em JCAP} {\bf 2012}, {\em 05},~034,  \href{http://arxiv.org/abs/1112.0493}{{\normalfont [arXiv:hep-ph/1112.0493]}}.
\newblock {\url{https://doi.org/10.1088/1475-7516/2012/05/034}}.

\bibitem[Hall et~al.(2010)Hall, Jedamzik, March-Russell, and West]{Hall:2009bx}
Hall, L.J.; Jedamzik, K.; March-Russell, J.; West, S.M.
\newblock {Freeze-In Production of FIMP Dark Matter}.
\newblock {\em JHEP} {\bf 2010}, {\em 03},~080,  \href{http://arxiv.org/abs/0911.1120}{{\normalfont [arXiv:hep-ph/0911.1120]}}.
\newblock {\url{https://doi.org/10.1007/JHEP03(2010)080}}.

\bibitem[Abi et~al.(2020)]{DUNE:2020lwj}
Abi, B.;  et~al.
\newblock {Deep Underground Neutrino Experiment (DUNE), Far Detector Technical Design Report, Volume I Introduction to DUNE}.
\newblock {\em JINST} {\bf 2020}, {\em 15},~T08008,  \href{http://arxiv.org/abs/2002.02967}{{\normalfont [arXiv:physics.ins-det/2002.02967]}}.
\newblock {\url{https://doi.org/10.1088/1748-0221/15/08/T08008}}.

\bibitem[Feng(2010)]{Feng:2010gw}
Feng, J.L.
\newblock {Dark Matter Candidates from Particle Physics and Methods of Detection}.
\newblock {\em Ann. Rev. Astron. Astrophys.} {\bf 2010}, {\em 48},~495--545,  \href{http://arxiv.org/abs/1003.0904}{{\normalfont [arXiv:astro-ph.CO/1003.0904]}}.
\newblock {\url{https://doi.org/10.1146/annurev-astro-082708-101659}}.

\bibitem[Alves et~al.(2008)]{LHCb:2008vvz}
Alves, Jr., A.A.;  et~al.
\newblock {The LHCb Detector at the LHC}.
\newblock {\em JINST} {\bf 2008}, {\em 3},~S08005.
\newblock {\url{https://doi.org/10.1088/1748-0221/3/08/S08005}}.

\bibitem[Steigman et~al.(2012)Steigman, Dasgupta, and Beacom]{Steigman:2012nb}
Steigman, G.; Dasgupta, B.; Beacom, J.F.
\newblock {Precise Relic WIMP Abundance and its Impact on Searches for Dark Matter Annihilation}.
\newblock {\em Phys. Rev. D} {\bf 2012}, {\em 86},~023506,  \href{http://arxiv.org/abs/1204.3622}{{\normalfont [arXiv:hep-ph/1204.3622]}}.
\newblock {\url{https://doi.org/10.1103/PhysRevD.86.023506}}.

\bibitem[Frigerio et~al.(2011)Frigerio, Hambye, and Masso]{Frigerio:2011in}
Frigerio, M.; Hambye, T.; Masso, E.
\newblock {Sub-GeV dark matter as pseudo-Goldstone from the seesaw scale}.
\newblock {\em Phys. Rev. X} {\bf 2011}, {\em 1},~021026,  \href{http://arxiv.org/abs/1107.4564}{{\normalfont [arXiv:hep-ph/1107.4564]}}.
\newblock {\url{https://doi.org/10.1103/PhysRevX.1.021026}}.

\bibitem[Wang et~al.(2024)Wang, Reyimuaji, and Yalikun]{Wang:2024qhe}
Wang, Z.; Reyimuaji, Y.; Yalikun, N.
\newblock {A $Z_4$ symmetric inverse seesaw model for neutrino masses and FIMP dark matter},  2024,  \href{http://arxiv.org/abs/2412.15672}{{\normalfont [arXiv:hep-ph/2412.15672]}}.

\bibitem[Babu et~al.(2023)Babu, Chakdar, Das, Ghosh, and Ghosh]{Babu:2023zni}
Babu, K.S.; Chakdar, S.; Das, N.; Ghosh, D.K.; Ghosh, P.
\newblock {FIMP dark matter from flavon portals}.
\newblock {\em JHEP} {\bf 2023}, {\em 07},~143,  \href{http://arxiv.org/abs/2305.03167}{{\normalfont [arXiv:hep-ph/2305.03167]}}.
\newblock {\url{https://doi.org/10.1007/JHEP07(2023)143}}.

\bibitem[Junius(2022)]{Junius:2022vzl}
Junius, S.
\newblock {Searching for feebly interacting dark matter in colliders and beam-dump experiments}.
\newblock PhD thesis, U. Brussels, U. Brussels (main),  2022,  \href{http://arxiv.org/abs/2212.09432}{{\normalfont [arXiv:hep-ph/2212.09432]}}.

\bibitem[Barman et~al.(2024)Barman, Bhattacharya, Jahedi, Pradhan, and Sarkar]{Barman:2024nhr}
Barman, B.; Bhattacharya, S.; Jahedi, S.; Pradhan, D.; Sarkar, A.
\newblock {Lepton Collider as a window to Reheating},  2024,  \href{http://arxiv.org/abs/2406.11963}{{\normalfont [arXiv:hep-ph/2406.11963]}}.

\bibitem[B\'elanger et~al.(2024)B\'elanger, Chakraborti, and Pukhov]{Belanger:2023azf}
B\'elanger, G.; Chakraborti, S.; Pukhov, A.
\newblock {Feebly-interacting dark matter}.
\newblock {\em Eur. Phys. J. ST} {\bf 2024}, {\em 233},~11--12,  \href{http://arxiv.org/abs/2309.00491}{{\normalfont [arXiv:hep-ph/2309.00491]}}.
\newblock {\url{https://doi.org/10.1140/epjs/s11734-024-01134-1}}.

\bibitem[Blinnikov and Khlopov(1982)]{Blinnikov:1982eh}
Blinnikov, S.I.; Khlopov, M.Y.
\newblock {ON POSSIBLE EFFECTS OF 'MIRROR' PARTICLES}.
\newblock {\em Sov. J. Nucl. Phys.} {\bf 1982}, {\em 36},~472.

\bibitem[Blinnikov and Khlopov(1983)]{Blinnikov:1983gh}
Blinnikov, S.I.; Khlopov, M.
\newblock {Possible astronomical effects of mirror particles}.
\newblock {\em Sov. Astron.} {\bf 1983}, {\em 27},~371--375.

\bibitem[Khlopov et~al.(1991)Khlopov, Beskin, Bochkarev, Pustylnik, and Pustylnik]{Khlopov:1989fj}
Khlopov, M.Y.; Beskin, G.M.; Bochkarev, N.E.; Pustylnik, L.A.; Pustylnik, S.A.
\newblock {Observational Physics of Mirror World}.
\newblock {\em Sov. Astron.} {\bf 1991}, {\em 35},~21.

\bibitem[D'Eramo and Lenoci(2021)]{DEramo:2020gpr}
D'Eramo, F.; Lenoci, A.
\newblock {Lower mass bounds on FIMP dark matter produced via freeze-in}.
\newblock {\em JCAP} {\bf 2021}, {\em 10},~045,  \href{http://arxiv.org/abs/2012.01446}{{\normalfont [arXiv:hep-ph/2012.01446]}}.
\newblock {\url{https://doi.org/10.1088/1475-7516/2021/10/045}}.

\bibitem[McDonald(2016)]{McDonald:2015ljz}
McDonald, J.
\newblock {Warm Dark Matter via Ultra-Violet Freeze-In: Reheating Temperature and Non-Thermal Distribution for Fermionic Higgs Portal Dark Matter}.
\newblock {\em JCAP} {\bf 2016}, {\em 08},~035,  \href{http://arxiv.org/abs/1512.06422}{{\normalfont [arXiv:hep-ph/1512.06422]}}.
\newblock {\url{https://doi.org/10.1088/1475-7516/2016/08/035}}.

\bibitem[Klasen and Yaguna(2013)]{Klasen:2013ypa}
Klasen, M.; Yaguna, C.E.
\newblock {Warm and cold fermionic dark matter via freeze-in}.
\newblock {\em JCAP} {\bf 2013}, {\em 11},~039,  \href{http://arxiv.org/abs/1309.2777}{{\normalfont [arXiv:hep-ph/1309.2777]}}.
\newblock {\url{https://doi.org/10.1088/1475-7516/2013/11/039}}.

\bibitem[de~Giorgi and Vogl(2023)]{deGiorgi:2022yha}
de~Giorgi, A.; Vogl, S.
\newblock {Warm dark matter from a gravitational freeze-in in extra dimensions}.
\newblock {\em JHEP} {\bf 2023}, {\em 04},~032,  \href{http://arxiv.org/abs/2208.03153}{{\normalfont [arXiv:hep-ph/2208.03153]}}.
\newblock {\url{https://doi.org/10.1007/JHEP04(2023)032}}.

\bibitem[Kuo et~al.(2018)Kuo, Lattanzi, Cheung, and Valle]{Kuo:2018fgw}
Kuo, J.L.; Lattanzi, M.; Cheung, K.; Valle, J.W.F.
\newblock {Decaying warm dark matter and structure formation}.
\newblock {\em JCAP} {\bf 2018}, {\em 12},~026,  \href{http://arxiv.org/abs/1803.05650}{{\normalfont [arXiv:astro-ph.CO/1803.05650]}}.
\newblock {\url{https://doi.org/10.1088/1475-7516/2018/12/026}}.

\bibitem[Bode et~al.(2001)Bode, Ostriker, and Turok]{Bode:2000gq}
Bode, P.; Ostriker, J.P.; Turok, N.
\newblock {Halo formation in warm dark matter models}.
\newblock {\em Astrophys. J.} {\bf 2001}, {\em 556},~93--107,  \href{http://arxiv.org/abs/astro-ph/0010389}{{\normalfont [astro-ph/0010389]}}.
\newblock {\url{https://doi.org/10.1086/321541}}.

\bibitem[de~Vega and Sanchez(2011)]{deVega:2011si}
de~Vega, H.J.; Sanchez, N.G.
\newblock {Warm dark matter in the galaxies:theoretical and observational progresses. Highlights and conclusions of the chalonge meudon workshop 2011},  2011,  \href{http://arxiv.org/abs/1109.3187}{{\normalfont [arXiv:astro-ph.CO/1109.3187]}}.

\bibitem[Klypin et~al.(1999)Klypin, Kravtsov, Valenzuela, and Prada]{Klypin:1999uc}
Klypin, A.A.; Kravtsov, A.V.; Valenzuela, O.; Prada, F.
\newblock {Where are the missing Galactic satellites?}
\newblock {\em Astrophys. J.} {\bf 1999}, {\em 522},~82--92,  \href{http://arxiv.org/abs/astro-ph/9901240}{{\normalfont [astro-ph/9901240]}}.
\newblock {\url{https://doi.org/10.1086/307643}}.

\bibitem[Moore et~al.(1999)Moore, Ghigna, Governato, Lake, Quinn, Stadel, and Tozzi]{Moore:1999nt}
Moore, B.; Ghigna, S.; Governato, F.; Lake, G.; Quinn, T.R.; Stadel, J.; Tozzi, P.
\newblock {Dark matter substructure within galactic halos}.
\newblock {\em Astrophys. J. Lett.} {\bf 1999}, {\em 524},~L19--L22,  \href{http://arxiv.org/abs/astro-ph/9907411}{{\normalfont [astro-ph/9907411]}}.
\newblock {\url{https://doi.org/10.1086/312287}}.

\bibitem[Weinberg et~al.(2015)Weinberg, Bullock, Governato, Kuzio~de Naray, and Peter]{Weinberg:2013aya}
Weinberg, D.H.; Bullock, J.S.; Governato, F.; Kuzio~de Naray, R.; Peter, A.H.G.
\newblock {Cold dark matter: controversies on small scales}.
\newblock {\em Proc. Nat. Acad. Sci.} {\bf 2015}, {\em 112},~12249--12255,  \href{http://arxiv.org/abs/1306.0913}{{\normalfont [arXiv:astro-ph.CO/1306.0913]}}.
\newblock {\url{https://doi.org/10.1073/pnas.1308716112}}.

\bibitem[Viel et~al.(2005)Viel, Lesgourgues, Haehnelt, Matarrese, and Riotto]{Viel:2005qj}
Viel, M.; Lesgourgues, J.; Haehnelt, M.G.; Matarrese, S.; Riotto, A.
\newblock {Constraining warm dark matter candidates including sterile neutrinos and light gravitinos with WMAP and the Lyman-alpha forest}.
\newblock {\em Phys. Rev. D} {\bf 2005}, {\em 71},~063534,  \href{http://arxiv.org/abs/astro-ph/0501562}{{\normalfont [astro-ph/0501562]}}.
\newblock {\url{https://doi.org/10.1103/PhysRevD.71.063534}}.

\bibitem[Newton et~al.(2021)Newton, Leo, Cautun, Jenkins, Frenk, Lovell, Helly, Benson, and Cole]{Newton:2020cog}
Newton, O.; Leo, M.; Cautun, M.; Jenkins, A.; Frenk, C.S.; Lovell, M.R.; Helly, J.C.; Benson, A.J.; Cole, S.
\newblock {Constraints on the properties of warm dark matter using the satellite galaxies of the Milky Way}.
\newblock {\em JCAP} {\bf 2021}, {\em 08},~062,  \href{http://arxiv.org/abs/2011.08865}{{\normalfont [arXiv:astro-ph.CO/2011.08865]}}.
\newblock {\url{https://doi.org/10.1088/1475-7516/2021/08/062}}.

\bibitem[Maartens(1996)]{Maartens:1996vi}
Maartens, R.
\newblock {Causal thermodynamics in relativity}.
\newblock  9 1996,  \href{http://arxiv.org/abs/astro-ph/9609119}{{\normalfont [astro-ph/9609119]}}.

\bibitem[Pandey et~al.(2020)Pandey, Karwal, and Das]{Pandey:2019plg}
Pandey, K.L.; Karwal, T.; Das, S.
\newblock {Alleviating the $H_0$ and $\sigma_8$ anomalies with a decaying dark matter model}.
\newblock {\em JCAP} {\bf 2020}, {\em 07},~026,  \href{http://arxiv.org/abs/1902.10636}{{\normalfont [arXiv:astro-ph.CO/1902.10636]}}.
\newblock {\url{https://doi.org/10.1088/1475-7516/2020/07/026}}.

\bibitem[Yang et~al.(2019)Yang, Pan, Di~Valentino, Paliathanasis, and Lu]{Yang:2019qza}
Yang, W.; Pan, S.; Di~Valentino, E.; Paliathanasis, A.; Lu, J.
\newblock {Challenging bulk viscous unified scenarios with cosmological observations}.
\newblock {\em Phys. Rev. D} {\bf 2019}, {\em 100},~103518,  \href{http://arxiv.org/abs/1906.04162}{{\normalfont [arXiv:astro-ph.CO/1906.04162]}}.
\newblock {\url{https://doi.org/10.1103/PhysRevD.100.103518}}.

\bibitem[Di~Valentino et~al.(2021)Di~Valentino, Mena, Pan, Visinelli, Yang, Melchiorri, Mota, Riess, and Silk]{DiValentino:2021izs}
Di~Valentino, E.; Mena, O.; Pan, S.; Visinelli, L.; Yang, W.; Melchiorri, A.; Mota, D.F.; Riess, A.G.; Silk, J.
\newblock {In the realm of the Hubble tension\textemdash{}a review of solutions}.
\newblock {\em Class. Quant. Grav.} {\bf 2021}, {\em 38},~153001,  \href{http://arxiv.org/abs/2103.01183}{{\normalfont [arXiv:astro-ph.CO/2103.01183]}}.
\newblock {\url{https://doi.org/10.1088/1361-6382/ac086d}}.

\bibitem[Normann and Brevik(2021)]{Normann:2021bjy}
Normann, B.D.; Brevik, I.H.
\newblock {Can the Hubble tension be resolved by bulk viscosity?}
\newblock {\em Mod. Phys. Lett. A} {\bf 2021}, {\em 36},~2150198,  \href{http://arxiv.org/abs/2107.13533}{{\normalfont [arXiv:gr-qc/2107.13533]}}.
\newblock {\url{https://doi.org/10.1142/S0217732321501984}}.

\bibitem[Brevik et~al.(2017)Brevik, Gr\o{}n, de~Haro, Odintsov, and Saridakis]{Brevik:2017msy}
Brevik, I.; Gr\o{}n, O.; de~Haro, J.; Odintsov, S.D.; Saridakis, E.N.
\newblock {Viscous Cosmology for Early- and Late-Time Universe}.
\newblock {\em Int. J. Mod. Phys. D} {\bf 2017}, {\em 26},~1730024,  \href{http://arxiv.org/abs/1706.02543}{{\normalfont [arXiv:gr-qc/1706.02543]}}.
\newblock {\url{https://doi.org/10.1142/S0218271817300245}}.

\bibitem[Eckart(1940{\natexlab{a}})]{Eckart:1940zz}
Eckart, C.
\newblock {The Thermodynamics of Irreversible Processes. 1. The Simple Fluid}.
\newblock {\em Phys. Rev.} {\bf 1940}, {\em 58},~267--269.
\newblock {\url{https://doi.org/10.1103/PhysRev.58.267}}.

\bibitem[Eckart(1940{\natexlab{b}})]{PhysRev.58.919}
Eckart, C.
\newblock The Thermodynamics of Irreversible Processes. III. Relativistic Theory of the Simple Fluid.
\newblock {\em Phys. Rev.} {\bf 1940}, {\em 58},~919--924.
\newblock {\url{https://doi.org/10.1103/PhysRev.58.919}}.

\bibitem[{Stewart}(1977)]{1977RSPSA.357...59S}
{Stewart}, J.M.
\newblock {On transient relativistic thermodynamics and kinetic theory}.
\newblock {\em Proc. R. Soc. A} {\bf 1977}, {\em 357},~59--75.
\newblock {\url{https://doi.org/10.1098/rspa.1977.0155}}.

\bibitem[{Israel} and {Stewart}(1979)]{1979RSPSA.365...43I}
{Israel}, W.; {Stewart}, J.M.
\newblock {On transient relativistic thermodynamics and kinetic theory. II}.
\newblock {\em Proc. R. Soc. A} {\bf 1979}, {\em 365},~43--52.
\newblock {\url{https://doi.org/10.1098/rspa.1979.0005}}.

\bibitem[Israel(1976)]{Israel:1976tn}
Israel, W.
\newblock {Nonstationary irreversible thermodynamics: A Causal relativistic theory}.
\newblock {\em Annals Phys.} {\bf 1976}, {\em 100},~310--331.
\newblock {\url{https://doi.org/10.1016/0003-4916(76)90064-6}}.

\bibitem[Maartens(1995)]{Maartens:1995wt}
Maartens, R.
\newblock {Dissipative cosmology}.
\newblock {\em Class. Quant. Grav.} {\bf 1995}, {\em 12},~1455--1465.
\newblock {\url{https://doi.org/10.1088/0264-9381/12/6/011}}.

\bibitem[Floerchinger et~al.(2015)Floerchinger, Tetradis, and Wiedemann]{Floerchinger:2014jsa}
Floerchinger, S.; Tetradis, N.; Wiedemann, U.A.
\newblock {Accelerating Cosmological Expansion from Shear and Bulk Viscosity}.
\newblock {\em Phys. Rev. Lett.} {\bf 2015}, {\em 114},~091301,  \href{http://arxiv.org/abs/1411.3280}{{\normalfont [arXiv:gr-qc/1411.3280]}}.
\newblock {\url{https://doi.org/10.1103/PhysRevLett.114.091301}}.

\bibitem[Zimdahl(1996)]{Zimdahl:1996fj}
Zimdahl, W.
\newblock {'Understanding' cosmological bulk viscosity}.
\newblock {\em Mon. Not. Roy. Astron. Soc.} {\bf 1996}, {\em 280},~1239,  \href{http://arxiv.org/abs/astro-ph/9602128}{{\normalfont [astro-ph/9602128]}}.
\newblock {\url{https://doi.org/10.1093/mnras/280.4.1239}}.

\bibitem[Wilson et~al.(2007)Wilson, Mathews, and Fuller]{Wilson:2006gf}
Wilson, J.R.; Mathews, G.J.; Fuller, G.M.
\newblock {Bulk Viscosity, Decaying Dark Matter, and the Cosmic Acceleration}.
\newblock {\em Phys. Rev. D} {\bf 2007}, {\em 75},~043521,  \href{http://arxiv.org/abs/astro-ph/0609687}{{\normalfont [astro-ph/0609687]}}.
\newblock {\url{https://doi.org/10.1103/PhysRevD.75.043521}}.

\bibitem[Mathews et~al.(2008)Mathews, Lan, and Kolda]{Mathews:2008hk}
Mathews, G.J.; Lan, N.Q.; Kolda, C.
\newblock {Late Decaying Dark Matter, Bulk Viscosity and the Cosmic Acceleration}.
\newblock {\em Phys. Rev. D} {\bf 2008}, {\em 78},~043525,  \href{http://arxiv.org/abs/0801.0853}{{\normalfont [arXiv:astro-ph/0801.0853]}}.
\newblock {\url{https://doi.org/10.1103/PhysRevD.78.043525}}.

\bibitem[Hofmann et~al.(2001)Hofmann, Schwarz, and Stoecker]{Hofmann:2001bi}
Hofmann, S.; Schwarz, D.J.; Stoecker, H.
\newblock {Damping scales of neutralino cold dark matter}.
\newblock {\em Phys. Rev. D} {\bf 2001}, {\em 64},~083507,  \href{http://arxiv.org/abs/astro-ph/0104173}{{\normalfont [astro-ph/0104173]}}.
\newblock {\url{https://doi.org/10.1103/PhysRevD.64.083507}}.

\bibitem[Foot and Vagnozzi(2015)]{Foot:2014uba}
Foot, R.; Vagnozzi, S.
\newblock {Dissipative hidden sector dark matter}.
\newblock {\em Phys. Rev. D} {\bf 2015}, {\em 91},~023512,  \href{http://arxiv.org/abs/1409.7174}{{\normalfont [arXiv:hep-ph/1409.7174]}}.
\newblock {\url{https://doi.org/10.1103/PhysRevD.91.023512}}.

\bibitem[Foot and Vagnozzi(2016)]{Foot:2016wvj}
Foot, R.; Vagnozzi, S.
\newblock {Solving the small-scale structure puzzles with dissipative dark matter}.
\newblock {\em JCAP} {\bf 2016}, {\em 07},~013,  \href{http://arxiv.org/abs/1602.02467}{{\normalfont [arXiv:astro-ph.CO/1602.02467]}}.
\newblock {\url{https://doi.org/10.1088/1475-7516/2016/07/013}}.

\bibitem[Goswami et~al.(2017)Goswami, Chakravarty, Mohanty, and Prasanna]{Goswami:2016tsu}
Goswami, G.; Chakravarty, G.K.; Mohanty, S.; Prasanna, A.R.
\newblock {Constraints on cosmological viscosity and self interacting dark matter from gravitational wave observations}.
\newblock {\em Phys. Rev. D} {\bf 2017}, {\em 95},~103509,  \href{http://arxiv.org/abs/1603.02635}{{\normalfont [arXiv:hep-ph/1603.02635]}}.
\newblock {\url{https://doi.org/10.1103/PhysRevD.95.103509}}.

\bibitem[Alford et~al.(2018)Alford, Bovard, Hanauske, Rezzolla, and Schwenzer]{Alford:2017rxf}
Alford, M.G.; Bovard, L.; Hanauske, M.; Rezzolla, L.; Schwenzer, K.
\newblock {Viscous Dissipation and Heat Conduction in Binary Neutron-Star Mergers}.
\newblock {\em Phys. Rev. Lett.} {\bf 2018}, {\em 120},~041101,  \href{http://arxiv.org/abs/1707.09475}{{\normalfont [arXiv:gr-qc/1707.09475]}}.
\newblock {\url{https://doi.org/10.1103/PhysRevLett.120.041101}}.

\bibitem[Nojiri and Odintsov(2004)]{Nojiri:2004pf}
Nojiri, S.; Odintsov, S.D.
\newblock {The Final state and thermodynamics of dark energy universe}.
\newblock {\em Phys. Rev. D} {\bf 2004}, {\em 70},~103522,  \href{http://arxiv.org/abs/hep-th/0408170}{{\normalfont [hep-th/0408170]}}.
\newblock {\url{https://doi.org/10.1103/PhysRevD.70.103522}}.

\bibitem[Capozziello et~al.(2006)Capozziello, Cardone, Elizalde, Nojiri, and Odintsov]{Capozziello:2005pa}
Capozziello, S.; Cardone, V.F.; Elizalde, E.; Nojiri, S.; Odintsov, S.D.
\newblock {Observational constraints on dark energy with generalized equations of state}.
\newblock {\em Phys. Rev. D} {\bf 2006}, {\em 73},~043512,  \href{http://arxiv.org/abs/astro-ph/0508350}{{\normalfont [astro-ph/0508350]}}.
\newblock {\url{https://doi.org/10.1103/PhysRevD.73.043512}}.

\bibitem[Nojiri and Odintsov(2005)]{Nojiri:2005sr}
Nojiri, S.; Odintsov, S.D.
\newblock {Inhomogeneous equation of state of the universe: Phantom era, future singularity and crossing the phantom barrier}.
\newblock {\em Phys. Rev. D} {\bf 2005}, {\em 72},~023003,  \href{http://arxiv.org/abs/hep-th/0505215}{{\normalfont [hep-th/0505215]}}.
\newblock {\url{https://doi.org/10.1103/PhysRevD.72.023003}}.

\bibitem[Cataldo et~al.(2005)Cataldo, Cruz, and Lepe]{Cataldo:2005qh}
Cataldo, M.; Cruz, N.; Lepe, S.
\newblock {Viscous dark energy and phantom evolution}.
\newblock {\em Phys. Lett. B} {\bf 2005}, {\em 619},~5--10,  \href{http://arxiv.org/abs/hep-th/0506153}{{\normalfont [hep-th/0506153]}}.
\newblock {\url{https://doi.org/10.1016/j.physletb.2005.05.029}}.

\bibitem[Brevik(2006)]{Brevik:2006wa}
Brevik, I.H.
\newblock {Crossing of the W = -1 barrier in two-fluid viscous modified gravity}.
\newblock {\em Gen. Rel. Grav.} {\bf 2006}, {\em 38},~1317--1328,  \href{http://arxiv.org/abs/gr-qc/0603025}{{\normalfont [gr-qc/0603025]}}.
\newblock {\url{https://doi.org/10.1007/s10714-006-0309-y}}.

\bibitem[Cruz et~al.(2017)Cruz, Lepe, and Pe\~na]{Cruz:2016rqi}
Cruz, N.; Lepe, S.; Pe\~na, F.
\newblock {Crossing the phantom divide with dissipative normal matter in the Israel\textendash{}Stewart formalism}.
\newblock {\em Phys. Lett. B} {\bf 2017}, {\em 767},~103--109,  \href{http://arxiv.org/abs/1607.04192}{{\normalfont [arXiv:gr-qc/1607.04192]}}.
\newblock {\url{https://doi.org/10.1016/j.physletb.2017.01.035}}.

\bibitem[Velten and Schwarz(2012)]{Velten:2012uv}
Velten, H.; Schwarz, D.
\newblock {Dissipation of dark matter}.
\newblock {\em Phys. Rev. D} {\bf 2012}, {\em 86},~083501,  \href{http://arxiv.org/abs/1206.0986}{{\normalfont [arXiv:astro-ph.CO/1206.0986]}}.
\newblock {\url{https://doi.org/10.1103/PhysRevD.86.083501}}.

\bibitem[Acquaviva and Beesham(2014)]{Acquaviva:2014vga}
Acquaviva, G.; Beesham, A.
\newblock {Nonlinear bulk viscosity and the stability of accelerated expansion in FRW spacetime}.
\newblock {\em Phys. Rev. D} {\bf 2014}, {\em 90},~023503,  \href{http://arxiv.org/abs/1405.3459}{{\normalfont [arXiv:gr-qc/1405.3459]}}.
\newblock {\url{https://doi.org/10.1103/PhysRevD.90.023503}}.

\bibitem[Cruz et~al.(2017)Cruz, Cruz, and Lepe]{Cruz:2017bcv}
Cruz, M.; Cruz, N.; Lepe, S.
\newblock {Accelerated and decelerated expansion in a causal dissipative cosmology}.
\newblock {\em Phys. Rev. D} {\bf 2017}, {\em 96},~124020,  \href{http://arxiv.org/abs/1710.02607}{{\normalfont [arXiv:gr-qc/1710.02607]}}.
\newblock {\url{https://doi.org/10.1103/PhysRevD.96.124020}}.

\bibitem[Cruz et~al.(2018)Cruz, Gonz\'alez, Lepe, and S\'aez-Chill\'on~G\'omez]{Cruz:2018yrr}
Cruz, N.; Gonz\'alez, E.; Lepe, S.; S\'aez-Chill\'on~G\'omez, D.
\newblock {Analysing dissipative effects in the $\Lambda$CDM model}.
\newblock {\em JCAP} {\bf 2018}, {\em 12},~017,  \href{http://arxiv.org/abs/1807.10729}{{\normalfont [arXiv:gr-qc/1807.10729]}}.
\newblock {\url{https://doi.org/10.1088/1475-7516/2018/12/017}}.

\bibitem[Cruz et~al.(2022)Cruz, Gonz\'alez, and Jovel]{Cruz:2022wme}
Cruz, N.; Gonz\'alez, E.; Jovel, J.
\newblock {Study of a Viscous \ensuremath{\Lambda}WDM Model: Near-Equilibrium Condition, Entropy Production, and Cosmological Constraints}.
\newblock {\em Symmetry} {\bf 2022}, {\em 14},~1866,  \href{http://arxiv.org/abs/2202.02362}{{\normalfont [arXiv:gr-qc/2202.02362]}}.
\newblock {\url{https://doi.org/10.3390/sym14091866}}.

\bibitem[Cruz et~al.(2023)Cruz, Gonz\'alez, and Jovel]{Cruz:2022zxe}
Cruz, N.; Gonz\'alez, E.; Jovel, J.
\newblock {A non-singular early-time viscous cosmological model}.
\newblock {\em Mod. Phys. Lett. A} {\bf 2023}, {\em 38},~2350088,  \href{http://arxiv.org/abs/2207.14244}{{\normalfont [arXiv:gr-qc/2207.14244]}}.
\newblock {\url{https://doi.org/10.1142/S0217732323500888}}.

\bibitem[G\'omez et~al.(2023)G\'omez, Palma, Gonz\'alez, Rinc\'on, and Cruz]{Gomez:2022qcu}
G\'omez, G.; Palma, G.; Gonz\'alez, E.; Rinc\'on, A.; Cruz, N.
\newblock {A new parametrization for bulk viscosity cosmology as extension of the $\Lambda $CDM model}.
\newblock {\em Eur. Phys. J. Plus} {\bf 2023}, {\em 138},~738,  \href{http://arxiv.org/abs/2210.09429}{{\normalfont [arXiv:gr-qc/2210.09429]}}.
\newblock {\url{https://doi.org/10.1140/epjp/s13360-023-04367-6}}.

\bibitem[Cruz et~al.(2023)Cruz, Gomez, Gonz\'alez, Palma, and Rincon]{Cruz:2023dzn}
Cruz, N.; Gomez, G.; Gonz\'alez, E.; Palma, G.; Rincon, A.
\newblock {Exploring models of running vacuum energy with viscous dark matter from a dynamical system perspective}.
\newblock {\em Phys. Dark Univ.} {\bf 2023}, {\em 42},~101351,  \href{http://arxiv.org/abs/2304.12407}{{\normalfont [arXiv:gr-qc/2304.12407]}}.
\newblock {\url{https://doi.org/10.1016/j.dark.2023.101351}}.

\bibitem[Fabris et~al.(2006)Fabris, Goncalves, and de~Sa~Ribeiro]{Fabris:2005ts}
Fabris, J.C.; Goncalves, S.V.B.; de~Sa~Ribeiro, R.
\newblock {Bulk viscosity driving the acceleration of the Universe}.
\newblock {\em Gen. Rel. Grav.} {\bf 2006}, {\em 38},~495--506,  \href{http://arxiv.org/abs/astro-ph/0503362}{{\normalfont [astro-ph/0503362]}}.
\newblock {\url{https://doi.org/10.1007/s10714-006-0236-y}}.

\bibitem[Avelino and Nucamendi(2009)]{Avelino:2008ph}
Avelino, A.; Nucamendi, U.
\newblock {Can a matter-dominated model with constant bulk viscosity drive the accelerated expansion of the universe?}
\newblock {\em JCAP} {\bf 2009}, {\em 04},~006,  \href{http://arxiv.org/abs/0811.3253}{{\normalfont [arXiv:gr-qc/0811.3253]}}.
\newblock {\url{https://doi.org/10.1088/1475-7516/2009/04/006}}.

\bibitem[Li and Barrow(2009)]{Li:2009mf}
Li, B.; Barrow, J.D.
\newblock {Does Bulk Viscosity Create a Viable Unified Dark Matter Model?}
\newblock {\em Phys. Rev. D} {\bf 2009}, {\em 79},~103521,  \href{http://arxiv.org/abs/0902.3163}{{\normalfont [arXiv:gr-qc/0902.3163]}}.
\newblock {\url{https://doi.org/10.1103/PhysRevD.79.103521}}.

\bibitem[Avelino and Nucamendi(2010)]{Avelino:2010pb}
Avelino, A.; Nucamendi, U.
\newblock {Exploring a matter-dominated model with bulk viscosity to drive the accelerated expansion of the Universe}.
\newblock {\em JCAP} {\bf 2010}, {\em 08},~009,  \href{http://arxiv.org/abs/1002.3605}{{\normalfont [arXiv:gr-qc/1002.3605]}}.
\newblock {\url{https://doi.org/10.1088/1475-7516/2010/08/009}}.

\bibitem[Hipolito-Ricaldi et~al.(2010)Hipolito-Ricaldi, Velten, and Zimdahl]{Hipolito-Ricaldi:2010wrq}
Hipolito-Ricaldi, W.S.; Velten, H.E.S.; Zimdahl, W.
\newblock {The Viscous Dark Fluid Universe}.
\newblock {\em Phys. Rev. D} {\bf 2010}, {\em 82},~063507,  \href{http://arxiv.org/abs/1007.0675}{{\normalfont [arXiv:astro-ph.CO/1007.0675]}}.
\newblock {\url{https://doi.org/10.1103/PhysRevD.82.063507}}.

\bibitem[Velten and Schwarz(2011)]{Velten_2011}
Velten, H.; Schwarz, D.J.
\newblock {Constraints on dissipative unified dark matter}.
\newblock {\em JCAP} {\bf 2011}, {\em 09},~016,  \href{http://arxiv.org/abs/1107.1143}{{\normalfont [arXiv:astro-ph.CO/1107.1143]}}.
\newblock {\url{https://doi.org/10.1088/1475-7516/2011/09/016}}.

\bibitem[Gagnon and Lesgourgues(2011)]{Gagnon:2011id}
Gagnon, J.S.; Lesgourgues, J.
\newblock {Dark goo: Bulk viscosity as an alternative to dark energy}.
\newblock {\em JCAP} {\bf 2011}, {\em 09},~026,  \href{http://arxiv.org/abs/1107.1503}{{\normalfont [arXiv:astro-ph.CO/1107.1503]}}.
\newblock {\url{https://doi.org/10.1088/1475-7516/2011/09/026}}.

\bibitem[Bruni et~al.(2013)Bruni, Lazkoz, and Rozas-Fernandez]{Bruni:2012sn}
Bruni, M.; Lazkoz, R.; Rozas-Fernandez, A.
\newblock {Phenomenological models for Unified Dark Matter with fast transition}.
\newblock {\em Mon. Not. Roy. Astron. Soc.} {\bf 2013}, {\em 431},~2907--2916,  \href{http://arxiv.org/abs/1210.1880}{{\normalfont [arXiv:astro-ph.CO/1210.1880]}}.
\newblock {\url{https://doi.org/10.1093/mnras/stt395}}.

\bibitem[Cruz et~al.(2017)Cruz, Cruz, and Lepe]{Cruz:2017lbu}
Cruz, M.; Cruz, N.; Lepe, S.
\newblock {Phantom solution in a non-linear Israel\textendash{}Stewart theory}.
\newblock {\em Phys. Lett. B} {\bf 2017}, {\em 769},~159--165,  \href{http://arxiv.org/abs/1701.06724}{{\normalfont [arXiv:gr-qc/1701.06724]}}.
\newblock {\url{https://doi.org/10.1016/j.physletb.2017.03.065}}.

\bibitem[Cruz et~al.(2020)Cruz, Gonz\'alez, and Palma]{Cruz:2018psw}
Cruz, N.; Gonz\'alez, E.; Palma, G.
\newblock {Exact analytical solution for an Israel\textendash{}Stewart cosmology}.
\newblock {\em Gen. Rel. Grav.} {\bf 2020}, {\em 52},~62,  \href{http://arxiv.org/abs/1812.05009}{{\normalfont [arXiv:gr-qc/1812.05009]}}.
\newblock {\url{https://doi.org/10.1007/s10714-020-02712-z}}.

\bibitem[Cruz et~al.(2021)Cruz, Gonz\'alez, and Palma]{Cruz:2019uya}
Cruz, N.; Gonz\'alez, E.; Palma, G.
\newblock {Testing dissipative dark matter in causal thermodynamics}.
\newblock {\em Mod. Phys. Lett. A} {\bf 2021}, {\em 36},~2150032,  \href{http://arxiv.org/abs/1906.04570}{{\normalfont [arXiv:gr-qc/1906.04570]}}.
\newblock {\url{https://doi.org/10.1142/S0217732321500322}}.

\bibitem[Mak and Harko(2004)]{Mak:2003gw}
Mak, M.K.; Harko, T.
\newblock {Full causal dissipative cosmologies with stiff matter}.
\newblock {\em Int. J. Mod. Phys. D} {\bf 2004}, {\em 13},~273--280,  \href{http://arxiv.org/abs/gr-qc/0311050}{{\normalfont [gr-qc/0311050]}}.
\newblock {\url{https://doi.org/10.1142/S0218271804004645}}.

\bibitem[Padmanabhan and Chitre(1987)]{Padmanabhan:1987dg}
Padmanabhan, T.; Chitre, S.M.
\newblock {Viscous universes}.
\newblock {\em Phys. Lett. A} {\bf 1987}, {\em 120},~433--436.
\newblock {\url{https://doi.org/10.1016/0375-9601(87)90104-6}}.

\bibitem[Barrow(1988)]{Barrow:1988yc}
Barrow, J.D.
\newblock {String-Driven Inflationary and Deflationary Cosmological Models}.
\newblock {\em Nucl. Phys. B} {\bf 1988}, {\em 310},~743--763.
\newblock {\url{https://doi.org/10.1016/0550-3213(88)90101-0}}.

\bibitem[Maartens and Mendez(1997)]{Maartens:1996dk}
Maartens, R.; Mendez, V.
\newblock {Nonlinear bulk viscosity and inflation}.
\newblock {\em Phys. Rev. D} {\bf 1997}, {\em 55},~1937--1942,  \href{http://arxiv.org/abs/astro-ph/9611205}{{\normalfont [astro-ph/9611205]}}.
\newblock {\url{https://doi.org/10.1103/PhysRevD.55.1937}}.

\bibitem[Avelino et~al.(2013)Avelino, Leyva, and Urena-Lopez]{Avelino:2013wea}
Avelino, A.; Leyva, Y.; Urena-Lopez, L.A.
\newblock {Interacting viscous dark fluids}.
\newblock {\em Phys. Rev. D} {\bf 2013}, {\em 88},~123004,  \href{http://arxiv.org/abs/1306.3270}{{\normalfont [arXiv:astro-ph.CO/1306.3270]}}.
\newblock {\url{https://doi.org/10.1103/PhysRevD.88.123004}}.

\bibitem[Hern\'andez-Almada et~al.(2020)Hern\'andez-Almada, Garc\'\i{}a-Aspeitia, Maga\~na, and Motta]{Hernandez-Almada:2020ulm}
Hern\'andez-Almada, A.; Garc\'\i{}a-Aspeitia, M.A.; Maga\~na, J.; Motta, V.
\newblock {Stability analysis and constraints on interacting viscous cosmology}.
\newblock {\em Phys. Rev. D} {\bf 2020}, {\em 101},~063516,  \href{http://arxiv.org/abs/2001.08667}{{\normalfont [arXiv:astro-ph.CO/2001.08667]}}.
\newblock {\url{https://doi.org/10.1103/PhysRevD.101.063516}}.

\bibitem[Brevik et~al.(2005)Brevik, Gorbunova, and Shaido]{Brevik:2005ue}
Brevik, I.H.; Gorbunova, O.; Shaido, Y.A.
\newblock {Viscous FRW cosmology in modified gravity}.
\newblock {\em Int. J. Mod. Phys. D} {\bf 2005}, {\em 14},~1899--1906,  \href{http://arxiv.org/abs/gr-qc/0508038}{{\normalfont [gr-qc/0508038]}}.
\newblock {\url{https://doi.org/10.1142/S0218271805007450}}.

\bibitem[Brevik and Gorbunova(2005)]{Brevik:2005bj}
Brevik, I.H.; Gorbunova, O.
\newblock {Dark energy and viscous cosmology}.
\newblock {\em Gen. Rel. Grav.} {\bf 2005}, {\em 37},~2039--2045,  \href{http://arxiv.org/abs/gr-qc/0504001}{{\normalfont [gr-qc/0504001]}}.
\newblock {\url{https://doi.org/10.1007/s10714-005-0178-9}}.

\bibitem[Brevik and Gorbunova(2008)]{Brevik:2008xv}
Brevik, I.H.; Gorbunova, O.
\newblock {Viscous Dark Cosmology with Account of Quantum Effects}.
\newblock {\em Eur. Phys. J. C} {\bf 2008}, {\em 56},~425--428,  \href{http://arxiv.org/abs/0806.1399}{{\normalfont [arXiv:gr-qc/0806.1399]}}.
\newblock {\url{https://doi.org/10.1140/epjc/s10052-008-0664-9}}.

\bibitem[Brevik et~al.(2010{\natexlab{a}})Brevik, Gorbunova, and Saez-Gomez]{Brevik:2010okp}
Brevik, I.; Gorbunova, O.; Saez-Gomez, D.
\newblock {Casimir Effects Near the Big Rip Singularity in Viscous Cosmology}.
\newblock {\em Gen. Rel. Grav.} {\bf 2010}, {\em 42},~1513--1522,  \href{http://arxiv.org/abs/0908.2882}{{\normalfont [arXiv:gr-qc/0908.2882]}}.
\newblock {\url{https://doi.org/10.1007/s10714-009-0923-6}}.

\bibitem[Brevik et~al.(2010{\natexlab{b}})Brevik, Nojiri, Odintsov, and Saez-Gomez]{Brevik:2010jv}
Brevik, I.; Nojiri, S.; Odintsov, S.D.; Saez-Gomez, D.
\newblock {Cardy-Verlinde formula in FRW Universe with inhomogeneous generalized fluid and dynamical entropy bounds near the future singularity}.
\newblock {\em Eur. Phys. J. C} {\bf 2010}, {\em 69},~563--574,  \href{http://arxiv.org/abs/1002.1942}{{\normalfont [arXiv:hep-th/1002.1942]}}.
\newblock {\url{https://doi.org/10.1140/epjc/s10052-010-1425-0}}.

\bibitem[Brevik et~al.(2011)Brevik, Elizalde, Nojiri, and Odintsov]{Brevik:2011mm}
Brevik, I.; Elizalde, E.; Nojiri, S.; Odintsov, S.D.
\newblock {Viscous Little Rip Cosmology}.
\newblock {\em Phys. Rev. D} {\bf 2011}, {\em 84},~103508,  \href{http://arxiv.org/abs/1107.4642}{{\normalfont [arXiv:hep-th/1107.4642]}}.
\newblock {\url{https://doi.org/10.1103/PhysRevD.84.103508}}.

\bibitem[Contreras et~al.(2016)Contreras, Cruz, and Gonz\'alez]{Contreras:2015ooa}
Contreras, F.; Cruz, N.; Gonz\'alez, E.
\newblock {Generalized equations of state and regular universes}.
\newblock {\em J. Phys. Conf. Ser.} {\bf 2016}, {\em 720},~012014,  \href{http://arxiv.org/abs/1510.02782}{{\normalfont [arXiv:gr-qc/1510.02782]}}.
\newblock {\url{https://doi.org/10.1088/1742-6596/720/1/012014}}.

\bibitem[Contreras et~al.(2018)Contreras, Cruz, Elizalde, Gonz\'alez, and Odintsov]{Contreras:2018two}
Contreras, F.; Cruz, N.; Elizalde, E.; Gonz\'alez, E.; Odintsov, S.
\newblock {Linking little rip cosmologies with regular early universes}.
\newblock {\em Phys. Rev. D} {\bf 2018}, {\em 98},~123520,  \href{http://arxiv.org/abs/1808.06546}{{\normalfont [arXiv:gr-qc/1808.06546]}}.
\newblock {\url{https://doi.org/10.1103/PhysRevD.98.123520}}.

\bibitem[Cruz et~al.(2022)Cruz, Gonz\'alez, and Jovel]{Cruz:2021knz}
Cruz, N.; Gonz\'alez, E.; Jovel, J.
\newblock {Singularities and soft-Big Bang in a viscous \ensuremath{\Lambda}CDM model}.
\newblock {\em Phys. Rev. D} {\bf 2022}, {\em 105},~024047,  \href{http://arxiv.org/abs/2109.09865}{{\normalfont [arXiv:gr-qc/2109.09865]}}.
\newblock {\url{https://doi.org/10.1103/PhysRevD.105.024047}}.

\bibitem[Barta(2019)]{Barta:2019tpv}
Barta, D.
\newblock {Effect of viscosity and thermal conductivity on the radial oscillation and relaxation of relativistic stars}.
\newblock {\em Class. Quant. Grav.} {\bf 2019}, {\em 36},~215012,  \href{http://arxiv.org/abs/1904.00907}{{\normalfont [arXiv:gr-qc/1904.00907]}}.
\newblock {\url{https://doi.org/10.1088/1361-6382/ab472e}}.

\bibitem[Bravo~Medina et~al.(2019)Bravo~Medina, Nowakowski, and Batic]{BravoMedina:2019han}
Bravo~Medina, S.; Nowakowski, M.; Batic, D.
\newblock {Viscous Cosmologies}.
\newblock {\em Class. Quant. Grav.} {\bf 2019}, {\em 36},~215002,  \href{http://arxiv.org/abs/1901.09787}{{\normalfont [arXiv:gr-qc/1901.09787]}}.
\newblock {\url{https://doi.org/10.1088/1361-6382/ab45bb}}.

\bibitem[Gonz\'alez et~al.(2024)Gonz\'alez, Maldonado, Mite, and Salinas]{Gonzalez:2024dtb}
Gonz\'alez, E.; Maldonado, C.; Mite, N.S.; Salinas, R.
\newblock {WIMP Dark Matter in bulk viscous non-standard cosmologies}.
\newblock {\em JCAP} {\bf 2024}, {\em 10},~088,  \href{http://arxiv.org/abs/2409.03083}{{\normalfont [arXiv:hep-ph/2409.03083]}}.
\newblock {\url{https://doi.org/10.1088/1475-7516/2024/10/088}}.

\bibitem[Giudice et~al.(2001)Giudice, Kolb, and Riotto]{Giudice:2000ex}
Giudice, G.F.; Kolb, E.W.; Riotto, A.
\newblock {Largest temperature of the radiation era and its cosmological implications}.
\newblock {\em Phys. Rev. D} {\bf 2001}, {\em 64},~023508,  \href{http://arxiv.org/abs/hep-ph/0005123}{{\normalfont [hep-ph/0005123]}}.
\newblock {\url{https://doi.org/10.1103/PhysRevD.64.023508}}.

\bibitem[Hamdan and Unwin(2018)]{Hamdan:2017psw}
Hamdan, S.; Unwin, J.
\newblock {Dark Matter Freeze-out During Matter Domination}.
\newblock {\em Mod. Phys. Lett. A} {\bf 2018}, {\em 33},~1850181,  \href{http://arxiv.org/abs/1710.03758}{{\normalfont [arXiv:hep-ph/1710.03758]}}.
\newblock {\url{https://doi.org/10.1142/S021773231850181X}}.

\bibitem[D'Eramo et~al.(2018)D'Eramo, Fernandez, and Profumo]{DEramo:2017ecx}
D'Eramo, F.; Fernandez, N.; Profumo, S.
\newblock {Dark Matter Freeze-in Production in Fast-Expanding Universes}.
\newblock {\em JCAP} {\bf 2018}, {\em 02},~046,  \href{http://arxiv.org/abs/1712.07453}{{\normalfont [arXiv:hep-ph/1712.07453]}}.
\newblock {\url{https://doi.org/10.1088/1475-7516/2018/02/046}}.

\bibitem[D'Eramo et~al.(2017)D'Eramo, Fernandez, and Profumo]{DEramo:2017gpl}
D'Eramo, F.; Fernandez, N.; Profumo, S.
\newblock {When the Universe Expands Too Fast: Relentless Dark Matter}.
\newblock {\em JCAP} {\bf 2017}, {\em 05},~012,  \href{http://arxiv.org/abs/1703.04793}{{\normalfont [arXiv:hep-ph/1703.04793]}}.
\newblock {\url{https://doi.org/10.1088/1475-7516/2017/05/012}}.

\bibitem[Visinelli(2018)]{Visinelli:2017qga}
Visinelli, L.
\newblock {(Non-)thermal production of WIMPs during kination}.
\newblock {\em Symmetry} {\bf 2018}, {\em 10},~546,  \href{http://arxiv.org/abs/1710.11006}{{\normalfont [arXiv:astro-ph.CO/1710.11006]}}.
\newblock {\url{https://doi.org/10.3390/sym10110546}}.

\bibitem[Drees and Hajkarim(2018)]{Drees:2018dsj}
Drees, M.; Hajkarim, F.
\newblock {Neutralino Dark Matter in Scenarios with Early Matter Domination}.
\newblock {\em JHEP} {\bf 2018}, {\em 12},~042,  \href{http://arxiv.org/abs/1808.05706}{{\normalfont [arXiv:hep-ph/1808.05706]}}.
\newblock {\url{https://doi.org/10.1007/JHEP12(2018)042}}.

\bibitem[Bernal et~al.(2019{\natexlab{a}})Bernal, Cosme, and Tenkanen]{Bernal:2018ins}
Bernal, N.; Cosme, C.; Tenkanen, T.
\newblock {Phenomenology of Self-Interacting Dark Matter in a Matter-Dominated Universe}.
\newblock {\em Eur. Phys. J. C} {\bf 2019}, {\em 79},~99,  \href{http://arxiv.org/abs/1803.08064}{{\normalfont [arXiv:hep-ph/1803.08064]}}.
\newblock {\url{https://doi.org/10.1140/epjc/s10052-019-6608-8}}.

\bibitem[Bernal et~al.(2019{\natexlab{b}})Bernal, Cosme, Tenkanen, and Vaskonen]{Bernal:2018kcw}
Bernal, N.; Cosme, C.; Tenkanen, T.; Vaskonen, V.
\newblock {Scalar singlet dark matter in non-standard cosmologies}.
\newblock {\em Eur. Phys. J. C} {\bf 2019}, {\em 79},~30,  \href{http://arxiv.org/abs/1806.11122}{{\normalfont [arXiv:hep-ph/1806.11122]}}.
\newblock {\url{https://doi.org/10.1140/epjc/s10052-019-6550-9}}.

\bibitem[Maldonado and Unwin(2019)]{Maldonado:2019qmp}
Maldonado, C.; Unwin, J.
\newblock {Establishing the Dark Matter Relic Density in an Era of Particle Decays}.
\newblock {\em JCAP} {\bf 2019}, {\em 06},~037,  \href{http://arxiv.org/abs/1902.10746}{{\normalfont [arXiv:hep-ph/1902.10746]}}.
\newblock {\url{https://doi.org/10.1088/1475-7516/2019/06/037}}.

\bibitem[Arias et~al.(2019)Arias, Bernal, Herrera, and Maldonado]{Arias:2019uol}
Arias, P.; Bernal, N.; Herrera, A.; Maldonado, C.
\newblock {Reconstructing Non-standard Cosmologies with Dark Matter}.
\newblock {\em JCAP} {\bf 2019}, {\em 10},~047,  \href{http://arxiv.org/abs/1906.04183}{{\normalfont [arXiv:hep-ph/1906.04183]}}.
\newblock {\url{https://doi.org/10.1088/1475-7516/2019/10/047}}.

\bibitem[Bernal et~al.(2019)Bernal, Elahi, Maldonado, and Unwin]{Bernal:2019mhf}
Bernal, N.; Elahi, F.; Maldonado, C.; Unwin, J.
\newblock {Ultraviolet Freeze-in and Non-Standard Cosmologies}.
\newblock {\em JCAP} {\bf 2019}, {\em 11},~026,  \href{http://arxiv.org/abs/1909.07992}{{\normalfont [arXiv:hep-ph/1909.07992]}}.
\newblock {\url{https://doi.org/10.1088/1475-7516/2019/11/026}}.

\bibitem[Arias et~al.(2021)Arias, Bernal, Karamitros, Maldonado, Roszkowski, and Venegas]{Arias:2021rer}
Arias, P.; Bernal, N.; Karamitros, D.; Maldonado, C.; Roszkowski, L.; Venegas, M.
\newblock {New opportunities for axion dark matter searches in nonstandard cosmological models}.
\newblock {\em JCAP} {\bf 2021}, {\em 11},~003,  \href{http://arxiv.org/abs/2107.13588}{{\normalfont [arXiv:hep-ph/2107.13588]}}.
\newblock {\url{https://doi.org/10.1088/1475-7516/2021/11/003}}.

\bibitem[Bernal and Xu(2022)]{Bernal:2022wck}
Bernal, N.; Xu, Y.
\newblock {WIMPs during reheating}.
\newblock {\em JCAP} {\bf 2022}, {\em 12},~017,  \href{http://arxiv.org/abs/2209.07546}{{\normalfont [arXiv:hep-ph/2209.07546]}}.
\newblock {\url{https://doi.org/10.1088/1475-7516/2022/12/017}}.

\bibitem[Silva-Malpartida et~al.(2024)Silva-Malpartida, Bernal, Jones-P\'erez, and Lineros]{Silva-Malpartida:2024emu}
Silva-Malpartida, J.; Bernal, N.; Jones-P\'erez, J.; Lineros, R.A.
\newblock {From WIMPs to FIMPs: Impact of Early Matter Domination},  2024,  \href{http://arxiv.org/abs/2408.08950}{{\normalfont [arXiv:hep-ph/2408.08950]}}.

\bibitem[Drees and Hajkarim(2018)]{Drees:2017iod}
Drees, M.; Hajkarim, F.
\newblock {Dark Matter Production in an Early Matter Dominated Era}.
\newblock {\em JCAP} {\bf 2018}, {\em 02},~057,  \href{http://arxiv.org/abs/1711.05007}{{\normalfont [arXiv:hep-ph/1711.05007]}}.
\newblock {\url{https://doi.org/10.1088/1475-7516/2018/02/057}}.

\bibitem[Sarkar(1996)]{Sarkar:1995dd}
Sarkar, S.
\newblock {Big bang nucleosynthesis and physics beyond the standard model}.
\newblock {\em Rept. Prog. Phys.} {\bf 1996}, {\em 59},~1493--1610,  \href{http://arxiv.org/abs/hep-ph/9602260}{{\normalfont [hep-ph/9602260]}}.
\newblock {\url{https://doi.org/10.1088/0034-4885/59/12/001}}.

\bibitem[Hannestad(2004)]{Hannestad:2004px}
Hannestad, S.
\newblock {What is the lowest possible reheating temperature?}
\newblock {\em Phys. Rev. D} {\bf 2004}, {\em 70},~043506,  \href{http://arxiv.org/abs/astro-ph/0403291}{{\normalfont [astro-ph/0403291]}}.
\newblock {\url{https://doi.org/10.1103/PhysRevD.70.043506}}.

\bibitem[De~Bernardis et~al.(2008)De~Bernardis, Pagano, and Melchiorri]{DeBernardis:2008zz}
De~Bernardis, F.; Pagano, L.; Melchiorri, A.
\newblock {New constraints on the reheating temperature of the universe after WMAP-5}.
\newblock {\em Astropart. Phys.} {\bf 2008}, {\em 30},~192--195.
\newblock {\url{https://doi.org/10.1016/j.astropartphys.2008.09.005}}.

\bibitem[Kofman et~al.(1997)Kofman, Linde, and Starobinsky]{Kofman:1997yn}
Kofman, L.; Linde, A.D.; Starobinsky, A.A.
\newblock {Towards the theory of reheating after inflation}.
\newblock {\em Phys. Rev. D} {\bf 1997}, {\em 56},~3258--3295,  \href{http://arxiv.org/abs/hep-ph/9704452}{{\normalfont [hep-ph/9704452]}}.
\newblock {\url{https://doi.org/10.1103/PhysRevD.56.3258}}.

\bibitem[Ackerman et~al.(2011)Ackerman, Fischler, Kundu, and Sivanandam]{Ackerman:2010he}
Ackerman, L.; Fischler, W.; Kundu, S.; Sivanandam, N.
\newblock {Constraining the Inflationary Equation of State}.
\newblock {\em JCAP} {\bf 2011}, {\em 05},~024,  \href{http://arxiv.org/abs/1007.3511}{{\normalfont [arXiv:astro-ph.CO/1007.3511]}}.
\newblock {\url{https://doi.org/10.1088/1475-7516/2011/05/024}}.

\bibitem[Garcia et~al.(2020)Garcia, Kaneta, Mambrini, and Olive]{Garcia:2020eof}
Garcia, M.A.G.; Kaneta, K.; Mambrini, Y.; Olive, K.A.
\newblock {Reheating and Post-inflationary Production of Dark Matter}.
\newblock {\em Phys. Rev. D} {\bf 2020}, {\em 101},~123507,  \href{http://arxiv.org/abs/2004.08404}{{\normalfont [arXiv:hep-ph/2004.08404]}}.
\newblock {\url{https://doi.org/10.1103/PhysRevD.101.123507}}.

\bibitem[Garcia et~al.(2021)Garcia, Kaneta, Mambrini, and Olive]{Garcia:2020wiy}
Garcia, M.A.G.; Kaneta, K.; Mambrini, Y.; Olive, K.A.
\newblock {Inflaton Oscillations and Post-Inflationary Reheating}.
\newblock {\em JCAP} {\bf 2021}, {\em 04},~012,  \href{http://arxiv.org/abs/2012.10756}{{\normalfont [arXiv:hep-ph/2012.10756]}}.
\newblock {\url{https://doi.org/10.1088/1475-7516/2021/04/012}}.

\bibitem[Silva-Malpartida et~al.(2023)Silva-Malpartida, Bernal, Jones-P\'erez, and Lineros]{Silva-Malpartida:2023yks}
Silva-Malpartida, J.; Bernal, N.; Jones-P\'erez, J.; Lineros, R.A.
\newblock {From WIMPs to FIMPs with low~reheating~temperatures}.
\newblock {\em JCAP} {\bf 2023}, {\em 09},~015,  \href{http://arxiv.org/abs/2306.14943}{{\normalfont [arXiv:hep-ph/2306.14943]}}.
\newblock {\url{https://doi.org/10.1088/1475-7516/2023/09/015}}.

\bibitem[Haque et~al.(2024)Haque, Maity, and Mondal]{Haque:2024zdq}
Haque, M.R.; Maity, D.; Mondal, R.
\newblock {Thermal and non-thermal dark matters with gravitational neutrino reheating},  2024,  \href{http://arxiv.org/abs/2408.12450}{{\normalfont [arXiv:hep-ph/2408.12450]}}.

\bibitem[Gelmini and Gondolo(2006)]{Gelmini:2006pw}
Gelmini, G.B.; Gondolo, P.
\newblock {Neutralino with the right cold dark matter abundance in (almost) any supersymmetric model}.
\newblock {\em Phys. Rev. D} {\bf 2006}, {\em 74},~023510,  \href{http://arxiv.org/abs/hep-ph/0602230}{{\normalfont [hep-ph/0602230]}}.
\newblock {\url{https://doi.org/10.1103/PhysRevD.74.023510}}.

\bibitem[Gelmini et~al.(2006)Gelmini, Gondolo, Soldatenko, and Yaguna]{Gelmini:2006pq}
Gelmini, G.; Gondolo, P.; Soldatenko, A.; Yaguna, C.E.
\newblock {The Effect of a late decaying scalar on the neutralino relic density}.
\newblock {\em Phys. Rev. D} {\bf 2006}, {\em 74},~083514,  \href{http://arxiv.org/abs/hep-ph/0605016}{{\normalfont [hep-ph/0605016]}}.
\newblock {\url{https://doi.org/10.1103/PhysRevD.74.083514}}.

\bibitem[Freese et~al.(2024)Freese, Montefalcone, and Shams Es~Haghi]{Freese:2024ogj}
Freese, K.; Montefalcone, G.; Shams Es~Haghi, B.
\newblock {Dark Matter production during Warm Inflation via Freeze-In},  2024,  \href{http://arxiv.org/abs/2401.17371}{{\normalfont [arXiv:hep-ph/2401.17371]}}.

\bibitem[Chowdhury et~al.(2019)Chowdhury, Dudas, Dutra, and Mambrini]{Chowdhury:2018tzw}
Chowdhury, D.; Dudas, E.; Dutra, M.; Mambrini, Y.
\newblock {Moduli Portal Dark Matter}.
\newblock {\em Phys. Rev. D} {\bf 2019}, {\em 99},~095028,  \href{http://arxiv.org/abs/1811.01947}{{\normalfont [arXiv:hep-ph/1811.01947]}}.
\newblock {\url{https://doi.org/10.1103/PhysRevD.99.095028}}.

\bibitem[Elahi et~al.(2015)Elahi, Kolda, and Unwin]{Elahi:2014fsa}
Elahi, F.; Kolda, C.; Unwin, J.
\newblock {UltraViolet Freeze-in}.
\newblock {\em JHEP} {\bf 2015}, {\em 03},~048,  \href{http://arxiv.org/abs/1410.6157}{{\normalfont [arXiv:hep-ph/1410.6157]}}.
\newblock {\url{https://doi.org/10.1007/JHEP03(2015)048}}.

\bibitem[Biswas et~al.(2020)Biswas, Ganguly, and Roy]{Biswas:2019iqm}
Biswas, A.; Ganguly, S.; Roy, S.
\newblock {Fermionic dark matter via UV and IR freeze-in and its possible X-ray signature}.
\newblock {\em JCAP} {\bf 2020}, {\em 03},~043,  \href{http://arxiv.org/abs/1907.07973}{{\normalfont [arXiv:hep-ph/1907.07973]}}.
\newblock {\url{https://doi.org/10.1088/1475-7516/2020/03/043}}.

\bibitem[Shakya(2016)]{Shakya:2015xnx}
Shakya, B.
\newblock {Sterile Neutrino Dark Matter from Freeze-In}.
\newblock {\em Mod. Phys. Lett. A} {\bf 2016}, {\em 31},~1630005,  \href{http://arxiv.org/abs/1512.02751}{{\normalfont [arXiv:hep-ph/1512.02751]}}.
\newblock {\url{https://doi.org/10.1142/S0217732316300056}}.

\bibitem[McDonald and Sahu(2009)]{McDonald:2008ua}
McDonald, J.; Sahu, N.
\newblock {keV Warm Dark Matter via the Supersymmetric Higgs Portal}.
\newblock {\em Phys. Rev. D} {\bf 2009}, {\em 79},~103523,  \href{http://arxiv.org/abs/0809.0247}{{\normalfont [arXiv:hep-ph/0809.0247]}}.
\newblock {\url{https://doi.org/10.1103/PhysRevD.79.103523}}.

\bibitem[Lebedev et~al.(2012)Lebedev, Lee, and Mambrini]{Lebedev:2011iq}
Lebedev, O.; Lee, H.M.; Mambrini, Y.
\newblock {Vector Higgs-portal dark matter and the invisible Higgs}.
\newblock {\em Phys. Lett. B} {\bf 2012}, {\em 707},~570--576,  \href{http://arxiv.org/abs/1111.4482}{{\normalfont [arXiv:hep-ph/1111.4482]}}.
\newblock {\url{https://doi.org/10.1016/j.physletb.2012.01.029}}.

\bibitem[Bernal et~al.(2018)Bernal, Dutra, Mambrini, Olive, Peloso, and Pierre]{Bernal:2018qlk}
Bernal, N.; Dutra, M.; Mambrini, Y.; Olive, K.; Peloso, M.; Pierre, M.
\newblock {Spin-2 Portal Dark Matter}.
\newblock {\em Phys. Rev. D} {\bf 2018}, {\em 97},~115020,  \href{http://arxiv.org/abs/1803.01866}{{\normalfont [arXiv:hep-ph/1803.01866]}}.
\newblock {\url{https://doi.org/10.1103/PhysRevD.97.115020}}.

\bibitem[Krauss et~al.(2014)Krauss, Morisi, Porod, and Winter]{Krauss:2013wfa}
Krauss, M.B.; Morisi, S.; Porod, W.; Winter, W.
\newblock {Higher Dimensional Effective Operators for Direct Dark Matter Detection}.
\newblock {\em JHEP} {\bf 2014}, {\em 02},~056,  \href{http://arxiv.org/abs/1312.0009}{{\normalfont [arXiv:hep-ph/1312.0009]}}.
\newblock {\url{https://doi.org/10.1007/JHEP02(2014)056}}.

\bibitem[Barman et~al.(2022)Barman, Bernal, Xu, and Zapata]{Barman:2022tzk}
Barman, B.; Bernal, N.; Xu, Y.; Zapata, O.
\newblock {Ultraviolet freeze-in with a time-dependent inflaton decay}.
\newblock {\em JCAP} {\bf 2022}, {\em 07},~019,  \href{http://arxiv.org/abs/2202.12906}{{\normalfont [arXiv:hep-ph/2202.12906]}}.
\newblock {\url{https://doi.org/10.1088/1475-7516/2022/07/019}}.

\bibitem[Bernal et~al.(2025)Bernal, Deka, and Losada]{Bernal:2025fdr}
Bernal, N.; Deka, K.; Losada, M.
\newblock {Dark Matter Ultraviolet Freeze-in in General Reheating Scenarios},  2025,  \href{http://arxiv.org/abs/2501.04774}{{\normalfont [arXiv:hep-ph/2501.04774]}}.

\bibitem[Weinberg and Dicke()]{SWeinberg}
Weinberg, S.; Dicke, R.H.
\newblock {\em Gravitation and Cosmology: Principles and Applications of the General Theory of Relativity}.

\bibitem[Barranco et~al.(2018)Barranco, Bernal, and Delepine]{Barranco:2017lug}
Barranco, J.; Bernal, A.; Delepine, D.
\newblock {Diffuse neutrino supernova background as a cosmological test}.
\newblock {\em J. Phys. G} {\bf 2018}, {\em 45},~055201,  \href{http://arxiv.org/abs/1706.03834}{{\normalfont [arXiv:astro-ph.CO/1706.03834]}}.
\newblock {\url{https://doi.org/10.1088/1361-6471/aab8ae}}.

\bibitem[Lambiase et~al.(2018)Lambiase, Mohanty, and Stabile]{Lambiase:2018yql}
Lambiase, G.; Mohanty, S.; Stabile, A.
\newblock {PeV IceCube signals and Dark Matter relic abundance in modified cosmologies}.
\newblock {\em Eur. Phys. J. C} {\bf 2018}, {\em 78},~350,  \href{http://arxiv.org/abs/1804.07369}{{\normalfont [arXiv:astro-ph.CO/1804.07369]}}.
\newblock {\url{https://doi.org/10.1140/epjc/s10052-018-5821-1}}.

\end{thebibliography}

\isPreprints{}{
\end{adjustwidth}
} 
\end{document}